\documentclass[letterpaper,aps,prd,twocolumn,tightenlines,preprintnumbers,nofootinbib,showkeys,superscriptaddress,longbibliography]{revtex4-1}

\usepackage{lipsum} %%%%%%%%%%%%%%%%%%%%%%%%%%%%%%%%%%%%%%%%%%%%%%%%%%%%%%%%%%%%%%%%%%

\usepackage{XCharter}
\usepackage[T1]{fontenc}
\usepackage{bm}
\usepackage{bbold}
\usepackage{pifont}
\usepackage{rotating}
\usepackage{fullpage}
\usepackage{amsfonts,shorthand}
\usepackage{amsmath}
\usepackage{slashed}
\usepackage{siunitx}
\usepackage{amssymb}
\usepackage{graphicx}
\usepackage{epic}
\usepackage{eepic}
\usepackage{epsfig}
\usepackage{latexsym}
\usepackage[dvipsnames]{xcolor}
\usepackage[export]{adjustbox}
\usepackage{float}
\usepackage{multirow, makecell,changepage}
\usepackage{hyperref}
\usepackage{enumitem}
\hypersetup{colorlinks=true,citecolor=red,linkcolor=NavyBlue,urlcolor=NavyBlue}
\usepackage[caption=false]{subfig}
 
\usepackage{natbib}
\usepackage{relsize}
\usepackage{wrapfig}
\usepackage[left=1.65cm,right=1.65cm,top=1.9cm,bottom=1.9cm]{geometry}
\usepackage{slashed}

\begin{document}
\relscale{1.05}
%%%%%%%%%%%%%%%%%%%%%%%%%%%%%%%%%%

%\title{Single in Pair: Effects on Leptoquark Exclusion Limits}

%\title{Accurate exclusion limits on scalar leptoquarks: combination brings precision}

\title{Fresh look at the LHC limits on vector leptoquarks}

% \author{Arvind Bhaskar}
% \email{arvind.bhaskar@iopb.res.in}
% \affiliation{Center for Computational Natural Sciences and Bioinformatics, International Institute of Information Technology, Hyderabad 500 032, India}
% \affiliation{Institute of Physics, Sachivalaya Marg, Bhubaneswar 751005, India}

\author{Arijit Das}
\email{arijit21@iisertvm.ac.in}
\affiliation{Indian Institute of Science Education and Research Thiruvananthapuram, Vithura, Kerala, 695 551, India}

\author{Tanumoy Mandal}
\email{tanumoy@iisertvm.ac.in}
\affiliation{Indian Institute of Science Education and Research Thiruvananthapuram, Vithura, Kerala, 695 551, India}

\author{Subhadip Mitra}
\email{subhadip.mitra@iiit.ac.in}
\affiliation{Center for Computational Natural Science and Bioinformatics, 
International Institute of Information Technology, Hyderabad 500 032, India}
\affiliation{Center for Quantum Science and Technology,
International Institute of Information Technology, Hyderabad 500 032, India.}

\author{Rachit Sharma}
\email{rachit21@iisertvm.ac.in}
\affiliation{Indian Institute of Science Education and Research Thiruvananthapuram, Vithura, Kerala, 695 551, India}

\begin{abstract}
\noindent
Vector leptoquarks (vLQs) are popular candidates for searching for physics beyond the Standard Model. In this paper, we present updated exclusion limits on various vLQ species, accounting for all the relevant production mechanisms at the LHC. In particular, we highlight the critical role of indirect production and its interference with the Standard Model Drell-Yan process. This interference can be constructive or destructive, depending on the specific quantum numbers of the vLQ, and significantly impacts the sensitivity of current searches. Furthermore, we demonstrate that including QCD-QED mixed pair production channels leads to a noticeable shift in model-independent mass limits. Additionally, we examine the validity of the full theory with vLQs and corresponding effective operators in the high mass regime. Overall, our analysis yields a substantial improvement in the exclusion limits on vLQs compared to the existing results in the literature.
\end{abstract}

\maketitle 

\section{Introduction}
\noindent
Since their emergence within the unified theories~\cite{Pati:1974yy,Georgi:1974sy,Fritzsch:1974nn,Farhi:1980xs,Schrempp:1984nj,Wudka:1985ef,Barbier:2004ez}, Leptoquarks (LQs) have been some of the most widely studied hypothetical particles by particle physicists. They are colour-triplet, hypercharged bosons, some of which can also carry a weak charge. LQs appear naturally around the TeV scale in many top-down new physics theories (see Ref.~\cite{Dorsner:2016wpm,Goncalves:2023qpz,DaRold:2023hmx} for a review on LQ phenomenology and relevant references therein). Their unique ability to interact simultaneously with a lepton and a quark makes the LQ search program a prominent one at the Large Hadron Collider (LHC) (see the public pages of the ATLAS~\cite{ATLASPublic} and CMS~\cite{CMSPublic} collaborations for the recent LQ searches at the LHC). The collider search strategies of these particles is one of the active areas at LHC phenomenology~\cite{Blumlein:1996qp,Dorsner:2014axa,Diaz:2017lit,Dey:2017ede,Bandyopadhyay:2018syt,Schmaltz:2018nls,Bhaskar:2020kdr,Greljo:2020tgv,Dorsner:2021chv,Crivellin:2021egp,Bandyopadhyay:2021pld,Bhaskar:2022ygp,Parashar:2022wrd,Ghosh:2023ocz,Florez:2023jdb,Arganda:2023qni,Bhaskar:2023ftn,Bertenstam:2025jvd,Ghosh:2025gue}. The main reason behind their recent popularity is their potential role in resolving various flavour and low-energy anomalies (for example, the $(g-2)_\mu$ anomaly; see Ref. ~\cite{Athron:2021iuf,Muong-2:2021vma,Muong-2:2021ojo,Muong-2:2023cdq} for a recent review and related references).

For LQs, many interesting signatures and search strategies involving their resonant and non-resonant productions have been proposed in the literature~\cite{Mandal:2015vfa,Mandal:2018kau,Aydemir:2019ynb,Chandak:2019iwj,Bhaskar:2021pml,Bhaskar:2021gsy,Bhaskar:2022vgk,Aydemir:2022lrq,Bhaskar:2023ftn,Bhaskar:2024snl}. Several studies have explored LQ phenomenology alongside right-handed neutrinos, vector-like leptons etc., in the literature~\cite{Ghosh:2022vpb,Bhaskar:2023xkm,Duraikandan:2024kcy,De:2024foq,Kumar:2025aek}. The direct searches for their resonant productions at the LHC have put very stringent constraints on LQ parameters, such as their masses, branching ratios (BRs), and couplings (see the summary plots from the ATLAS~\cite{ATL-PHYS-PUB-2024-012} and CMS~\cite{CMSPlot} collaborations for the current limits). Their pair production (PP) is helpful in setting model-independent limits on the masses and BRs as it is governed by QCD interactions and is practically insensitive to the new couplings. A mild model-dependence enters through $t$-channel lepton exchange diagrams present in the PP. Therefore, the PP channels are generally considered unsuitable for setting limits on unknown LQ couplings. On the other hand, single productions (SPs), indirect productions (IPs, $t$-channel LQ exchange processes), and indirect interferences [IIs, interference of the IPs with the Standard Model (SM) background] are all sensitive to the couplings, and hence, are used to set limits on their couplings. This method can also be applied in the context of other beyond the SM particles as shown in Refs.~\cite{Mandal:2012rx,Mandal:2016csb}.

Refs.~\cite{Mandal:2015vfa,Bhaskar:2023ftn} showed that it is possible to recast the cross-section upper limits in the dilepton-dijet final states from the direct LQ  PP searches and obtain competitive limits on the new LQ couplings. This is due to the coupling-dependent SP and IP processes also leading to dilepton-dijet events after jet radiation. Many of these events also pass the selection cuts used in the PP searches. Moreover, IP processes interfere heavily (constructively or destructively, depending on the LQ species) with the SM Drell-Yan ($qq\to\ell\ell$) background. Recently, Ref.~\cite{Bhaskar:2023ftn} has demonstrated that even though only a small fraction of the interference events pass the cuts optimised to select the PP events, the interference contribution is important for setting the exclusion limits on the unknown LQ couplings. The IP and II contributions can also alter the high-$p_T$ dilepton tail distributions seen in the dilepton resonance searches (e.g., searches for a heavy neutral gauge boson $Z'$). Hence, one can use, e.g., a $\chi^2$ method to set strong limits on LQ couplings for masses even beyond the reach of the resonance searches from the IP and II LQ contributions; this method has been used in Refs.~\cite{Mandal:2018kau,Bhaskar:2024swq,Bhaskar:2024wic} 
to constrain LQ couplings in various contexts. 

Previously, in Ref.~\cite{Bhaskar:2023ftn}, we obtained the exclusion limits on the masses and couplings of all scalar LQs (sLQs) by recasting the dimuon (and related) data. There, we also counted the PP and SP contributions along with the dominant IP and II contributions in our analysis to obtain precise limits. Here, we estimate similar limits on all vector LQs (vLQs) listed in Ref.~\cite{Dorsner:2016wpm}. As we found in the case of sLQs, a QCD-QED mixed contribution to the PP can significantly improve the model-independent mass limits on vLQs, especially for those with a high electric charge. If a LQ is heavy, the initial partons may not have enough energy to produce it resonantly; however, the indirect processes (i.e, IP and II) might still be within reach of the LHC. In such cases, it is possible to integrate the LQ in the $t$-channel out, and obtain an effective description of the $qq\to \ell\ell$ process in terms of six-dimensional operators. In the high mass limit, the actual process with LQs and its effective field theory (EFT) description should give close results. We compare these two descriptions and show the validity of the LQ-EFT in the high mass/scale.

The rest of the paper is organised as follows. We discuss various production channels of vLQs in Sec.~\ref{sec:production}. In Sec.~\ref{methodology}, we describe the recasting of experimental data. The resulting exclusion limits are presented in Sec.~\ref{sec:exclulim}. Finally, we summarise our findings in Sec.~\ref{sec:conclu}.

%%%%%%%%%%%%%%%%%%%%%%%%%%%%%%%%%%%%%%%%%%%%%%%%%%%%%%%%%%%%%%%%%%
\begin{table*}
\caption{Yukawa interactions for all vLQs in both up and down-aligned scenarios. Diquark interactions are excluded as they are not pertinent to our analysis. We employ a slightly modified yet more explicit notation for the couplings. The SM gauge group representations of the LQs are mentioned in the parentheses.\label{tab:vLQint}}
\centering{\small\renewcommand\baselinestretch{2}\selectfont
\begin{tabular*}{\textwidth}{l @{\extracolsep{\fill}}cc}
\hline \hline LQ Model & Down-aligned Interactions & Up-aligned Interactions \\ \hline\hline
{$U_1~(\mathbf{3},1,2/3)$ }     & $(Vx_{1}^{LL})_{1ij}\ \overline{u}_{L}^{i}\gamma^{\mu}\nu^{j}_{L}U_{1,\mu} + x_{1ij}^{LL} \overline{d}_{L}^{i}\gamma^{\mu}e^{j}_{L}U_{1,\mu}$            & $x^{LL}_{1ij}\ \overline{u}_{L}^{i}\gamma^{\mu}\nu^{j}_{L}U_{1,\mu} + (V^{\dagger}x^{LL}_{1})_{ij}\overline{d}_{L}^{i}\gamma_{\mu}e^{j}_{L}U_{1,\mu}$          \\ 
&$ + x^{RR}_{1ij}\ \overline{d}_R^{i}\gamma^{\mu}e^{j}_{R}U_{1,\mu} + x^{\overline{RR}}_{1ij}\ \overline{u}_R^{i}\gamma^{\mu}\nu^{j}_{R}U_{1,\mu}$&$ + x^{RR}_{1ij}\ \overline{d}_R^{i}\gamma^{\mu}e^{j}_{R}U_{1,\mu} + x^{\overline{RR}}_{1ij}\ \overline{u}_R^{i}\gamma^{\mu}\nu^{j}_{R}U_{1,\mu}$\\ 
$\widetilde{U}_{1}~(\mathbf{3},1,5/3)$     & \multicolumn{2}{c}{$\widetilde{x}^{RR}_{1ij}\ \overline{u_{R}}^{i}\gamma^{\mu}e^{j}_{R}\widetilde{U}_{1}$}           \\ 
%& $(\widetilde{y}^{RR}_{1})_{ij}\overline{d_{R}^{i}}^{C}\widetilde{S}_{1}e^{j}_{R}$          \\ 
\multirow{2}{*}{$V_{2}~(\mathbf{\Bar{3}},2,5/6)$}     & $-x^{RL}_{2ij}\ (\overline{d^{C}_{R}}^{i}\gamma^{\mu}\nu^{j}_{L}V^{1/3}_{2,\mu} - \overline{d^{C}_{R}}^{i}\gamma^{\mu}e^{j}_{L}V^{4/3}_{2,\mu})$            & $-x^{RL}_{2ij}\ (\overline{d^{C}_{R}}^{i}\gamma^{\mu}\nu^{j}_{L}V^{1/3}_{2,\mu} - \overline{d^{C}_{R}}^{i}\gamma^{\mu}e^{j}_{L}V^{4/3}_{2,\mu})$          \\ 
&$ + (V^{*}x_{2}^{LR})_{ij}\ \overline{u^{C}_L}^{i}\gamma^{\mu}e^{j}_{R}V^{1/3}_{2,\mu} - x^{{LR}}_{2ij}\ \overline{d^{C}_L}^{i}\gamma^{\mu}e^{j}_{R}V^{4/3}_{2,\mu}$&$ + x^{LR}_{2ij}\ \overline{u^{C}_L}^{i}\gamma^{\mu}e^{j}_{R}V^{1/3}_{2,\mu} - (V^{\dagger}x_{2}^{LR})_{ij}\ \overline{d^{C}_L}^{i}\gamma^{\mu}e^{j}_{R}V^{4/3}_{2,\mu}$\\
$\widetilde{V}_{2}~(\mathbf{\Bar{3}},2,-1/6)$     & \multicolumn{2}{c}{$-\widetilde{x}^{RL}_{2ij}\ 
(\overline{u^{C}_{R}}^{i}\gamma^{\mu}e^{j}_{L}\widetilde{V}^{1/3}_{2} - 
\overline{u^{C}_{R}}^{i}\gamma^{\mu}\nu^{j}_{L}\widetilde{V}^{-2/3}_{2})$} \\ 
 \multirow{2}{*}{$U_{3}~(\mathbf{3},3,2/3)$}     & $-x^{LL}_{3ij}\ \overline{d_{L}}^{i}\gamma^{\mu}e^{j}_{L}U^{2/3}_{\mu} + \sqrt{2}x^{LL}_{3ij}\ \overline{d_{L}}^{i}\gamma^{\mu}\nu^{j}_{L}U^{-1/3}_{3}$             & $x_{3ij}^{LL}\ \overline{u_{L}}^{i}\gamma^{\mu}\nu^{j}_{L}U^{2/3}_{\mu}+ \sqrt{2} x^{LL}_{3ij}\ \overline{u_{L}}^{i}\gamma^{\mu}e^{j}_{L}U^{5/3}_{\mu}$         \\ 
 &  $+(Vx^{LL}_3)_{ij}\ \overline{u_{L}}^{i}\gamma^{\mu}\nu^{j}_{L}U^{2/3}_{\mu}+\sqrt{2}(Vx^{LL}_{3})_{ij}\ \overline{u_{L}}^{i}\gamma^{\mu}e^{j}_{L}U^{5/3}_{\mu}$ & $+\sqrt{2}(V^{\dagger} x_{3ij}^{LL})\ \overline{d_{L}}^{i}\gamma^{\mu}\nu^{j}_{L}U^{-1/3}_{\mu}- (V^{\dagger}x^{LL}_{3})_{ij}\  \overline{d_{L}}^{i}\gamma^{\mu}e^{j}_{L}U^{2/3}_{\mu}$ \\
 
\hline
\hline
\end{tabular*}}
\end{table*}
%%%%%%%%%%%%%%%%%%%%%%%%%%%%%%%%%%%%%%%%%%%%%%%%%%%%%%%%%%%%%%%%%%
%%%%%%%%%%%%%%%%%%%%%%%%%%%%%%%%%%%%%%%%%%%%%%%%%%%%%%%%%%%%%%%%%%
\begin{figure*}
\centering
\captionsetup[subfigure]{labelformat=empty}
\subfloat[(a)]{\includegraphics[width=0.19\textwidth]{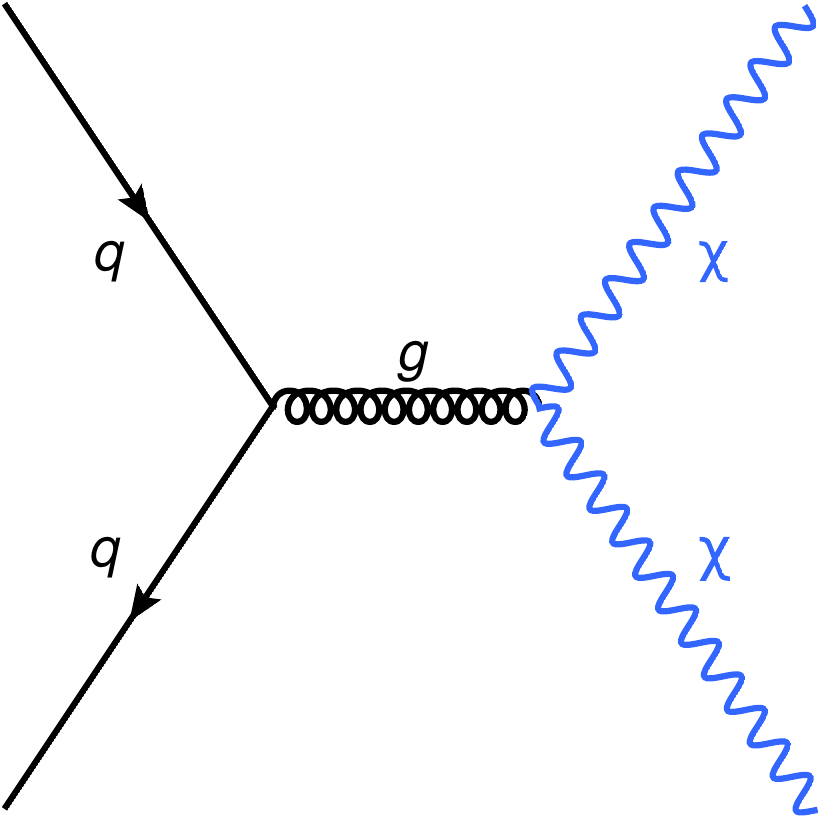}\label{qqlqlq}}\hfill
\subfloat[(b)]{\includegraphics[width=0.19\textwidth]{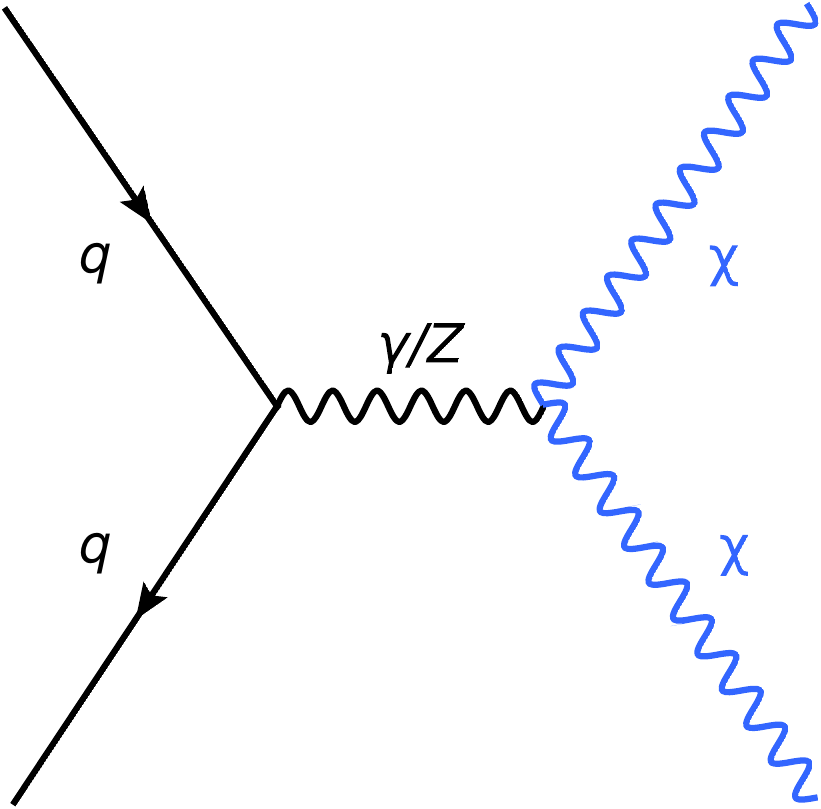}\label{qqzlqlq}}\hfill
\subfloat[(c)]{\includegraphics[width=0.19\textwidth]{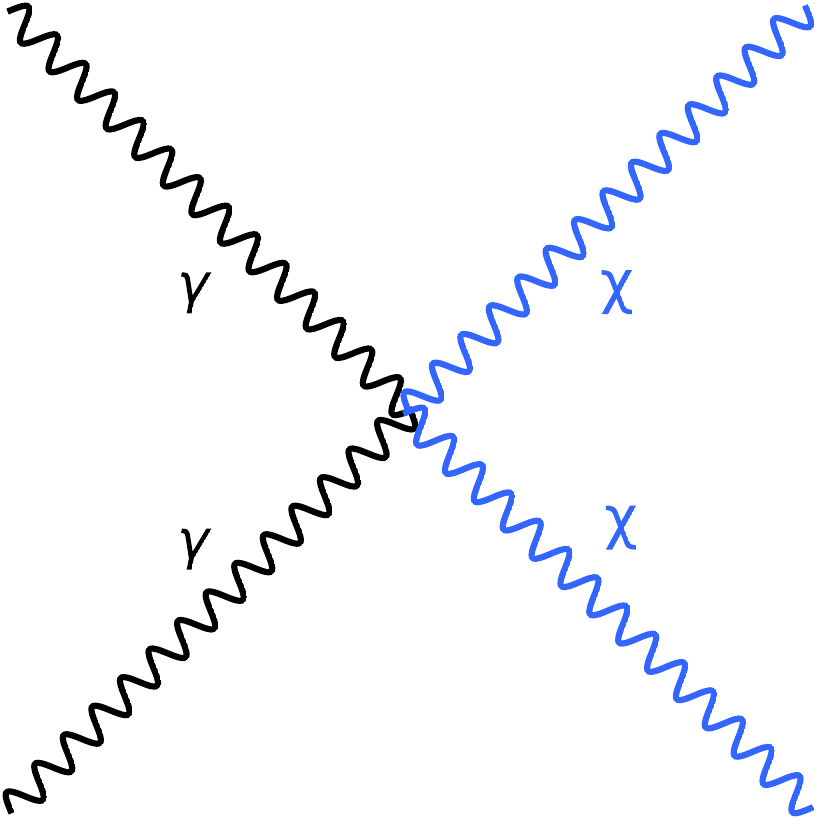}\label{pplqlq}}\hfill
\subfloat[(d)]{\includegraphics[width=0.19\textwidth]{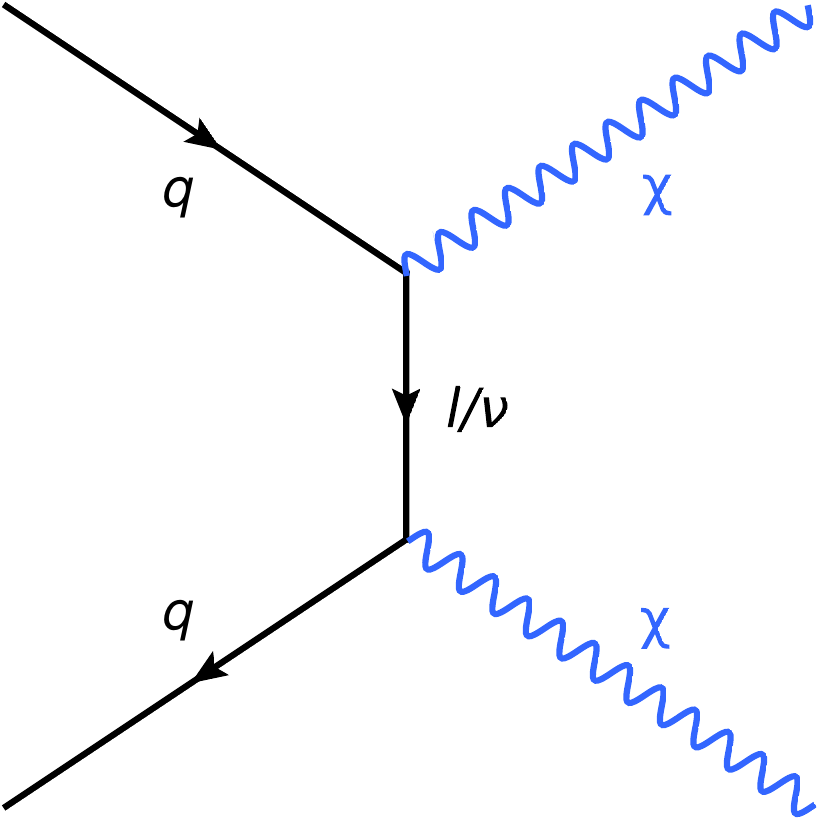}\label{qqtch}}
\\
\subfloat[(e)]{\includegraphics[width=0.19\textwidth]{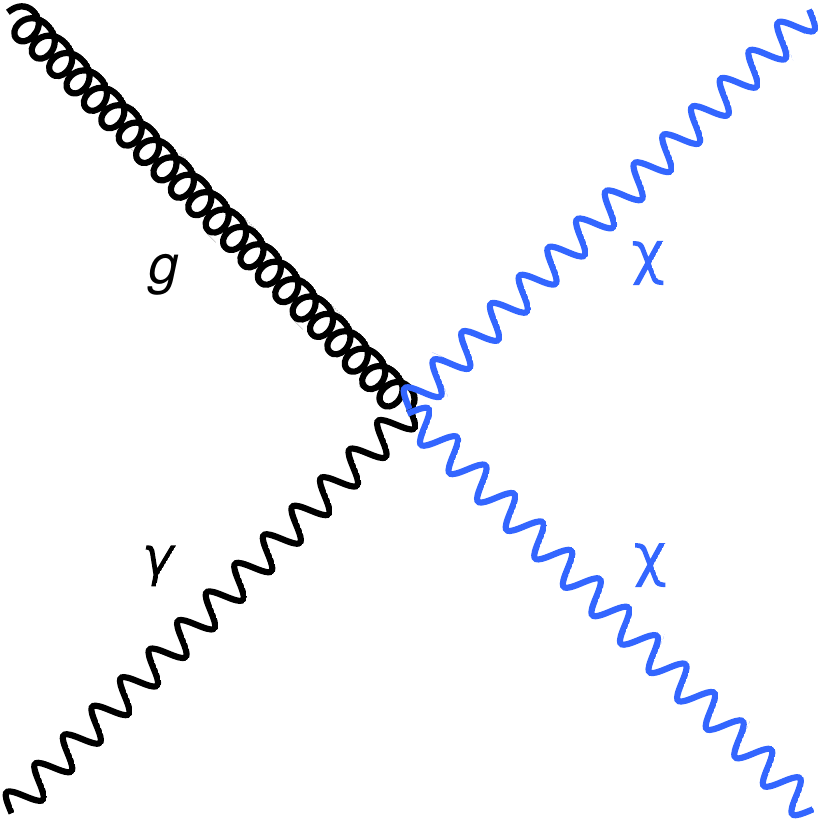}\label{gglqlq3pt}}\hfill
\subfloat[(f)]{\includegraphics[width=0.19\textwidth]{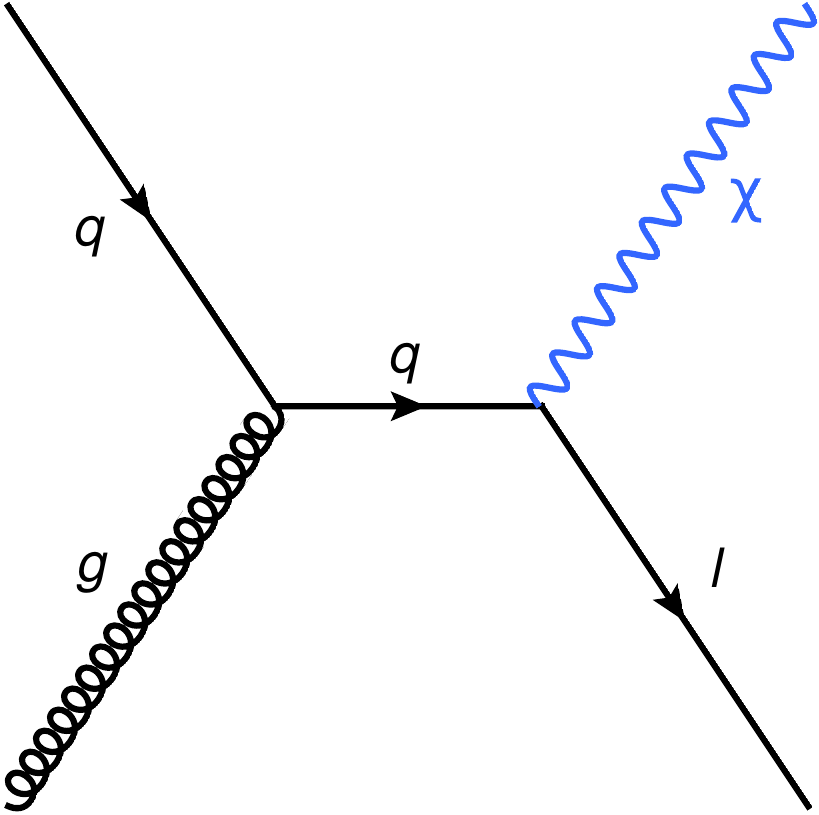}\label{2bspqg}}\hfill
\subfloat[(g)]{\includegraphics[width=0.19\textwidth]{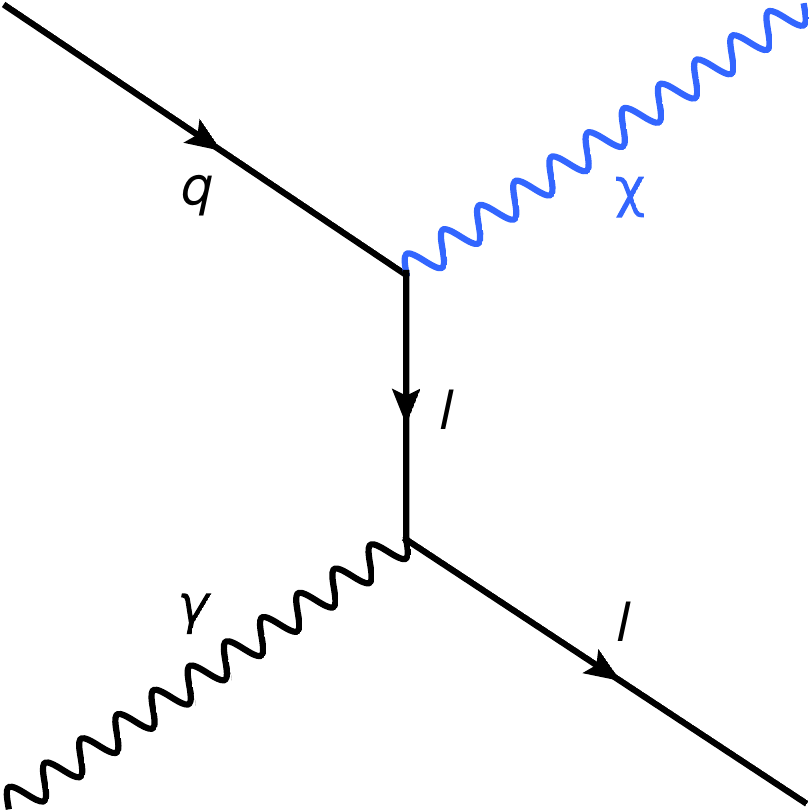}\label{2bspqgamma}}\hfill
\subfloat[(h)]{\includegraphics[width=0.19\textwidth]{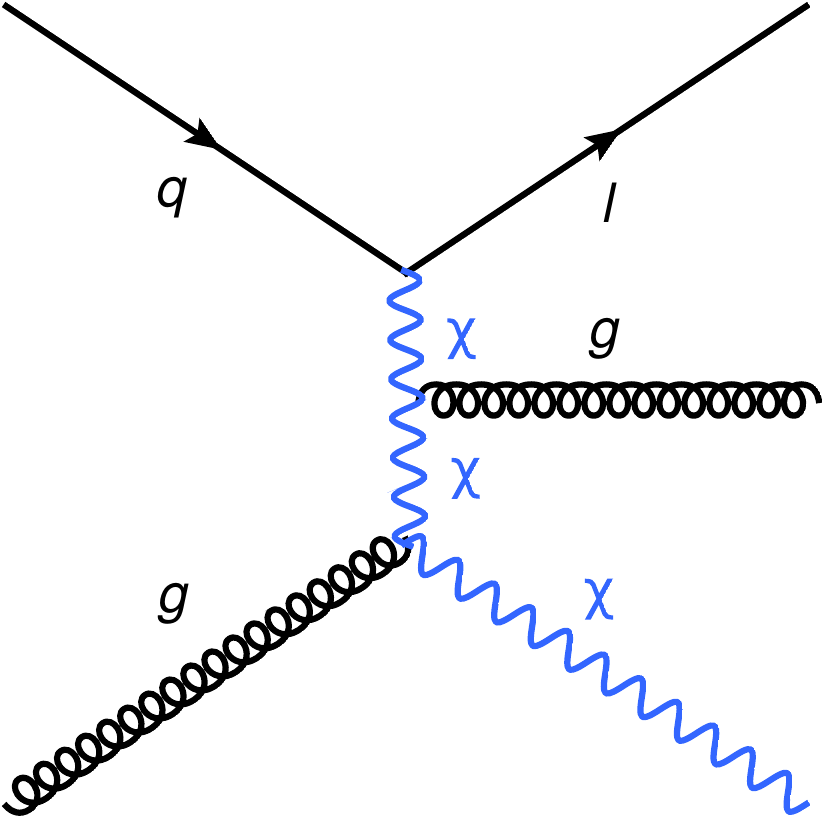}\label{SPqg}}
\\
\subfloat[(i)]{\includegraphics[width=0.19\textwidth]{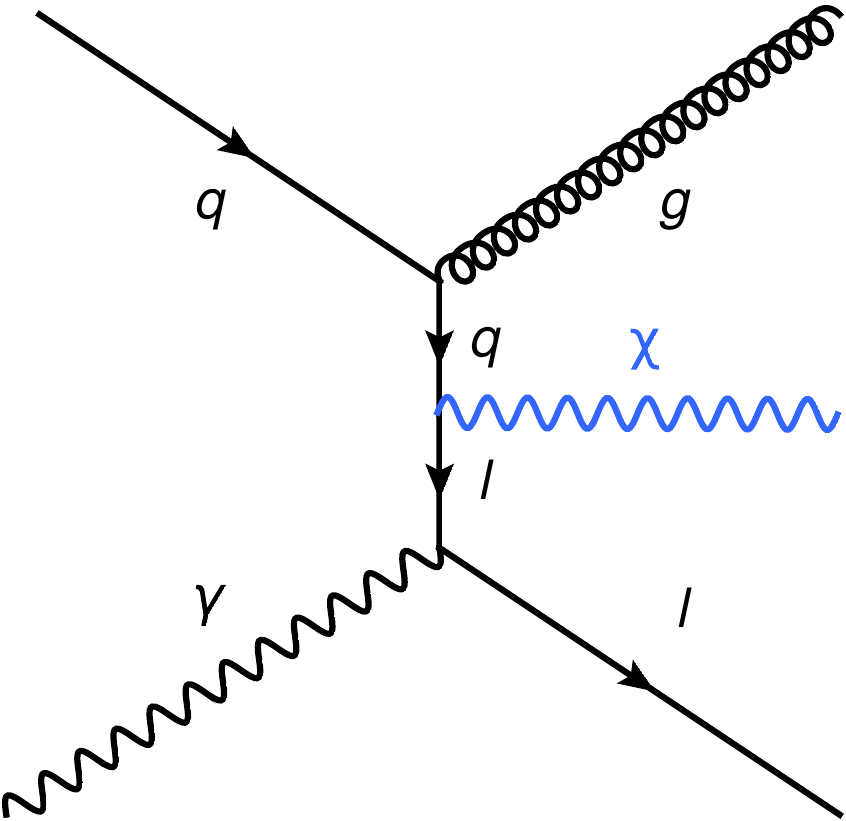}\label{SPqp}}\hfill
\subfloat[(j)]{\includegraphics[width=0.19\textwidth]{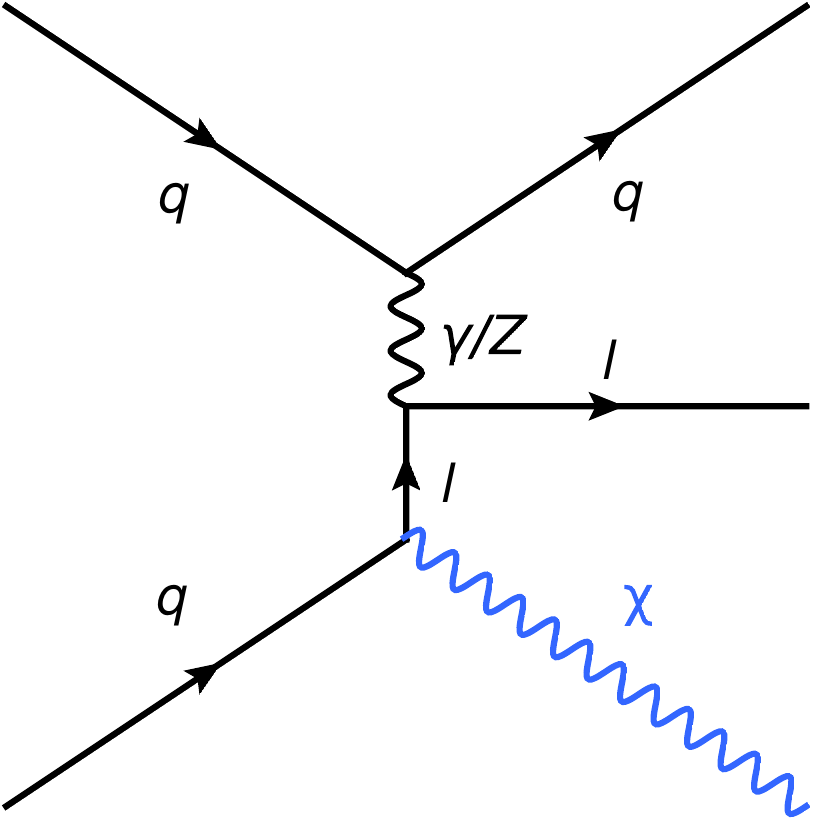}\label{SPqq}}\hfill
\subfloat[(k)]{\includegraphics[width=0.19\textwidth]{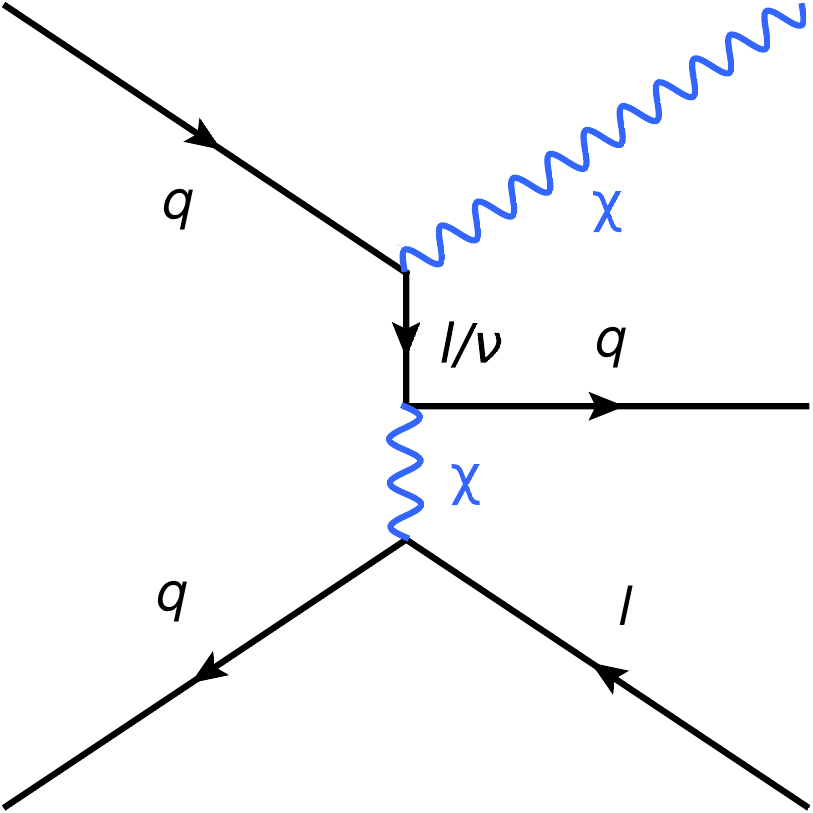}\label{SPSLQ3}}\hfill
\subfloat[(l)]{\includegraphics[width=0.19\textwidth]{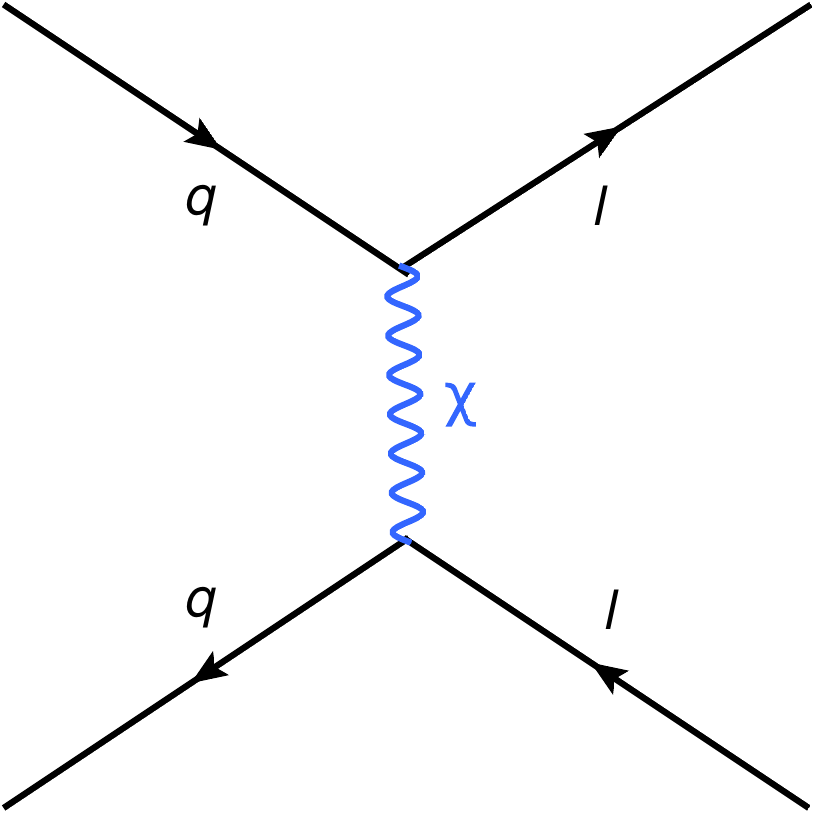}\label{indirect}}
\caption{Representative Feynman diagrams for pair, single and indirect productions of vLQs. A generic vLQ is denoted by $\chi$.\label{feyn_dgm}}
\end{figure*}
%%%%%%%%%%%%%%%%%%%%%%%%%%%%%%%%%%%%%%%%%%%%%%%%%%%%%%%%%%%%%%%%%%

\section{Production of vector leptoquarks at the LHC}\label{sec:production}
\noindent
Assuming baryon and lepton number conservation and the SM gauge symmetry for vLQ interactions, five LQ species can directly produce the $\ell\ell jj$ final state:  $U_1$, $\widetilde{U}_1$, $V_2$, $\widetilde{V}_2$, and $U_3$. Another vLQ, $\overline{U}_1$, that couples exclusively to right-handed neutrinos, cannot directly produce the $\ell\ell jj$ final state (though decays of right-handed neutrinos can lead to dilepton plus multi-jet final states). Following the notations of Ref.~\cite{Dorsner:2016wpm}, we summarise the interactions of vLQs with quarks and leptons in Table~\ref{tab:vLQint}. We consider two specific alignments of the vLQ mass basis. In the \emph{down-aligned} scenario, the vLQ mass eigenstates are aligned with the down-type quark masses, while in the \emph{up-aligned} scenario, they are aligned with the up-type quarks. We do not consider the most general scenario where vLQ mass eigenstates are aligned neither with the up not down quark sectors. Although the results for up-aligned and down-aligned scenarios can show minor differences in some cases, for illustration, we use the down-aligned scenario in our analysis. We assume $U_{\rm PMNS} = 1$ throughout our analysis, since the neutrino oscillation length is irrelevant to our collider study, and all neutrino flavours collectively contribute to the missing energy.

There is also an extra gluon coupling allowed by the gauge symmetries for vLQs that can influence the production at LHC~\cite{Blumlein:1996qp,Blumlein:1994tu}:
\begin{align}
\mathcal{L} \supset & 
 - i g_{s} (1-\kappa) \chi^{\dagger}_{\mu} T^{a} \chi_{\nu} G^{a \mu\nu},\label{eq:kappa}
\end{align}
where $\chi_{\mu}$ denotes a generic vLQ and $G^{a\mu\nu}$ denotes the field strength tensor of the gluon. 

At the LHC, vLQs produced via resonant (PP and SP) and nonresonant (IP and II) processes can contribute to the $\ell\ell jj$ final state. The interference can be constructive or destructive, depending on the type of vLQ involved. To understand how all production mechanisms contribute to the $\ell\ell jj$ final state, we systematically discuss all the production processes of vLQs (similar to the discussion presented in Ref.~\cite{PhysRevD.109.055018} for sLQs).
\medskip

\noindent
\textbf{Pair production:} Pair produced vLQs (same or different species) can decay to the $\ell \ell j j$ final state. The leading tree-level contributions are as follows:
\begin{enumerate}
\item $\mc{O}(\al_s^2\al_e^0\al_x^0)$: These model-independent, $gg$ or $qq$ fusion-initiated processes are purely QCD-driven. (Here, $\alpha_x = x^2/4\pi$ where $x$ denotes a vLQ-quark-lepton coupling.) Fig.~\ref{qqlqlq} shows a representative diagram. 

\item $\mc{O}(\al_s^0\al_e^2\al_x^0)$: These are $qq$ [Fig.~\ref{qqzlqlq}] or $\gamma\gamma$ [Fig.~\ref{pplqlq}] initiated processes. Their contributions depend on the electric charge of the vLQ involved.

\item $\mc{O}(\al_s^0\al_e^0\al_x^2)$: In these processes, two on-shell vLQs are produced via the exchange of a lepton in the $t$-channel [see Fig.~\ref{qqtch}]. The unknown vLQ-quark-lepton couplings control their contributions. The corresponding cross sections exhibit significant sensitivity to the initial-state quarks involved, owing to variations in the parton distribution functions (PDFs).

\item $\mc{O}(\al_s^1\al_e^0\al_x^1)$: These contributions arise from the interference between diagrams of $\mc{O}(\alpha_s^2\alpha_e^0\alpha_x^0)$ and $\mc{O}(\alpha_s^0\alpha_e^0\alpha_x^2)$. This interference is destructive for all vLQ species.

\item $\mc{O}(\al_s^0\al_e^1\al_x^1)$: Contributions at this order stem from the interference between diagrams of $\mc{O}(\alpha_s^0\alpha_e^2\alpha_x^0)$ and $\mc{O}(\alpha_s^0\alpha_e^0\alpha_x^2)$. Their impact on the total PP cross-section is negligible.

\item $\mc{O}(\al_s^1\al_e^1\al_x^0)$: 
The $g\gamma$ fusion is a QCD-QED mixed process for producing a pair of vLQs [Fig.~\ref{gglqlq3pt}]. The interaction originates in the vLQ kinetic terms and is independent of NP couplings. These contributions become significant when the vLQ has a high electric charge.
\end{enumerate}

\noindent
The total PP cross-section is expressed by summing up all the contributions of different orders as follows:
\begin{align}
\sigma_{\rm PP}(M_{\ell_q},x) =&\ \sigma_{\rm PP}^{200}(M_{\ell_q}) + \sigma_{\rm PP}^{110}(M_{\ell_q}) + \sigma_{\rm PP}^{020}(M_{\ell_q}) \nn \\
&+ x^2\,\overline{\sigma_{\rm PP}^{101}(M_{\ell_q})} +  x^2\, \overline{\sigma_{\rm PP}^{011}(M_{\ell_q})} \nn\\
&+ x^4\,\overline{\sigma_{\rm PP}^{002}(M_{\ell_q})},
\end{align}
where $\sigma_{\text{PP}}^{ijk}$ represents the contribution of order $\alpha_s^i\alpha_e^j\alpha_x^k$ to the  PP process. The line over $\sigma_{\text{PP}}^{ijk}$ indicates that those contributions are evaluated for coupling $x=1$, i.e., $\overline{\sigma^{ijk}_{\text{PP}}(M_{\ell_q})}=\sigma^{ijk}_{\text{PP}}(M_{\ell_q},x=1)$. \medskip

\noindent
\textbf{Single production:} The resonant productions of a single vLQ either with a lepton [i.e., the two-body single production (2BSP)] or a lepton and a jet [i.e., the three-body single production (3BSP)] can result in the $\ell\ell jj$ final state.\smallskip

\begin{adjustwidth}{1.1em}{0pt}
\textbf{2BSP:} There are two leading-order processes of this type:      
\end{adjustwidth}
\begin{enumerate}
\item $\mathcal{O}(\al_s^1\al_e^0\al_x^1)$, initiated by the $qg$ fusion [Fig.~\ref{2bspqg}]. 

\item $\mathcal{O}(\al_s^0\al_e^1\al_x^1)$, initiated by $q\gamma$ fusion [Fig.~\ref{2bspqgamma}].
\end{enumerate}
\begin{adjustwidth}{1.1em}{0pt} 
\textbf{3BSP:} Two types of diagrams can lead to 3BSP. In one type, a hard separable jet -- either initial state radiation (ISR) or final state radiation (FSR) jet -- is emitted from a 2BSP diagram, thus qualifying as a 3BSP process:
\end{adjustwidth}
\begin{enumerate}

\item $\mc{O}(\al_s^{1+(1)}\al_e^0\al_x^1)$: This contribution originates from the emission of a hard jet, arising from the $\mathcal{O}(\alpha_s^1\alpha_e^0\alpha_x^1)$ term of the 2BSP process (Fig. \ref{SPqg}). The jet can be emitted by any colored particle present in the diagram.

\item$\mc{O}(\al_s^{(1)}\al_e^1\al_x^1)$: The contribution at this order arises from the emission of a hard jet from the $\mathcal{O}(\alpha_s^0\alpha_e^1\alpha_x^1)$ diagrams of the 2BSP process (Fig. \ref{SPqp}).

\item[3.] $\mc{O}(\al_s^0\al_e^{1+(1)}\al_x^1)$: This contribution arises when an initial-state quark undergoes a splitting process, producing a hard quark and an electroweak boson. The boson then interacts with another quark, resulting in the production of a vLQ and a lepton (Fig. \ref{SPqq}).
\end{enumerate}
\begin{adjustwidth}{1.1em}{0pt} 
Additionally, there are 3BSP processes that cannot be mapped to an ISR or FSR off a 2BSP process
(see ~\cite{Mandal:2015vfa} for details):
\end{adjustwidth}
\begin{enumerate}
\item[4.] $\mc{O}(\al_s^0\al_e^0\al_x^3)$: Contributions at this order are governed solely by the NP coupling. These processes originate from the $qq$ initial state [Fig.~\ref{SPSLQ3}]. 

\item[5.] Another contribution in the 3BSP processes arises when a vLQ is produced along with a lepton-jet pair from an off-shell vLQ, i.e., $pp \to \ell_q \ell_q^* \to \ell_q \ell j$. This process occurs at one higher order in $\alpha_x$ than the six PP sub-categories and has a different kinematics than that of the PP process.
\end{enumerate}
%%%%%%%%%%%%%%%%%%%%%%%%%%%%%%%%%%%%%%%%%%%%%%%%%%%%%%%%%%%%%%%%%%
\begin{figure*}%[!t]
\centering
\captionsetup[subfigure]{labelformat=empty}
\subfloat[\quad\quad(a)]{\includegraphics[height=6cm,width=6.375cm]{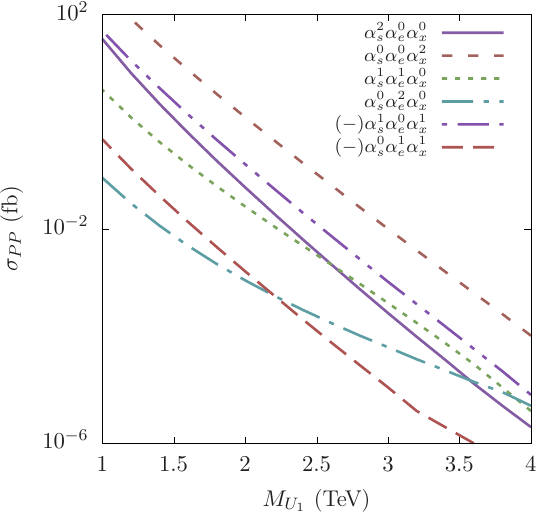}}\hspace{1cm}
\subfloat[\quad\quad(b)]{\includegraphics[height=6cm,width=6.375cm]{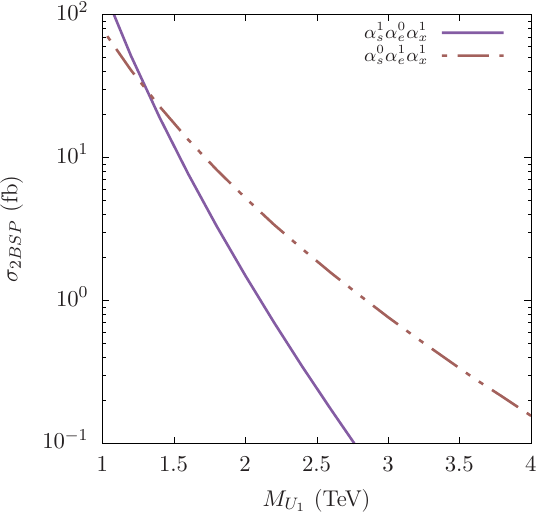}}\\
\subfloat[\quad\quad(c)]{\includegraphics[height=6cm,width=6.375cm]{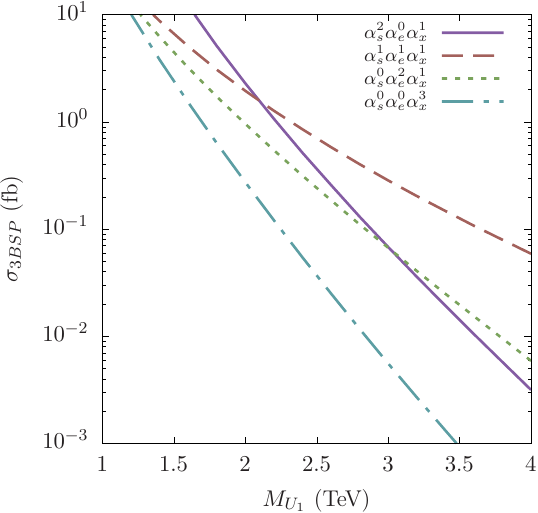}}\hspace{1cm}
\subfloat[\quad\quad(d)]{\includegraphics[height=6cm,width=6.375cm]{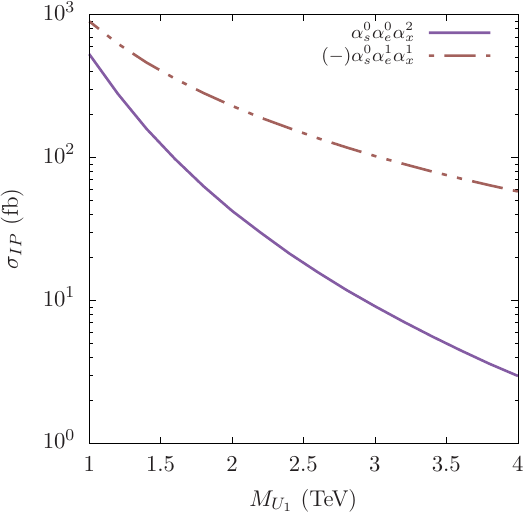}}
\caption{The PP, SP, IP, and II contributions for $U_1$ (subjected to a $M_{\mu^{+}\mu^{-}} \geq 120$ GeV cut) as functions of $M_{U_1}$ for $x^{LL}_{10,12}=1$ at different orders of 
$\alpha_{s}$, $\alpha_{e}$, and $\alpha_{x}$. The constructive/destructive interference contributions are indicated with $(\pm)$ signs in front. 
\label{partonxsec}}
\end{figure*}
%%%%%%%%%%%%%%%%%%%%%%%%%%%%%%%%%%%%%%%%%%%%%%%%%%%%%%%%%%%%%%%%%%

\noindent
To obtain the total SP contribution to the $\ell\ell jj$ final state, we follow the method described in Ref.~\cite{Mandal:2015vfa}. Radiations from the Born-level 2BSP processes can resemble the 3BSP processes, potentially leading to overcounting if both processes are included separately. The total SP cross section can be written as in the notation of Ref.~\cite{Mandal:2015vfa} as follows:
\begin{align}
\label{eq:incluSP}
\sg_{\rm SP}
=&\ \underbrace{\big(\sg^{\rm LO}_{\mathrm{2BB}(\ell_q\ell)} + \overbrace{\sg^{\rm virtual}_{(\ell_q\ell)} + \sg_{\mathrm{2BR}(\ell_q\ell j)}^{\rm soft+collinear}}^{\rm Divergent~terms} + \sg_{\mathrm{3BNS}(\ell_q\ell j)}^{\rm soft}\big)}_{\rm Negligible~contribution} \nn\\
&\ +\underbrace{\big(\sg_{\mathrm{2BR}(\ell_q\ell j)}^{\rm hard} + \sg_{\mathrm{3BNS}(\ell_q\ell j)}^{\rm hard}\big)}_{\rm Main~contribution \ (\approx\ 3BSP)}\ +\ \cdots,
\end{align}

The suffixes denote different contributions: 2BB represents two-body Born-level processes (i.e., 2BSP), 2BR indicates 2BSP with a radiation jet, and 3BNS denotes the NP contribution in 3BSP. The contribution to $\ell\ell jj$ final state from single productions mainly arises from the 3BSP processes. Therefore, we can write the total SP cross section as follows:
\begin{align}
\sigma_{\rm SP}(M_{\ell_q},x) =&\ x^2\left\{\overline{\sigma_{SP}^{021}(M_{\ell_q})} + \overline{\sigma_{SP}^{201}(M_{\ell_q})} + \overline{\sigma_{SP}^{111}(M_{\ell_q})}\right\} \nn \\
&\ + x^6\overline{\sigma_{SP}^{003}(M_{\ell_q})}.
\end{align}
\noindent
\textbf{Indirect production:} In this production channel, two leptons are generated via the $t$-channel exchange of a vLQ. Notably, the main contribution to $\ell \ell j j$ originates from a single order:
\begin{enumerate}
    \item $\mc{O}(\al_s^{(2)}\al_e^0\al_x^2)$: This process contributes to the $\ell \ell j j$ final state when two hard jets emit from the diagram shown in Fig.~\ref{indirect}. 
\end{enumerate}

\begin{figure}[t]
\centering
\includegraphics[height=6cm,width=6.375cm]{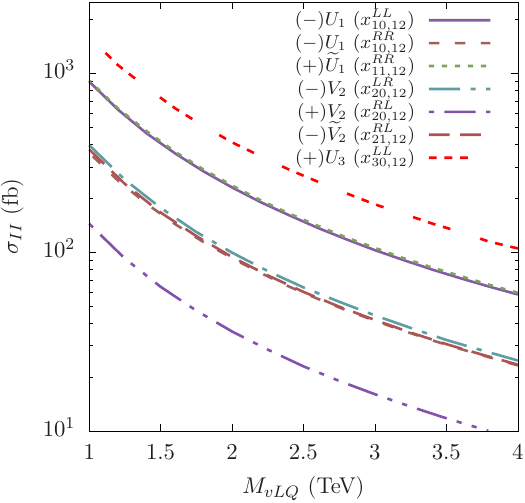}
\caption{Indirect interference contributions for different vLQs.\label{fig:ii}}
\end{figure}

\noindent
\textbf{Indirect interference:} This contribution arises from the interference between the IP processes and the SM $\ell\ell$ processes (via a $Z/\gamma$ boson). Note that the kinematics of the interference contribution is different from that of the IP process alone.
\begin{enumerate}
    \item $\mc{O}(\al_s^{(2)}\al_e^1\al_x^1)$: This contribution plays a decisive role in setting the exclusion limits, primarily due to its large interference with the SM. The nature of the interference -- whether destructive or constructive -- depends on the vLQ species involved.
\end{enumerate}
\noindent
The total nonresonant (NR) vLQ production cross section  from both IP and II can be expressed as:
\begin{equation}
\sigma_{\rm NR}(M_{\ell_q},x) = x^2\overline{\sigma_{\rm II}^{011}(M_{\ell_q})} + x^4\overline{\sigma_{\rm IP}^{002}(M_{\ell_q})}.
\end{equation}

In Fig.~\ref{partonxsec}, we show the cross sections of various production modes of $U_1$ at different orders as functions of its mass. The $U_1$ interfere destructively with the SM background for both $LL$ and $RR$-type couplings. This is shown in Fig.~\ref{fig:ii}.

%%%%%%%%%%%%%%%%%%%%%%%%%%%%%%%%%%%%%%%%%%%%%%%%%%%%%%%%%%%%%%%%%%
\begin{table*}
\caption{The cross sections ($\sigma$), the numbers of events ($\mc N$) surviving the $\m\m j j$-selection cuts~\cite{ATLAS:2020dsk} for $\mc L_{exp}= 139$ fb$^{-1}$ of integrated luminosity, and the cut efficiencies ($\epsilon$) for three benchmark masses of $U_1$. As in Fig.~\ref{partonxsec}, we have set $x_{10,12}^{LL}=1$ to obtain the numbers.  We see that the importance of the indirect modes increases with the mass of the LQ.\label{tab:table}}
\centering{\small\renewcommand\baselinestretch{1.8}\selectfont
\begin{tabular*}{\textwidth}{c @{\extracolsep{\fill}}rrr rrr rrr rrr}
\hline
\textbf{Mass} ($U_{1}$) & \multicolumn{3}{c}{\textbf{Pair Production}} & \multicolumn{3}{c}{\textbf{Single Production}} & \multicolumn{3}{c}{\textbf{Indirect Production}} & \multicolumn{3}{c}{\textbf{Indirect Interference}} \\ \cline{2-4}\cline{5-7}\cline{8-10}\cline{11-13}
(TeV) & $\sigma_{\rm PP}$ (fb) & $\epsilon_{\rm PP}$ & $\mathcal{N}_{\rm PP}$ & $\sigma_{\rm SP}$ (fb) & $\epsilon_{\rm SP}$ & $\mathcal{N}_{\rm SP}$ & $\sigma_{\rm IP}$ (fb) & $\epsilon_{\rm IP}$ & $\mathcal{N}_{\rm IP}$ & $\sigma_{\rm II}$ (fb) & $\epsilon_{\rm II}$ & $\mathcal{N}_{\rm II}$   \\ \hline\hline
$1.5$ & $14.72$ & $0.35$ & $716.00$ & $30.4$ & $0.15$ & $634.0$ & $129.4$ & $0.029$ & $521.6$ & $-406.2$ & $0.01$ & $-564.6$ \\ 
$2.5$ &  $1.015 \times 10^{-1}$ & $0.53$ & $7.48$& $1.3$ & $0.21$ & $38.0$ &  $18.86$ & $0.028$ & $73.4$& $-147.7$ & $0.01$ & $-205.3$ \\ 
$4.0$ & $1.045 \times10^{-4}$ &$0.62$ & $9.0 \times10^{-3}$ & $6.7 \times 10^{-2}$ & $0.27$ & $2.5$ & $3.1$ &$0.025$ & $11.0$ & $-57.8$ & $0.01$ & $-80.3$ \\ \hline
\end{tabular*}}
\end{table*}
%%%%%%%%%%%%%%%%%%%%%%%%%%%%%%%%%%%%%%%%%%%%%%%%%%%%%%%%%%%%%%%%%%

\section{Recasting LHC Searches}
\label{methodology}

\subsection{Direct-search data}
\noindent
Both ATLAS and CMS collaborations have searched for LQs in the PP channel. Here, we use the observed upper limit on the cross section in the $\mu\mu jj$ channel from the ATLAS measurement~\cite{ATLAS:2020dsk} to constrain the vLQ parameter space. Alternatively, the corresponding analysis by the CMS collaboration~\cite{CMS:2018lab} can also be used; we have verified that both yield similar bounds.

Our procedure to recast the direct bounds is as follows. 
We first pass all kinematically distinct production channels through the analysis cuts (tuned and validated to mimic those used in Ref.~\cite{ATLAS:2020dsk}) to determine the selection efficiencies. With those, we estimate the total number of $\mu\mu jj$ events from vLQ by adding the contributions from these channels. We get the exclusion limits by equating the total vLQ yield (as a function of model parameters) to the observed number of events reported by the experiment, using the following relation:
\begin{align}
    \dfrac{\mc N_{obs}(M_{\ell_q})}{\mathcal{L}_{exp}}  =&\ \sigma_{obs}(M_{\ell_q})\times\epsilon_{exp}(M_{\ell_q}) \nn\\
    =&\ \sum_{i~\in~{\rm topologies}}\sigma_i(M_{\ell_q},\vec x)\times \beta_i(\vec x)\times\epsilon_{i}(M_{\ell_q}).
\end{align}
The sum is performed over various topologies: PP, SP, IP, and II. The symbols $\epsilon$ denotes selection cut-efficiencies, $\beta$ denotes branching ratios (BRs), and $\mathcal{L}_{\text{exp}}$ denotes the integrated luminosity. We use ${\mc N_{obs}}$ for $M_{\ell_q}=2$~TeV for a conservative estimation, as choosing a lower mass point leads to weaker limits on vLQ parameters.

The vLQ interaction Lagrangians were implemented in \texttt{FeynRules}~\cite{Alloul:2013bka} in Ref.~\cite{Bhaskar:2024wic}; %(\href{https://github.com/rsrchtsm/LQ_Models}{GitHub}) 
we use the {\sc UFO} model files~\cite{Degrande:2011ua} for event generation. We generate all Monte Carlo events at the leading order using \texttt{MadGraph5}~\cite{Alwall:2014hca}, using the \texttt{NNPDF} PDFs~\cite{NNPDF:2021uiq} with dynamical renormalisation and factorisation scales. These events are then passed through \texttt{PYTHIA8}~\cite{Bierlich:2022pfr} for parton showering and hadronisation, and subsequently through \texttt{Delphes}~\cite{deFavereau:2013fsa} for detector simulation. Jets are clustered by the anti-$k_T$ algorithm~\cite{Cacciari:2008gp}, as implemented in \texttt{FastJet}~\cite{Cacciari:2011ma}.

\subsection{Dilepton-search data}
\noindent 
As previously noted, high-$p_T$ dilepton resonance searches also can constrain the vLQ parameter space. We use the latest CMS dimuon data~\cite{CMS:2021ctt} to derive bounds on vLQ couplings as functions of their masses. Both resonant and nonresonant vLQ production yield high-$p_T$ leptons, affecting the tail of the dimuon spectrum. Adopting the $\chi^2$ analysis method from Ref.~\cite{Mandal:2018kau}, we derive $2\sigma$ exclusion limits by fitting the dimuon invariant mass distribution for fixed $M_{\ell_q}$ values. The CMS analysis~\cite{CMS:2021ctt} does not impose any condition on the number of hard jets accompanying the lepton pair. Hence, all vLQ production modes -- which produce dileptons -- can contribute to the signal. We incorporate all such contributions within our statistical framework to derive exclusion limits. Since nonresonant production (IP and II) cross sections decrease at a slower rate with mass than the resonant channels (PP and SP), the exclusion limits in the high mass region are essentially determined by the IP and II contributions.

Recently, the CMS collaboration has performed a search for non-resonant production of LQs in the dilepton channel, taking into account the interference with the SM dilepton processes~\cite{CMS:2025iix}. Their analysis covers both sLQs and vLQs, and they presented exclusion limits in the mass-coupling plane. Our exclusion limits derived from the high-$p_T$ dimuon searches agree well with their results.

%%%%%%%%%%%%%%%%%%%%%%%%%%%%%%%%%%%%%%%%%%%%%%%%%%%%%%%%%%%%%%%%%%
\begin{table*}
\caption{Comparison of model-independent mass exclusion limits (in GeV) on various sLQ models with and without QED contributions, for $\kappa=0$ and $\kappa=1$. We assume sLQ decays through only one small coupling (shown in parentheses). The limits remain the same for second-generation quarks. \label{tab:SLQcombined}}
\centering
{\small\renewcommand\baselinestretch{1.5}\selectfont
\begin{tabular*}{0.8\textwidth}{l @{\extracolsep{\fill}}cccc}
\hline
\multirow{2}{*}{Model} & \multicolumn{2}{c}{$\kappa=1$} & \multicolumn{2}{c}{$\kappa=0$} \\
\cline{2-3}\cline{4-5}
 & QCD & QCD+QED & QCD & QCD+QED \\ \hline\hline
$U_1 (x^{LL}_{10,12})$ & $1746$ & $1806$ & $2038$ & $2051$ \\
$U_1 (x^{RR}_{10,12})$ & $1989$ & $2056$ & $2279$ & $2295$ \\
$\widetilde{U}_1 (x^{RR}_{11,12})$ & $1989$ & $2319$ & $2279$ & $2410$ \\
$V_2 (x^{LR}_{20,12})$ & $2091$ & $2275$ & $2383$ & $2451$ \\
$V_2 (x^{RL}_{20,12})$ & $1989$ & $2217$ & $2279$ & $2360$ \\
$\widetilde{V}_2 (x^{RL}_{21,12})$ & $1989$ & $2010$ & $2279$ & $2283$ \\
$U_3 (x^{LL}_{30,12})$ & $2027$ & $2337$ & $2310$ & $2436$ \\ \hline
\end{tabular*}
}
\end{table*}
%%%%%%%%%%%%%%%%%%%%%%%%%%%%%%%%%%%%%%%%%%%%%%%%%%%%%%%%%%%%%%%%%%
\begin{table*}
\caption{Effective operators obtained by integrating out the $t$-channel vLQ (in IP processes) producing a lepton pair and their Fierz transformed versions.
\label{tab:EFTLQ}}
\centering{\small\renewcommand\baselinestretch{2}\selectfont
\begin{tabular*}{\textwidth}{l @{\extracolsep{\fill}}cc}
\hline \hline LQ    & Effective Operator & Fierz transformed effective operator \\ \hline
$U_1$    &        $-\dfrac{C_1}{\Lambda^2} ([\overline{d} \gamma^{\mu} P_{L} \ell][\overline{\ell} \gamma^{\mu} P_{L} d] + [\overline{u} \gamma^{\mu} P_{L} \nu][\overline{\nu} \gamma^{\mu} P_{L} u] )$            &        $-\dfrac{C_1}{\Lambda^2} ([\overline{d} \gamma^{\mu} P_{L} d][\overline{\ell} \gamma^{\mu} P_{L} \ell] + [\overline{u} \gamma^{\mu} P_{L} u][\overline{\nu} \gamma^{\mu} P_{L} \nu])$                              \\

&        $ - \dfrac{C_2}{\Lambda^2} [\overline{d} \gamma^{\mu} P_{R} \ell][\overline{\ell} \gamma^{\mu} P_{R} d]$            &        $ - \dfrac{C_2}{\Lambda^2} [\overline{d} \gamma^{\mu} P_{R} d][\overline{\ell} \gamma^{\mu} P_{R} \ell]$                              \\

$\widetilde{U}_1$   &    $-\dfrac{C}{\Lambda^2} [\overline{u} \gamma^{\mu} P_{R} \ell][\overline{\ell} \gamma^{\mu} P_{R} u]$                &    $- \dfrac{C}{\Lambda^2} [\overline{u} \gamma^{\mu} P_{R} u][\overline{\ell} \gamma^{\mu} P_{R} \ell]$                                  \\
\multirow{2}{*}{$V_2$}    & $-\dfrac{C_1}{\Lambda^2}([\overline{d^C} \gamma^{\mu} P_{L} \ell][\overline{\ell} \gamma^{\mu} P_{L} d^C] + [\overline{d^C} \gamma^{\mu} P_{L} \nu][\overline{\nu} \gamma^{\mu} P_{L} d^C])$ & $\dfrac{C_1}{\Lambda^2} ([\overline{d} \gamma^{\mu} P_{R} d][\overline{\ell} \gamma^{\mu} P_{L} \ell] + [\overline{d} \gamma^{\mu} P_{R} d][\overline{\nu} \gamma^{\mu} P_{L} \nu])$

\\ & $-\dfrac{C_2}{\Lambda^2}([\overline{u^C} \gamma^{\mu} P_{R} \ell][\overline{\ell} \gamma^{\mu} P_{R} u^C] + [\overline{d^C} \gamma^{\mu} P_{R} \ell][\overline{\ell} \gamma^{\mu} P_{R} d^C])$                   &      $+\dfrac{C_2}{\Lambda^2} ([\overline{u} \gamma^{\mu} P_{L} u][\overline{\ell} \gamma^{\mu} P_{R} \ell] + [\overline{d} \gamma^{\mu} P_{L} d][\overline{\ell} \gamma^{\mu} P_{R} \ell])$ \\
$\widetilde{V}_2$   &     $-\dfrac{C}{\Lambda^2}([\overline{u^C} \gamma^{\mu} P_{L} \ell][\overline{\ell} \gamma^{\mu} P_{L} u^C] + [\overline{u^C} \gamma^{\mu} P_{L} \nu][\overline{\nu} \gamma^{\mu} P_{L} u^C])$               &     $\dfrac{C}{\Lambda^2} ([\overline{u} \gamma^{\mu} P_{R} u][\overline{\ell} \gamma^{\mu} P_{L} \ell] + [\overline{u} \gamma^{\mu} P_{R} u][\overline{\nu} \gamma^{\mu} P_{L} \nu])$                                 \\
\multirow{2}{*}{$U_3$}    & $-\dfrac{C_1}{\Lambda^2} ([\overline{d} \gamma^{\mu} P_{L} \ell][\overline{\ell} \gamma^{\mu} P_{L} d] + [\overline{u} \gamma^{\mu} P_{L} \nu][\overline{\nu} \gamma^{\mu} P_{L} u]$ & $-\dfrac{C_1}{\Lambda^2} ([\overline{d} \gamma^{\mu} P_{L} d][\overline{\ell} \gamma^{\mu} P_{L} \ell] + [\overline{u} \gamma^{\mu} P_{L} u][\overline{\nu} \gamma^{\mu} P_{L} \nu]$

\\ & $+2[\overline{u} \gamma^{\mu} P_{L} \ell][\overline{\ell} \gamma^{\mu} P_{L} u] + 2[\overline{d} \gamma^{\mu} P_{L} \nu][\overline{\nu} \gamma^{\mu} P_{L} d])$                  &  $+ 2[\overline{u} \gamma^{\mu} P_{L} u][\overline{\ell} \gamma^{\mu} P_{L} \ell] + 2[\overline{d} \gamma^{\mu} P_{L} d][\overline{\nu} \gamma^{\mu} P_{L} \nu])$    \\
\hline
\hline
\end{tabular*}}
\end{table*}
%%%%%%%%%%%%%%%%%%%%%%%%%%%%%%%%%%%%%%%%%%%%%%%%%%%%%%%%%%%%%%%%%%
%%%%%%%%%%%%%%%%%%%%%%%%%%%%%%%%%%%%%%%%%%%%%%%%%%%%%%%%%%%%%%%%%%
\begin{figure*}
\centering
\captionsetup[subfigure]{labelformat=empty}
\subfloat[\quad\quad(a)]{\includegraphics[height=4.3cm,width=4.3cm]{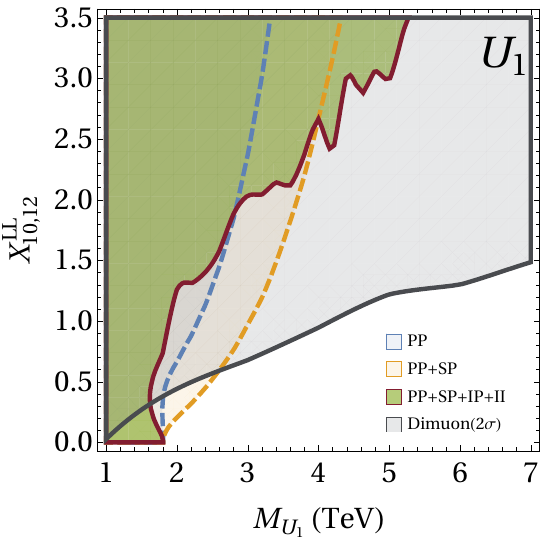}\label{fig:U1_XLL12_ExLim}}\hfill
\subfloat[\quad\quad(b)]{\includegraphics[height=4.3cm,width=4.3cm]{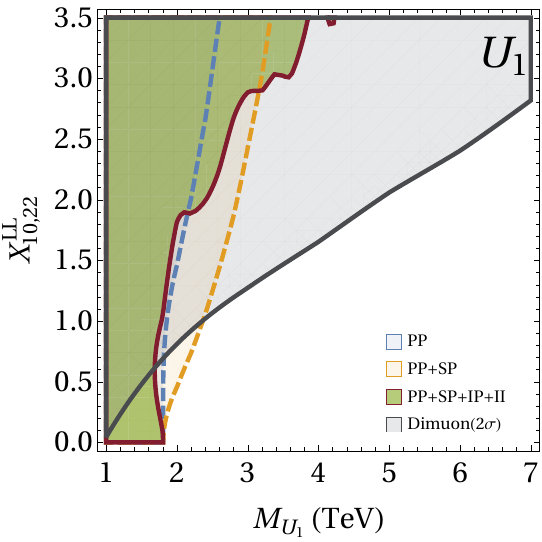}\label{fig:U1T_XTRR22_ExLim}}\hfill
\subfloat[\quad\quad(c)]{\includegraphics[height=4.3cm,width=4.3cm]{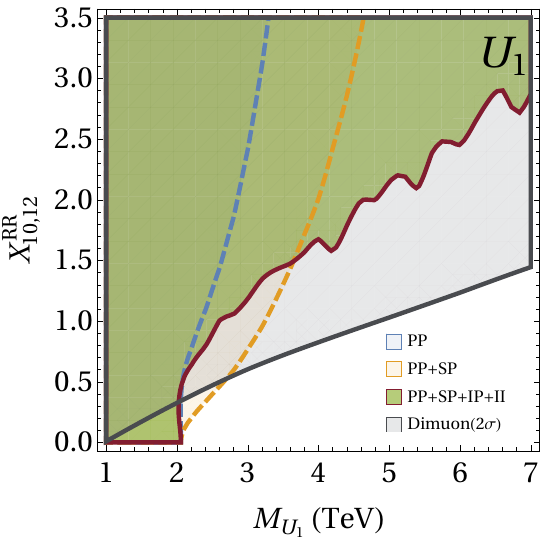}\label{fig:U1_XLL11_Combined_ExLim}}\hfill
\subfloat[\quad\quad(d)]{\includegraphics[height=4.3cm,width=4.3cm]{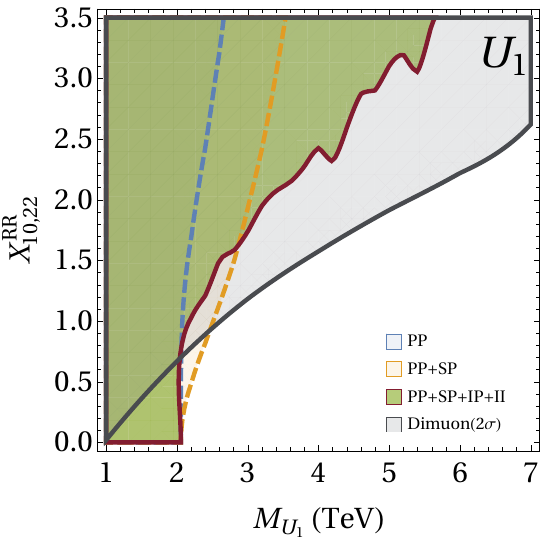}\label{fig:U1_XRR12_ExLim}}\\

\subfloat[\quad\quad(e)]{\includegraphics[height=4.3cm,width=4.3cm]{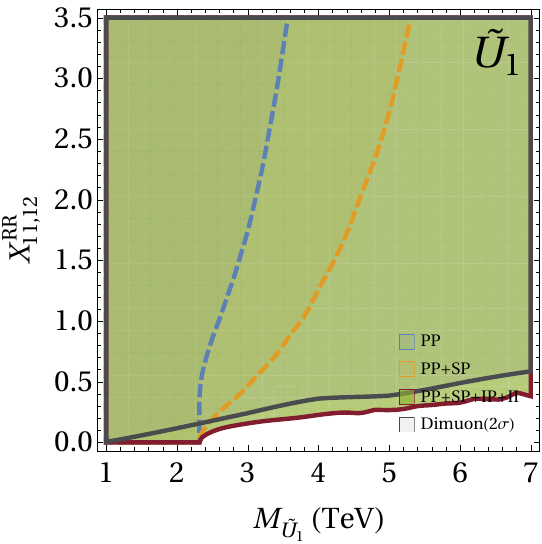}\label{fig:V2_XLR12_ExLim}}\hfill
\subfloat[\quad\quad(f)]{\includegraphics[height=4.3cm,width=4.3cm]{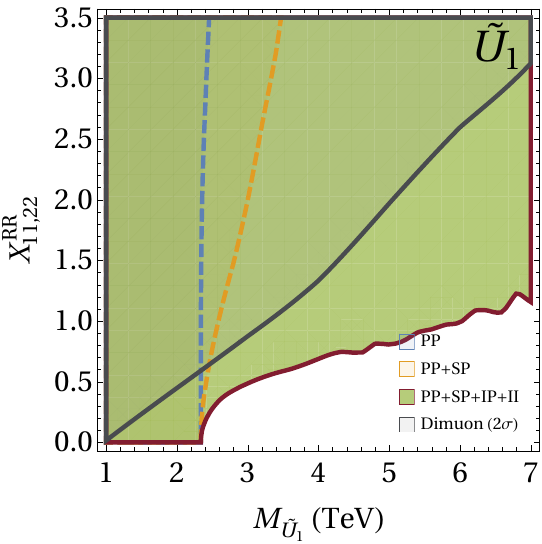}\label{fig:U1t_XRR22_ExLim}}\hfill
\subfloat[\quad\quad(g)]{\includegraphics[height=4.3cm,width=4.3cm]{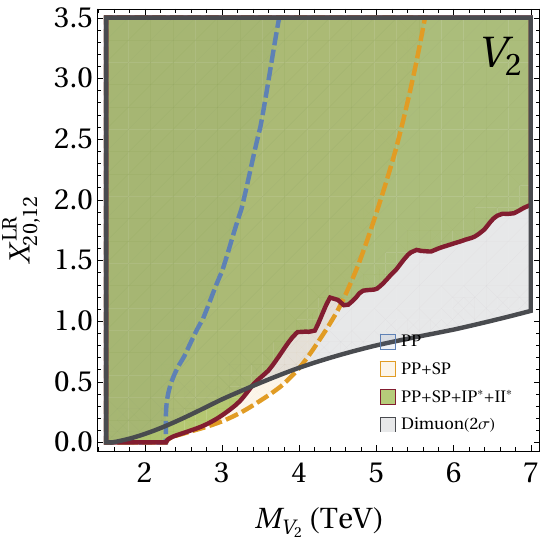}\label{fig:V2_XLR12_ExLim}}\hfill
\subfloat[\quad\quad(h)]{\includegraphics[height=4.3cm,width=4.3cm]{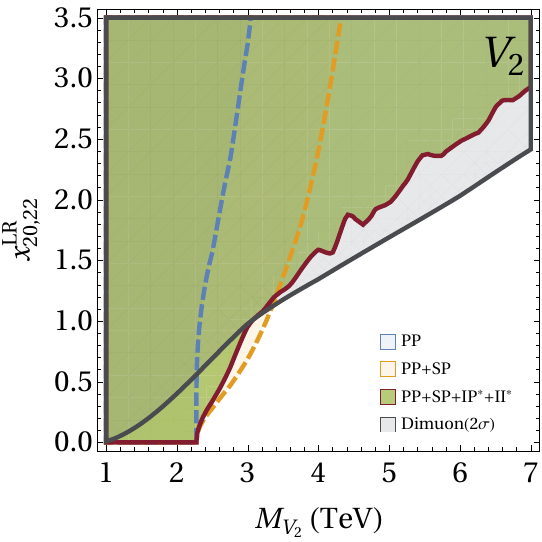}\label{fig:U3_XLL12_ExLim}}\\

\subfloat[\quad\quad(i)]{\includegraphics[height=4.3cm,width=4.3cm]{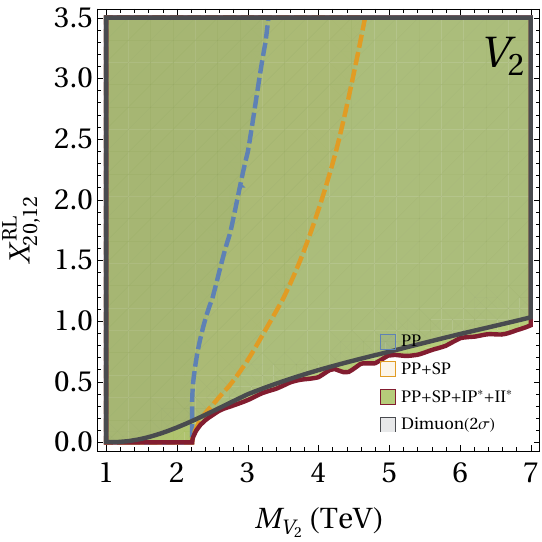}\label{fig:U3_XLL22_ExLim}}\hfill
\subfloat[\quad\quad(j)]{\includegraphics[height=4.3cm,width=4.3cm]{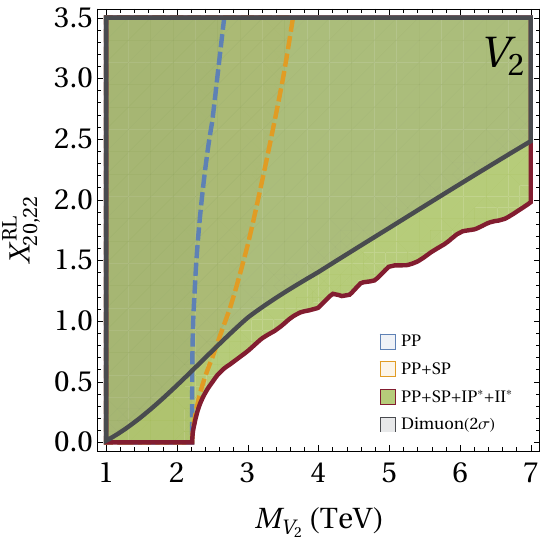}\label{fig:U3_XLL22_Ex_Lim_UA}}\hfill
\subfloat[\quad\quad(k)]{\includegraphics[height=4.3cm,width=4.3cm]{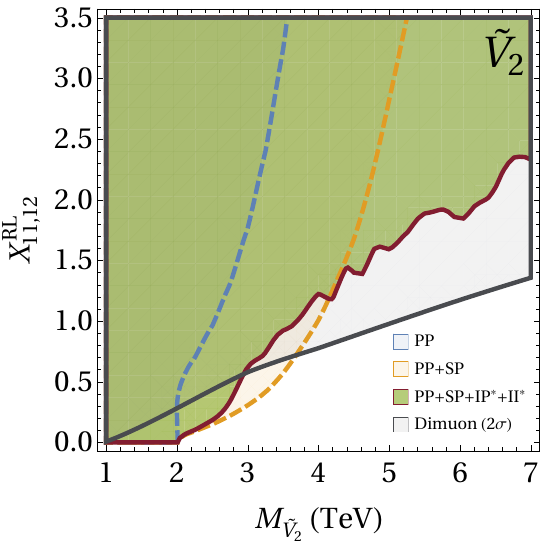}\label{fig:V2_XRL12_ExLim}}\hfill
\subfloat[\quad\quad(l)]{\includegraphics[height=4.3cm,width=4.3cm]{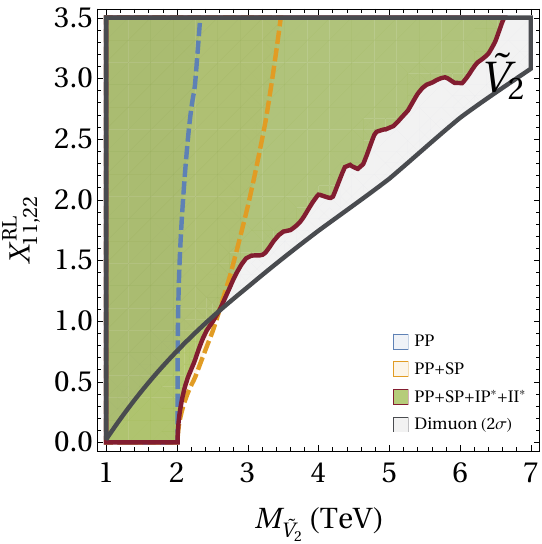}\label{fig:V2T_XTRL22_ExLim}}\\

\subfloat[\quad\quad(m)]{\includegraphics[height=4.3cm,width=4.3cm]{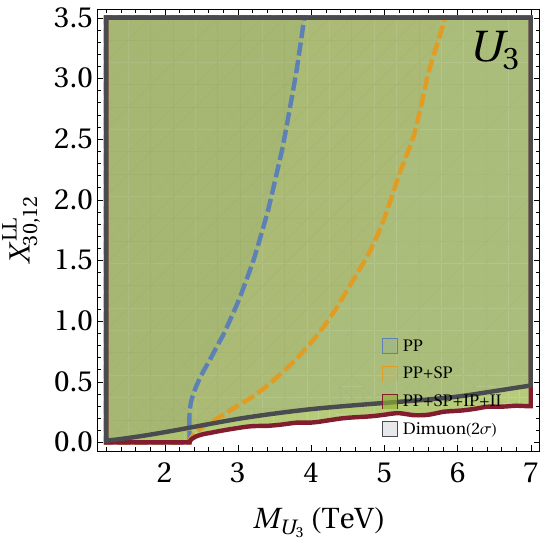}\label{fig:U3_XLL12_ExLim}}\hspace{0.17cm}
\subfloat[\quad\quad(n)]{\includegraphics[height=4.3cm,width=4.3cm]{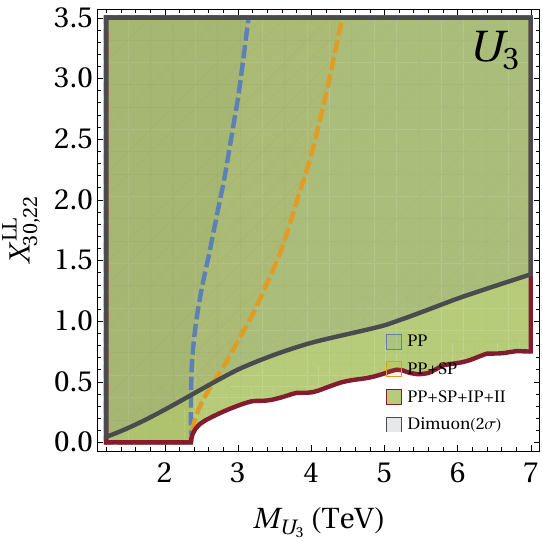}\label{fig:U3_XLL22_ExLim}}~\hfill~
\caption{Exclusion limits on all vLQs when only one coupling is nonzero (see Ref.~\cite{Bhaskar:2024wic} for notations). The contributions from the PP, PP+SP, and PP+SP+IP+II processes are separately marked; UA and DA indicate up and down-aligned scenarios, respectively (see Table~\ref{tab:vLQint}).}
\label{fig:MasslamEL}
\end{figure*}
%%%%%%%%%%%%%%%%%%%%%%%%%%%%%%%%%%%%%%%%%%%%%%%%%%%%%%%%%%%%%%%%%%
%%%%%%%%%%%%%%%%%%%%%%%%%%%%%%%%%%%%%%%%%%%%%%%%%%%%%%%%%%%%%%%%%%
\begin{figure*}
\centering
\captionsetup[subfigure]{labelformat=empty}
\subfloat[\quad\quad(a)]{\includegraphics[height=4.3cm,width=4.3cm]{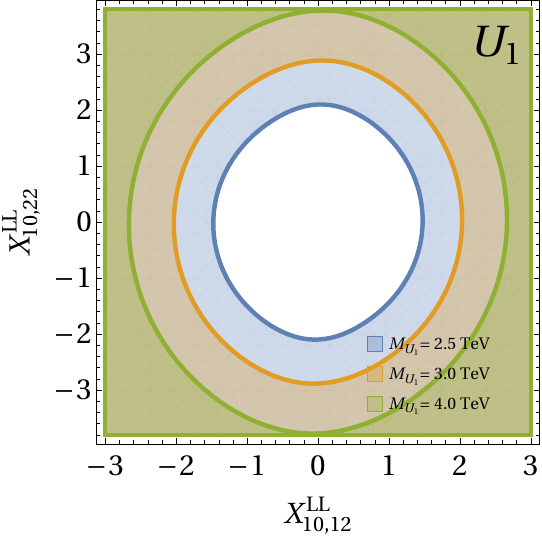}\label{fig:U1_XLL12_XLL22_ExLim}}\hfill
\subfloat[\quad\quad(b)]{\includegraphics[height=4.3cm,width=4.3cm]{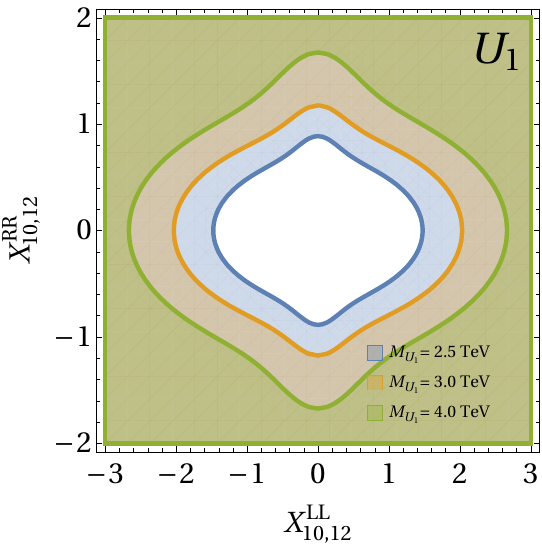}\label{fig:U1T_XTRR22_ExLim}}\hfill
\subfloat[\quad\quad(c)]{\includegraphics[height=4.3cm,width=4.3cm]{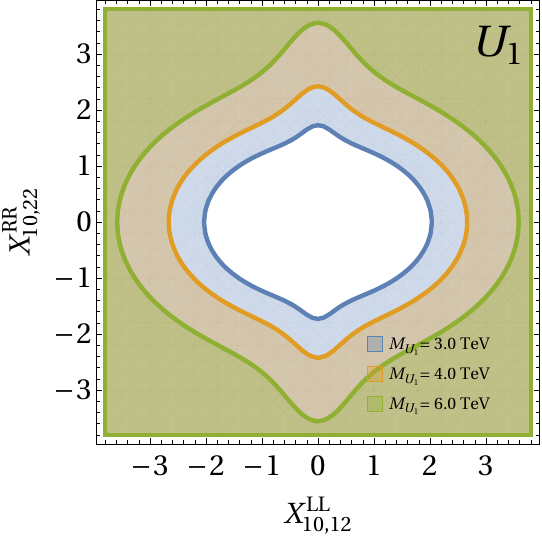}\label{fig:U1_XLL11_Combined_ExLim}}\hfill
\subfloat[\quad\quad(d)]{\includegraphics[height=4.3cm,width=4.3cm]{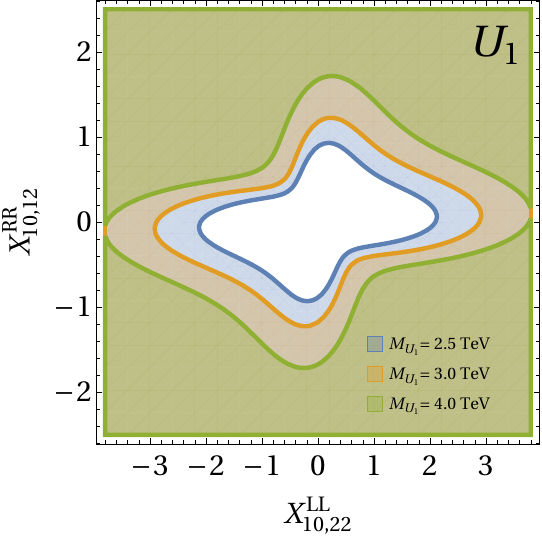}\label{fig:U1_XRR12_ExLim}}\\

\subfloat[\quad\quad(e)]{\includegraphics[height=4.3cm,width=4.3cm]{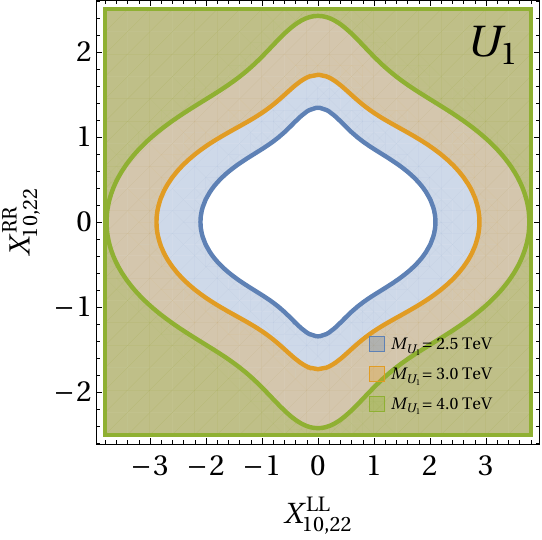}\label{fig:U1_XRR22_ExLim}}\hfill
\subfloat[\quad\quad(f)]{\includegraphics[height=4.3cm,width=4.3cm]{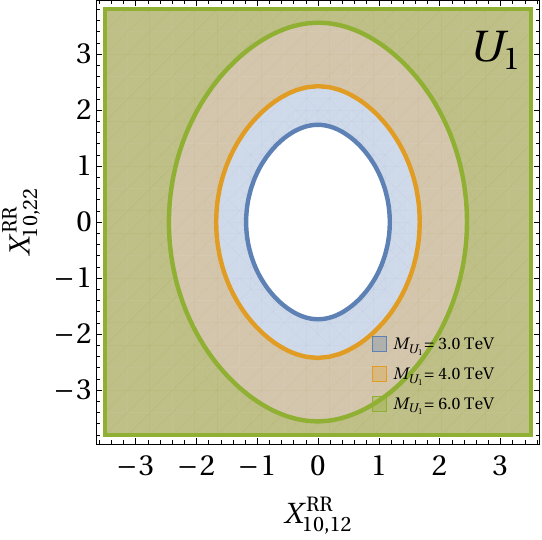}\label{fig:U1_XRR12_XRR22_Combined_ExLim}}\hfill
\subfloat[\quad\quad(g)]{\includegraphics[height=4.3cm,width=4.3cm]{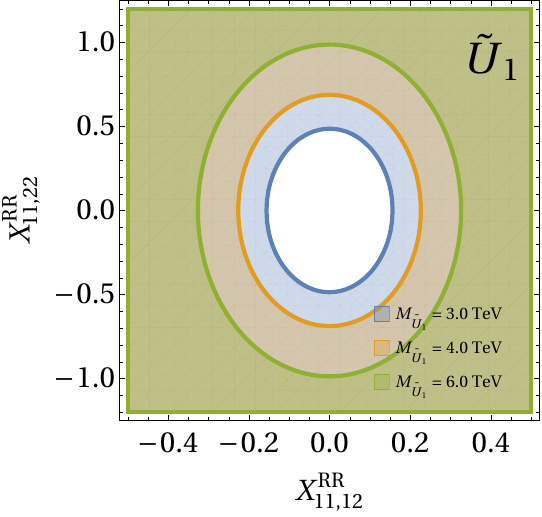}\label{fig:U1T_XTRR12_XTRR22_ExLim}}\hfill
\subfloat[\quad\quad(h)]{\includegraphics[height=4.3cm,width=4.3cm]{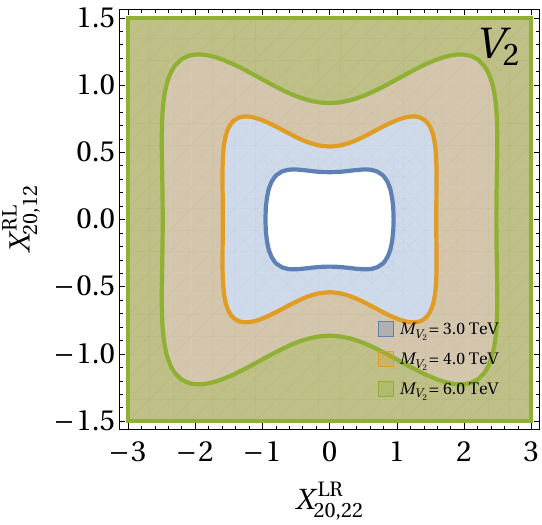}\label{fig:V2_XLR22_XRL12_ExLim}}\\

\subfloat[\quad\quad(i)]{\includegraphics[height=4.3cm,width=4.3cm]{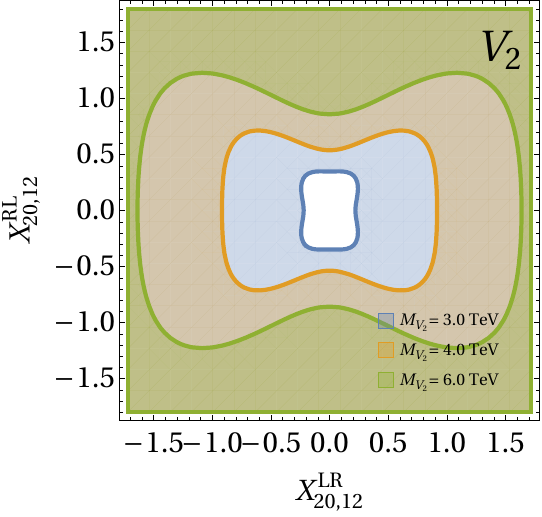}\label{fig:U3_XLL22_Ex_Lim_UA}}\hfill
\subfloat[\quad\quad(j)]{\includegraphics[height=4.3cm,width=4.3cm]{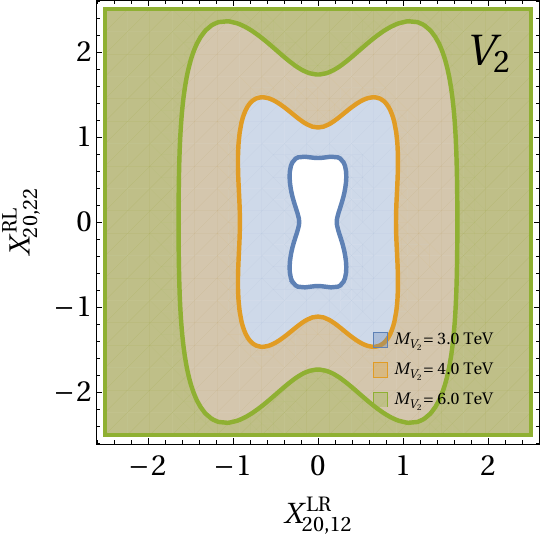}\label{fig:V2T_ExLim_XTRL12}}\hfill
\subfloat[\quad\quad(k)]{\includegraphics[height=4.3cm,width=4.3cm]{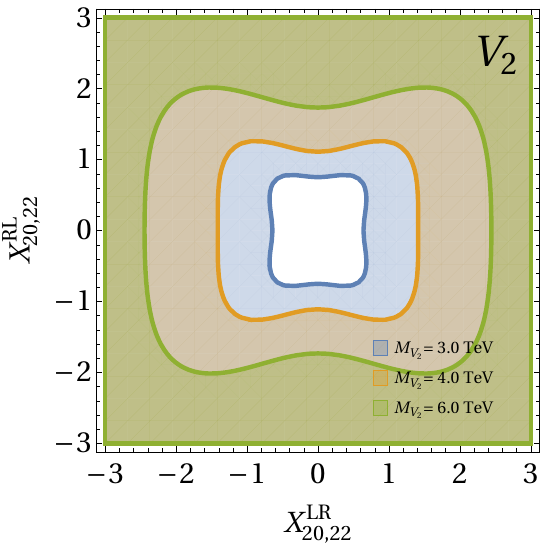}\label{fig:V2_XLR22_XRL22_Combined_ExLim}}\hfill
\subfloat[\quad\quad(l)]{\includegraphics[height=4.3cm,width=4.3cm]{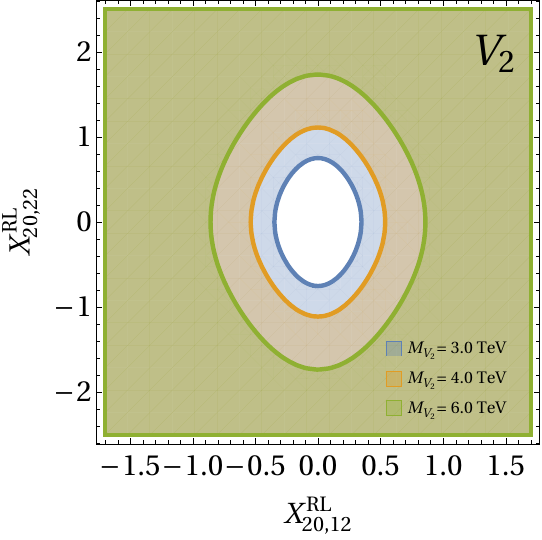}\label{fig:V2_XRL12_XRL22_Combined_ExLim}}\\

\subfloat[\quad\quad(m)]{\includegraphics[height=4.3cm,width=4.3cm]{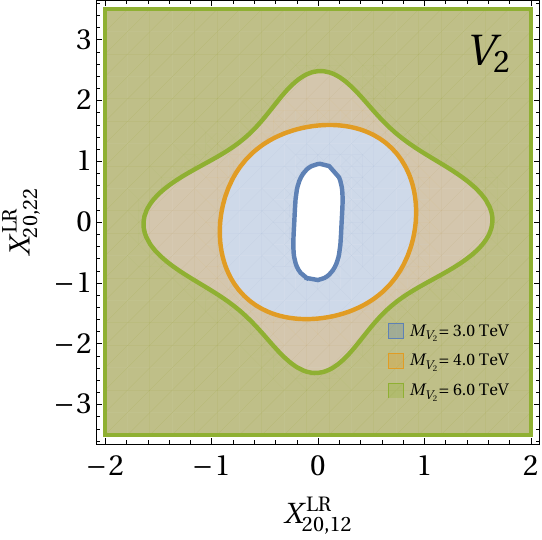}\label{fig:V2_XLR22_ExLim}}\hspace{0.17cm}
\subfloat[\quad\quad(n)]{\includegraphics[height=4.3cm,width=4.3cm]{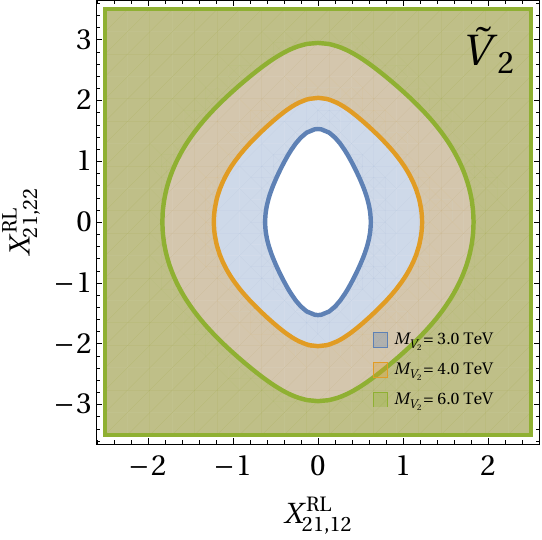}\label{fig:V2T_XTRL12_XTRL22_ExLim}}\hspace{0.17cm}
\subfloat[\quad\quad(o)]{\includegraphics[height=4.3cm,width=4.3cm]{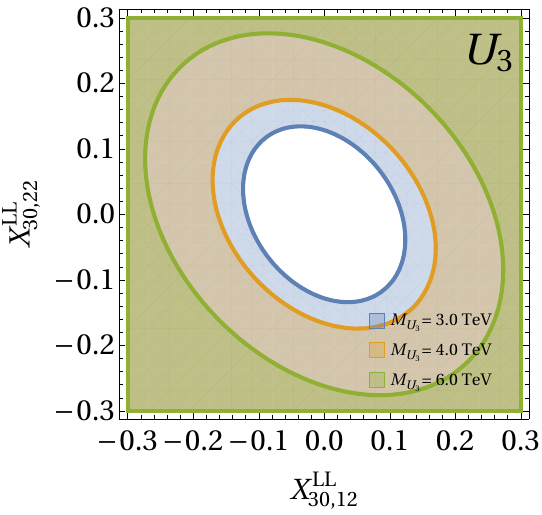}\label{fig:U3_XLL12_XLL22_Combined_ExLim}}~\hfill~
\caption{Same as Fig.~\ref{fig:MasslamEL}, except here, two couplings are nonzero.}
\label{fig:combined}
\end{figure*}
%%%%%%%%%%%%%%%%%%%%%%%%%%%%%%%%%%%%%%%%%%%%%%%%%%%%%%%%%%%%%%%%%%
%%%%%%%%%%%%%%%%%%%%%%%%%%%%%%%%%%%%%%%%%%%%%%%%%%%%%%%%%%%%%%%%%%
\begin{figure*}
\centering
\captionsetup[subfigure]{labelformat=empty}
\subfloat[\quad\quad(a)]{\includegraphics[height=4.3cm,width=4.3cm]{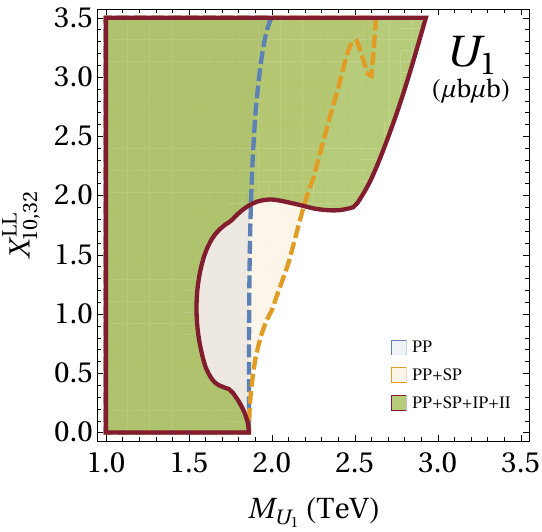}\label{fig:el1u1ll32}}\hfill
\subfloat[\quad\quad(b)]{\includegraphics[height=4.3cm,width=4.3cm]{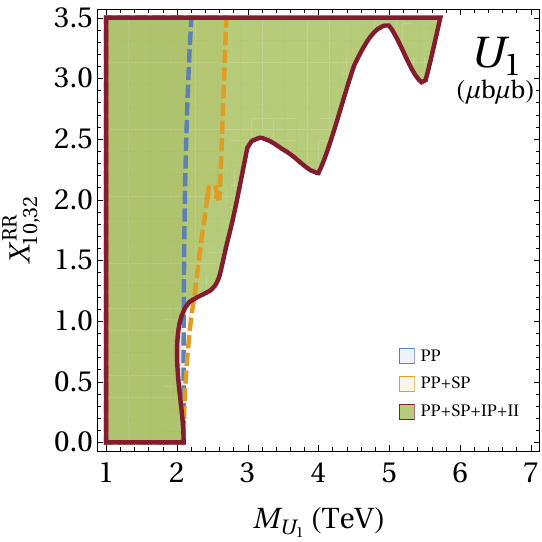}\label{fig:el1u1rr32}}\hfill
\subfloat[\quad\quad(c)]{\includegraphics[height=4.3cm,width=4.3cm]{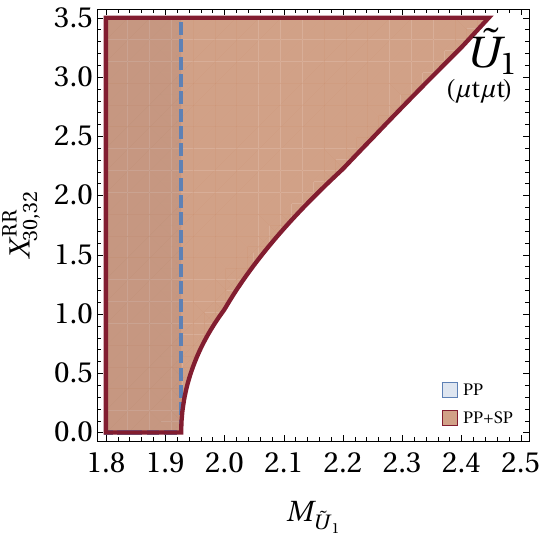}\label{fig:el1u1trr32}}\hfill
\subfloat[\quad\quad(d)]{\includegraphics[height=4.3cm,width=4.3cm]{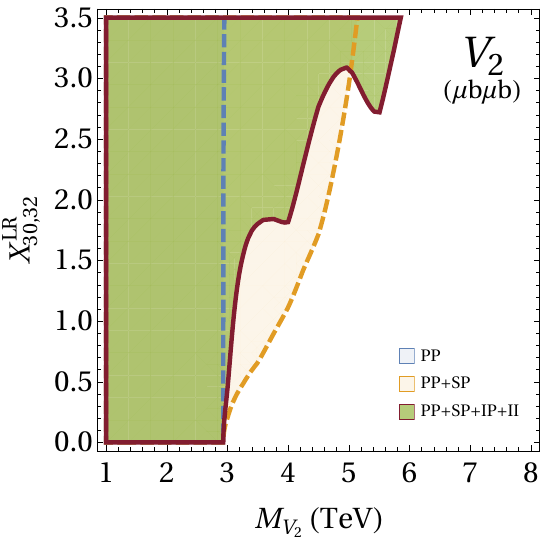}\label{fig:el1v2lr32bb}}\\
\subfloat[\quad\quad(e)]{\includegraphics[height=4.3cm,width=4.3cm]{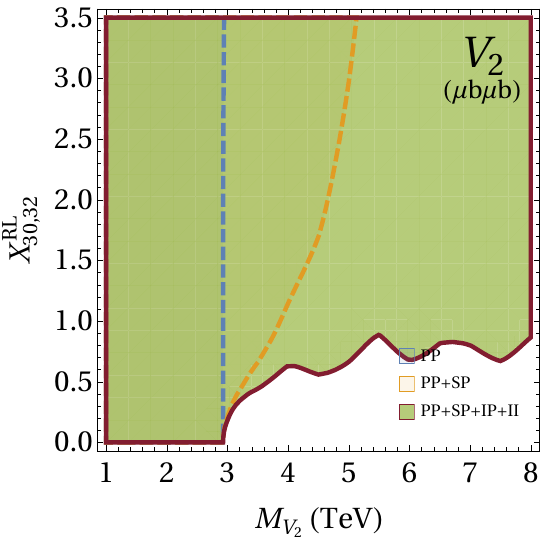}\label{fig:el1v2rl32bb}}\hfill
\subfloat[\quad\quad(f)]{\includegraphics[height=4.3cm,width=4.3cm]{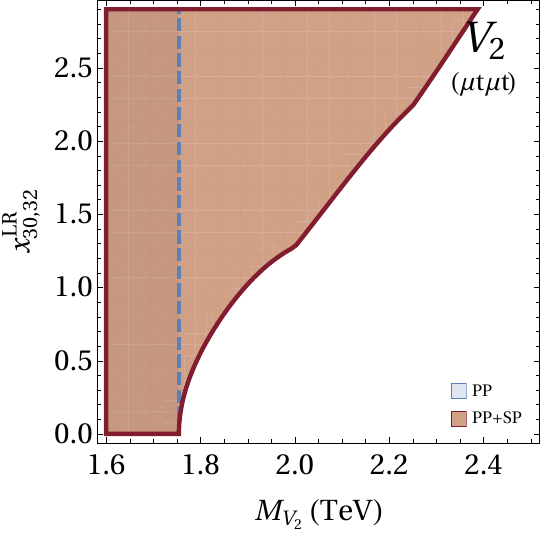}\label{fig:el1v2rl32tt}}\hfill
\subfloat[\quad\quad(g)]{\includegraphics[height=4.3cm,width=4.3cm]{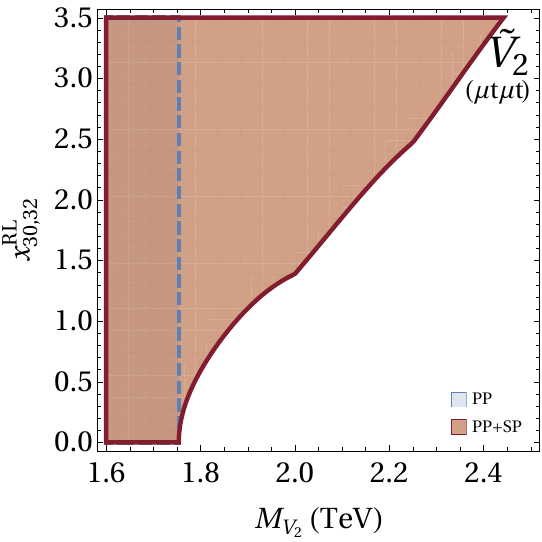}\label{fig:el1v2trl32tt}}\hfill
\subfloat[\quad\quad(h)]{\includegraphics[height=4.3cm,width=4.3cm]{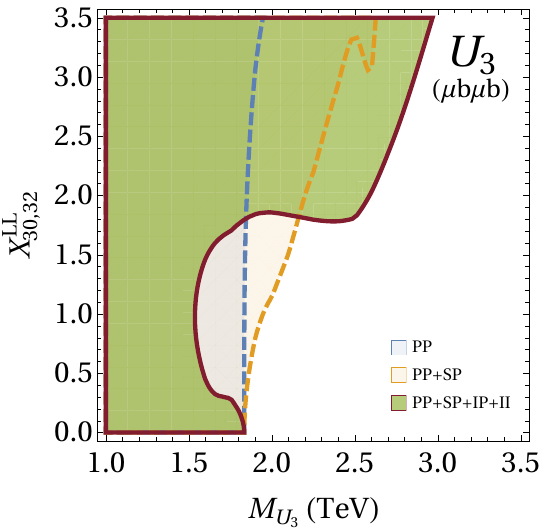}\label{fig:el1u3ll32bb}}\\
\subfloat[\quad\quad(i)]{\includegraphics[height=4.3cm,width=4.3cm]{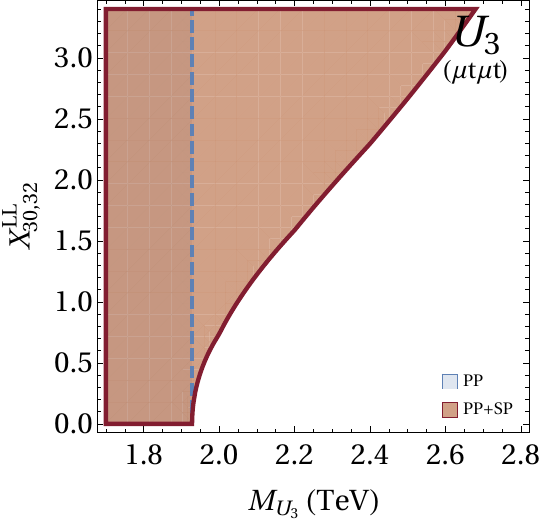}\label{fig:el1u3ll32}}
\hfill~
\caption{Single-coupling exclusion limits when the quark involved is a third-generation quark.  
\label{fig:ELOneCoup3rdgen} }
\end{figure*}
%%%%%%%%%%%%%%%%%%%%%%%%%%%%%%%%%%%%%%%%%%%%%%%%%%%%%%%%%%%%%%%%%%

%%%%%%%%%%%%%%%%%%%%%%%%%%%%%%%%%%%%%%%%%%%%%%%%%%%%%%%%%%%%%%%%%%
\begin{table}[!t]
\caption{Extrapolated maximum mass exclusion limit (TeV) for $x \approx \sqrt{4\pi}$ for different sLQs. To obtain these, we have assumed only one coupling is nonzero at a time.\label{tab:PertLim}}
\centering{\small\renewcommand\baselinestretch{1.6}\selectfont
\begin{tabular*}{\columnwidth}{l @{\extracolsep{\fill}}crcrcrr}
\hline
\multirow{2}{*}{sLQ}& \multirow{2}{*}{$x$} & Limit &\multirow{2}{*}{$x$} & Limit &\multirow{2}{*}{$x$} & \multicolumn{2}{c}{Limit (TeV)}\\\cline{7-8}
&& (TeV)&& (TeV)&& ($\mu b\m b$) &  ($\mu t \m t$) \\\hline\hline
\multirow{2}{*}{ $U_1$} &  $x^{LL}_{10,12}$& $5.3$  & $x^{LL}_{10,22}$ & $4.0$ & $x^{LL}_{10,32}$ & 2.8 & $-$\\ 
 & $x^{RR}_{10,12}$ & $8.1$ & $x^{RR}_{10,22}$ & $5.6$ & $x^{RR}_{10,32}$ & $6.0$ & $-$ \\\hline 
$\widetilde{U}_1$        & $x^{RR}_{11,12}$ & $57.0$  & $x^{RR}_{11,22}$  &  $21.5$ & $x^{RR}_{11,32}$ & $-$ & $2.4$ \\\hline
\multirow{2}{*}{$V_2$}     &  $x^{LR}_{20,12}$& $12.0$ & $x^{LR}_{20,22}$ & $8.2$  & $x^{LR}_{20,32}$ & $6.0$ & $2.4$\\ 
& $x^{RL}_{20,12}$& $27.0$ & $x^{RL}_{20,22}$ &$13.0$ & $x^{RL}_{20,32}$ & $40.0$ & $-$ \\\hline 
$\widetilde{V}_2$      & $x^{RL}_{21,12}$ & $9.5$ & $x^{RL}_{21,22}$ & $6.6$  & $x^{RL}_{21,32}$ & $-$ & $2.4$ \\\hline  
 $U_3$    & $x^{LL}_{30,12}$ & $85.0$  & $x^{LL}_{30,22}$ &  $33.0$ & $x^{LL}_{30,32}$ & $2.8$ & $2.7$\\ \hline
\end{tabular*}}
\end{table}
%%%%%%%%%%%%%%%%%%%%%%%%%%%%%%%%%%%%%%%%%%%%%%%%%%%%%%%%%%%%%%%%%%

\section{Exclusion limits}\label{sec:exclulim}
\noindent
In the limit $x \to 0$, vLQ mass exclusion limits are solely determined by the SM gauge couplings, and PP becomes the only production channel at the LHC. In Table~\ref{tab:SLQcombined}, we summarise such model-independent exclusion limits on the masses of various vLQ species. The QCD column shows the limits obtained from the $\mathcal{O}(\alpha_s^2 \alpha_e^0 \alpha_x^0)$ contribution to PP. The variation in limits across different vLQs arises from differences in their BRs. So far, in the literature, the QCD-QED mixed contributions to the vLQ mass exclusion limits were usually ignored. These additional contributions at $\mathcal{O}(\alpha_s^1 \alpha_e^1 \alpha_x^0)$ enhance the PP cross section, improving the bounds. We present results for two representative $\kappa$ values [Eq.~\eqref{eq:kappa}] --- the QCD+QED limits are much stronger than the pure QCD limits for vLQs with larger electric charges. However, due to the large uncertainty in the photon PDF in NNPDF, the limits can vary significantly. For example, for $\widetilde{U}_1$, they may range from $2016$ GeV to $2990$ GeV.\footnote{Currently, we are investigating the PDF dependence of this correction, which we will report elsewhere.}

\subsection{The singlets ($U_1$ and $\widetilde{U}_1$)}
\noindent
The $\mu\mu jj$ final state can emerge from $U_1$ interactions mediated by the following couplings: $x^{LL}_{10,12}, x^{LL}_{10,22}, x^{RR}_{10,12}$ and $x^{RR}_{10,22}$. Similarly, the couplings, $x^{RR}_{11,12}$ and $x^{RR}_{11,22}$ for $\widetilde{U}_1$ can also lead to the same final state. In Figs.~\ref{fig:U1_XLL12_ExLim}--\ref{fig:U1t_XRR22_ExLim}, we present the ($95$\% CL) exclusion limits on these couplings as functions of mass. These limits are obtained by recasting the LQ search results presented in Ref.~\cite{ATLAS:2020dsk} in the $\mu\mu jj$ channels. We make a comparison of how the single coupling limits change if we consider only the PP, PP+SP, or all contributions, i.e., PP+SP+IP+II together. We also present $2\sigma$ exclusion limits on couplings set using the high-$p_T$ dimuon (invariant mass distribution) tail on the same plot. 

Due to the destructive nature of the $U_1$ interference contribution, the dimuon-search limits are more stringent than the direct-search limits. A similar effect is also observed for the $S_1$ in Ref.~\cite{Bhaskar:2023ftn}. However, for $\widetilde{U}_1$, the direct-search limits are stronger, owing to the constructive interference, similar to the case of $\widetilde{S}_1$. For fixed values of $M_{\ell_q}$ and $x$, the IP and II cross sections are larger for those cases where first-generation quarks participate in the initial states, due to the larger PDFs compared to the second generation. As a result, vLQs coupling to first-generation quarks face tighter constraints. In the high-mass regime, where the limits are dominated by IP and II contributions—which depend on the coupling strength $x$ rather than on branching ratios—left- and right-handed couplings yield comparable exclusions. Some differences arise from the II terms, as the $Z$ boson couples differently to left- and right-handed fermions. The exclusion limits in the two-coupling plane for some fixed values of $M_{\ell_q}$ are presented in Figs.~\ref{fig:U1_XLL12_XLL22_ExLim}--\ref{fig:U1T_XTRR12_XTRR22_ExLim}.

\subsection{The doublets ($V_2$ and $\widetilde{V}_2$)}
\noindent
The $V_2$ can lead to the $\mu\mu jj$ final state through $x^{LR}_{20,12}$, $x^{LR}_{20,22}$, $x^{RL}_{20,12}$, and $x^{RL}_{20,22}$. With $LR$ couplings, both $V^{1/3}_{2}$ and $V^{4/3}_{2}$ components contribute to the $\mu\mu jj$ final state with a negative II. Whereas only $V^{4/3}_{2}$ contributes for $RL$ couplings with a positive II. In Figs.~\ref{fig:V2_XLR12_ExLim}--\ref{fig:V2T_XTRL22_ExLim}, we present the (95\% CL) exclusion limits~\cite{ATLAS:2020dsk} on these couplings as functions of mass. The exclusion limits in the two-coupling plane are presented in Figs.~\ref{fig:V2_XLR22_XRL12_ExLim}--\ref{fig:V2T_XTRL12_XTRL22_ExLim}.\footnote{
In the case of $V_2$ and $\widetilde{V}_2$, a straightforward evaluation of the interference contributions with our \texttt{FeynRules} models fails in \texttt{MadGraph}. While we are currently communicating with the authors of these packages to address the issue, in this paper, we bypassed the problem by taking an effective-operator approach by integrating out $V_2$ and $\widetilde{V}_2$ as described later in Sec.~\ref{sec:effoff}.}

\subsection{The triplet: $U_3$} 
\noindent
For $U_{3}$, the $x^{LL}_{30,12}$ and $x^{LL}_{30,22}$ couplings can lead to the $\mu\mu jj$ final state. Among the three components of $U_3$, only $U_{3}^{2/3}$ and $U_{3}^{5/3}$ interact with charged leptons. The contribution to II from the $U_{3}^{2/3}$ component is destructive, while that from $U_{3}^{5/3}$ is constructive. However, the overall effect of II remains positive for the involvement of quarks of the first and second generations in the initial states due to $\sqrt{2}$-times higher values of coupling for the positively interfering $U_{3}^{5/3}$ component (and also higher PDF for the first-generation case). For the third-generation case, the sign of II is negative, as the positively interfering contribution is missing due to the absence of the top quark PDF. The single coupling limits as functions of mass are shown in Figs.~\ref{fig:U3_XLL12_ExLim}--\ref{fig:U3_XLL22_ExLim}, whereas the two-coupling limits for some fixed masses are shown in Fig.~\ref{fig:U3_XLL12_XLL22_Combined_ExLim}. 

%%%%%%%%%%%%%%%%%%%%%%%%%%%%%%%%%%%%%%%%%%%%%%%%%%%%%%%%%%%%%%%%%%
\begin{figure*}%[!t]
\centering
\captionsetup[subfigure]{labelformat=empty}
\subfloat[\quad\quad(a)]{\includegraphics[height=6cm,width=6.375cm]{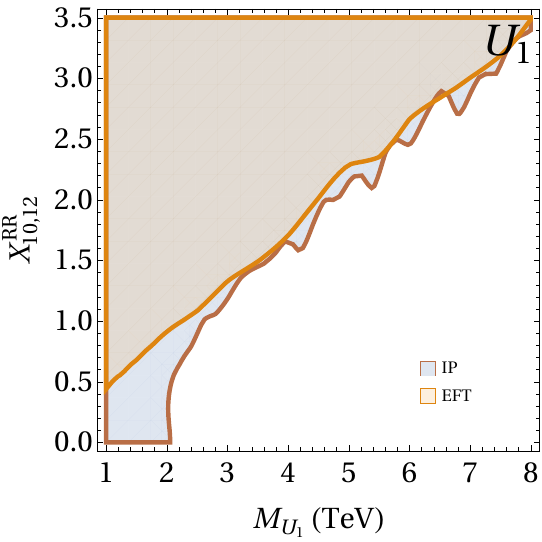}}\hspace{1cm}
\subfloat[\quad\quad(b)]{\includegraphics[height=6cm,width=6.375cm]{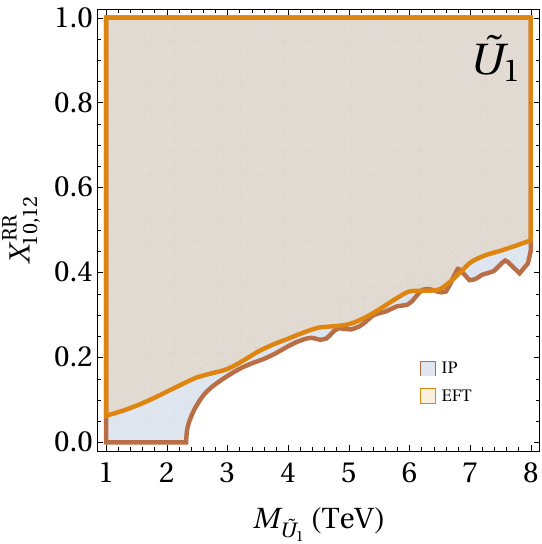}}
\caption{ Exclusion limits from the effective operator approach and the full theory in the high mass limit.
\label{EFTEL}}
\end{figure*}
%%%%%%%%%%%%%%%%%%%%%%%%%%%%%%%%%%%%%%%%%%%%%%%%%%%%%%%%%%%%%%%%%%
\subsection{Heavy jets: Third-generation quarks}
\noindent
So far, we have only considered light jets in the final states to obtain the exclusion limits. To obtain limits on the vLQ couplings with the third-generation quarks, we use the CMS $\mu\mu bb$~\cite{CMS:2024bnj} and $\mu\mu tt$~\cite{CMS:2022nty} searches in the context of LQ searches. If vLQs couple to the bottom quark (this happens for $U_1,~V_2$ and $U_3$), there is a possibility of IP and II productions (through the $t$-channel exchange) leading to a muon pair in the final state. Additionally, we need two $b$-tagged jets for recasting the $\mu\mu bb$ results. If bottom quarks are absent from the initial state of a process, it is unlikely to produce $b$ jets from QCD radiation. On the other hand, if a bottom quark is present in the initial state, which is the case here, the probability of having $b$ jets in the final state from the gluon splitting is higher. Therefore, substantial fractions of SP, IP and II would pass the selection cuts in the $\mu\mu bb$ search~\cite{CMS:2024bnj}. This leads to a stronger exclusion limit in the mass-coupling plane for vLQs [see Fig.~\ref{fig:el1v2rl32bb}] that constructively interfere with the SM $bb\to \mu\mu$ processes. On the other hand, in the case of destructive inference, the exclusion limits become much weaker [see Figs.~\ref{fig:el1u1ll32}, \ref{fig:el1u1rr32}, \ref{fig:el1v2lr32bb}, and \ref{fig:el1u3ll32bb}].

If vLQs couple to the top quark, we do not get any contribution from the coupling-dependent IP and II processes, as the top quark PDF essentially vanishes. However, 3BSP, like  $pp\to \ell_q\mu t\to \mu t\mu t$, can produce $\mu\mu tt$ final state. Combining PP and SP, we obtain the exclusion limits by recasting $\mu\mu tt$ results from Ref.~\cite{CMS:2022nty} for $\widetilde{U}_1, V_2, \widetilde{V}_2$, and $U_3$ as presented in Fig.~\ref{fig:ELOneCoup3rdgen}.

\subsection{Extrapolating the limits}
\noindent
The single coupling limits in Figs.~\ref{fig:MasslamEL} and \ref{fig:ELOneCoup3rdgen} show that the mass exclusion limits become stronger as the value of the couplings increases. Consequently, the LQ mass bound will settle at a maximum value as the coupling approaches $x \approx \sqrt{4 \pi}$ ($\sim$ the perturbative limit). Such a maximum bound will obviously depend on the LQ species, the coupling involved, the sign and the relative magnitudes of IP and II, etc. In Table~\ref{tab:PertLim}, we present the perturbative mass limits in the single coupling scenarios (see Ref.~\cite{Allwicher:2021rtd} for perturbative unitarity bounds on couplings). 

\subsection{Effective operator approach}\label{sec:effoff}
\noindent
A vLQ is exchanged in the $t$-channel in the IP and II channels, producing a lepton pair in the final state. If its mass is sufficiently high ($M_{\ell_q} \gg \sqrt{\hat{s}}$), one can integrate out the vLQs to obtain a description in terms of effective operators with appropriate Wilson coefficients (functions of vLQ parameters). This approach is closely related to that explored in Refs.~\cite{Allwicher:2024mzw,Aebischer:2025qhh}, where similar EFT operators are studied. The effective operator approach should give us the same bounds as the full theory for heavy vLQs. We consider only the dilepton final states in the one-coupling scenarios and write down the effective operators obtained by integrating out the $t$-channel vLQ propagator. In Table~\ref{tab:EFTLQ}, we present a list of these operators and, for convenience, their Fierz-transformed forms for all vLQs.  

Due to their small cross sections in the high mass region, the PP and SP processes hardly contribute to the exclusion limits, and the IP and II processes play the determining role. Hence, the limits from the effective operator approach agree with the ones obtained by combining only the IP and II contributions. In Fig.~\ref{EFTEL}, we see that the exclusion limits in the effective operator approach and the full theory are essentially the same in the high mass region, validating these two approaches in the high mass limits. From Fig.~\ref{EFTEL}, we see that both approaches give the same result for $M_{\ell_q}\approx \Lambda \gtrsim 3$~TeV for vLQs coupling to the first-generation quarks. For vLQs coupling to higher-generation quarks, we noticed that the point of matching shifts to the higher-mass side.

As mentioned before, we estimated the limits in Fig.~\ref{fig:MasslamEL} using the effective operator approach for $V_2$ and $\widetilde{V}_2$. This would give us a reasonable estimation of the exclusion limits for those vLQs, especially when they couple to the first and second generation quarks, for $M_{\ell_q}\gtrsim 3-4$~TeV. (Note that the PP and SP events for those vLQs are generated with \texttt{MadGraph} without any issue, and the cross sections are cross-checked with the other vLQs having similar interactions.) 

\section{Summary and Discussions}\label{sec:conclu}
\noindent
In this paper, we derived exclusion limits on vLQs using the latest LHC searches for LQs in the $\mu \mu jj$ channel. Pair production searches typically provide largely model-independent bounds on vLQ masses and branching ratios. In contrast, constraints on the LQ-quark-lepton couplings arise from the high-$p_T$ tail of the dilepton invariant mass spectrum (or the monolepton plus missing transverse energy, $\ell+\slashed E_{\rm T}$, distribution). This study extended our earlier work~\cite{Bhaskar:2023ftn}, where we had advocated for a systematic combination of events from various production modes -- PP, SP, IP and II -- for scalar LQs leading to the same (or experimentally indistinguishable) final states. To accurately determine the number of signal events at a given parameter point (specified by mass and couplings), our combined-signal approach offers a consistent method for estimating limits from processes contributing to the $\mu \mu jj$ channel or the $\mu\mu$ invariant mass spectrum.

For illustrations, we presented exclusion limits on all vLQs in the mass-coupling and two-coupling (for fixed masses) planes. Additionally, we used the LHC $\mu b\mu b$ and $\mu t\mu t$ data to constrain the vLQ couplings to third-generation quarks. Our analysis can be extended to include any number of couplings simultaneously for all vLQs listed in Table~\ref{tab:vLQint}. (This will be incorporated in a forthcoming version of our \texttt{TooLQit} package~\cite{Bhaskar:2024wic}.)

Typically, model-independent mass limits on vLQs are more stringent than those on sLQs, owing to the larger PP cross sections at the same mass. We derived such model-independent mass bounds using PP on all vLQ species, and these results were summarised in Table~\ref{tab:SLQcombined} for two benchmark choices of $\kappa$. Furthermore, we included the impact of mixed QCD+QED contributions to PP in our mass exclusion limits. We found these contributions particularly significant for vLQs with large electromagnetic charges. We have used the photon PDF from the NNPDF sets to estimate the QCD+QED effects. However, the photon PDF in NNPDF sets has significant uncertainties. A detailed study of the impact of the associated PDF uncertainties on the exclusion limits, including comparisons with photon PDFs from alternative PDF sets, is beyond the scope of the current paper. This will be investigated in future work.

A vLQ is exchanged in the $t$-channel in the IP and II processes. In the high-mass limit, the vLQ can be integrated out, yielding corresponding four-fermion (two quarks-two leptons) effective operators with appropriate Wilson coefficients. In the heavy regime, the full theory containing the vLQ and the EFT approach should produce consistent results, thereby validating the EFT framework. In Table~\ref{tab:EFTLQ}, we listed all relevant effective operators -- along with their Fierz-transformed versions -- that arise from integrating out $t$-channel vLQs in the dilepton channel. The EFT approach is particularly powerful when the resonant production of heavy particles lies just beyond the direct reach of the collider. Our analysis can be straightforwardly extended to include scalar and vector LQs in the dilepton (same or different flavour) or $\ell+\slashed E_{\rm T}$ channels. By recasting data from these channels, one can set bounds on the coefficients of the effective operators, which in turn constrain the underlying LQ parameters.

\section*{Acknowledgements}\label{sec:aknow}
\noindent
We thank Olivier Mattelaer for discussion on \texttt{MadGraph} related issues. T.M. acknowledges partial support from the SERB/ANRF,
Government of India, through the Core Research Grant
(CRG) No. CRG/2023/007031. 
R.S. acknowledges the Prime Minister's Research Fellowship (PMRF ID: 0802000).

\def\bibfont{\small}
\bibliography{References}{}

%merlin.mbs apsrev4-1.bst 2010-07-25 4.21a (PWD, AO, DPC) hacked
%Control: key (0)
%Control: author (0) dotless jnrlst
%Control: editor formatted (1) identically to author
%Control: production of article title (0) allowed
%Control: page (1) range
%Control: year (0) verbatim
%Control: production of eprint (0) enabled
\begin{thebibliography}{74}%
\makeatletter
\providecommand \@ifxundefined [1]{%
 \@ifx{#1\undefined}
}%
\providecommand \@ifnum [1]{%
 \ifnum #1\expandafter \@firstoftwo
 \else \expandafter \@secondoftwo
 \fi
}%
\providecommand \@ifx [1]{%
 \ifx #1\expandafter \@firstoftwo
 \else \expandafter \@secondoftwo
 \fi
}%
\providecommand \natexlab [1]{#1}%
\providecommand \enquote  [1]{``#1''}%
\providecommand \bibnamefont  [1]{#1}%
\providecommand \bibfnamefont [1]{#1}%
\providecommand \citenamefont [1]{#1}%
\providecommand \href@noop [0]{\@secondoftwo}%
\providecommand \href [0]{\begingroup \@sanitize@url \@href}%
\providecommand \@href[1]{\@@startlink{#1}\@@href}%
\providecommand \@@href[1]{\endgroup#1\@@endlink}%
\providecommand \@sanitize@url [0]{\catcode `\\12\catcode `\$12\catcode
  `\&12\catcode `\#12\catcode `\^12\catcode `\_12\catcode `\%12\relax}%
\providecommand \@@startlink[1]{}%
\providecommand \@@endlink[0]{}%
\providecommand \url  [0]{\begingroup\@sanitize@url \@url }%
\providecommand \@url [1]{\endgroup\@href {#1}{\urlprefix }}%
\providecommand \urlprefix  [0]{URL }%
\providecommand \Eprint [0]{\href }%
\providecommand \doibase [0]{http://dx.doi.org/}%
\providecommand \selectlanguage [0]{\@gobble}%
\providecommand \bibinfo  [0]{\@secondoftwo}%
\providecommand \bibfield  [0]{\@secondoftwo}%
\providecommand \translation [1]{[#1]}%
\providecommand \BibitemOpen [0]{}%
\providecommand \bibitemStop [0]{}%
\providecommand \bibitemNoStop [0]{.\EOS\space}%
\providecommand \EOS [0]{\spacefactor3000\relax}%
\providecommand \BibitemShut  [1]{\csname bibitem#1\endcsname}%
\let\auto@bib@innerbib\@empty
%</preamble>
\bibitem [{\citenamefont {Pati}\ and\ \citenamefont
  {Salam}(1974)}]{Pati:1974yy}%
  \BibitemOpen
  \bibfield  {author} {\bibinfo {author} {\bibfnamefont {Jogesh~C.}\
  \bibnamefont {Pati}}\ and\ \bibinfo {author} {\bibfnamefont {Abdus}\
  \bibnamefont {Salam}},\ }\bibfield  {title} {\enquote {\bibinfo {title}
  {{Lepton Number as the Fourth Color}},}\ }\href {\doibase
  10.1103/PhysRevD.10.275} {\bibfield  {journal} {\bibinfo  {journal} {Phys.
  Rev. D}\ }\textbf {\bibinfo {volume} {10}},\ \bibinfo {pages} {275--289}
  (\bibinfo {year} {1974})},\ \bibinfo {note} {[Erratum: Phys.Rev.D 11,
  703--703 (1975)]}\BibitemShut {NoStop}%
\bibitem [{\citenamefont {Georgi}\ and\ \citenamefont
  {Glashow}(1974)}]{Georgi:1974sy}%
  \BibitemOpen
  \bibfield  {author} {\bibinfo {author} {\bibfnamefont {H.}~\bibnamefont
  {Georgi}}\ and\ \bibinfo {author} {\bibfnamefont {S.~L.}\ \bibnamefont
  {Glashow}},\ }\bibfield  {title} {\enquote {\bibinfo {title} {{Unity of All
  Elementary Particle Forces}},}\ }\href {\doibase 10.1103/PhysRevLett.32.438}
  {\bibfield  {journal} {\bibinfo  {journal} {Phys. Rev. Lett.}\ }\textbf
  {\bibinfo {volume} {32}},\ \bibinfo {pages} {438--441} (\bibinfo {year}
  {1974})}\BibitemShut {NoStop}%
\bibitem [{\citenamefont {Fritzsch}\ and\ \citenamefont
  {Minkowski}(1975)}]{Fritzsch:1974nn}%
  \BibitemOpen
  \bibfield  {author} {\bibinfo {author} {\bibfnamefont {Harald}\ \bibnamefont
  {Fritzsch}}\ and\ \bibinfo {author} {\bibfnamefont {Peter}\ \bibnamefont
  {Minkowski}},\ }\bibfield  {title} {\enquote {\bibinfo {title} {{Unified
  Interactions of Leptons and Hadrons}},}\ }\href {\doibase
  10.1016/0003-4916(75)90211-0} {\bibfield  {journal} {\bibinfo  {journal}
  {Annals Phys.}\ }\textbf {\bibinfo {volume} {93}},\ \bibinfo {pages}
  {193--266} (\bibinfo {year} {1975})}\BibitemShut {NoStop}%
\bibitem [{\citenamefont {Farhi}\ and\ \citenamefont
  {Susskind}(1981)}]{Farhi:1980xs}%
  \BibitemOpen
  \bibfield  {author} {\bibinfo {author} {\bibfnamefont {Edward}\ \bibnamefont
  {Farhi}}\ and\ \bibinfo {author} {\bibfnamefont {Leonard}\ \bibnamefont
  {Susskind}},\ }\bibfield  {title} {\enquote {\bibinfo {title}
  {{Technicolor}},}\ }\href {\doibase 10.1016/0370-1573(81)90173-3} {\bibfield
  {journal} {\bibinfo  {journal} {Phys. Rept.}\ }\textbf {\bibinfo {volume}
  {74}},\ \bibinfo {pages} {277} (\bibinfo {year} {1981})}\BibitemShut
  {NoStop}%
\bibitem [{\citenamefont {Schrempp}\ and\ \citenamefont
  {Schrempp}(1985)}]{Schrempp:1984nj}%
  \BibitemOpen
  \bibfield  {author} {\bibinfo {author} {\bibfnamefont {Barbara}\ \bibnamefont
  {Schrempp}}\ and\ \bibinfo {author} {\bibfnamefont {Fridger}\ \bibnamefont
  {Schrempp}},\ }\bibfield  {title} {\enquote {\bibinfo {title} {{LIGHT
  LEPTOQUARKS}},}\ }\href {\doibase 10.1016/0370-2693(85)91450-9} {\bibfield
  {journal} {\bibinfo  {journal} {Phys. Lett. B}\ }\textbf {\bibinfo {volume}
  {153}},\ \bibinfo {pages} {101--107} (\bibinfo {year} {1985})}\BibitemShut
  {NoStop}%
\bibitem [{\citenamefont {Wudka}(1986)}]{Wudka:1985ef}%
  \BibitemOpen
  \bibfield  {author} {\bibinfo {author} {\bibfnamefont {Jose}\ \bibnamefont
  {Wudka}},\ }\bibfield  {title} {\enquote {\bibinfo {title} {{COMPOSITE
  LEPTOQUARKS}},}\ }\href {\doibase 10.1016/0370-2693(86)90356-4} {\bibfield
  {journal} {\bibinfo  {journal} {Phys. Lett. B}\ }\textbf {\bibinfo {volume}
  {167}},\ \bibinfo {pages} {337--342} (\bibinfo {year} {1986})}\BibitemShut
  {NoStop}%
\bibitem [{\citenamefont {Barbier}\ \emph {et~al.}(2005)\citenamefont {Barbier}
  \emph {et~al.}}]{Barbier:2004ez}%
  \BibitemOpen
  \bibfield  {author} {\bibinfo {author} {\bibfnamefont {R.}~\bibnamefont
  {Barbier}} \emph {et~al.},\ }\bibfield  {title} {\enquote {\bibinfo {title}
  {{R-parity violating supersymmetry}},}\ }\href {\doibase
  10.1016/j.physrep.2005.08.006} {\bibfield  {journal} {\bibinfo  {journal}
  {Phys. Rept.}\ }\textbf {\bibinfo {volume} {420}},\ \bibinfo {pages} {1--202}
  (\bibinfo {year} {2005})},\ \Eprint {http://arxiv.org/abs/hep-ph/0406039}
  {arXiv:hep-ph/0406039} \BibitemShut {NoStop}%
\bibitem [{\citenamefont {Dor\v{s}ner}\ \emph {et~al.}(2016)\citenamefont
  {Dor\v{s}ner}, \citenamefont {Fajfer}, \citenamefont {Greljo}, \citenamefont
  {Kamenik},\ and\ \citenamefont {Ko\v{s}nik}}]{Dorsner:2016wpm}%
  \BibitemOpen
  \bibfield  {author} {\bibinfo {author} {\bibfnamefont {I.}~\bibnamefont
  {Dor\v{s}ner}}, \bibinfo {author} {\bibfnamefont {S.}~\bibnamefont {Fajfer}},
  \bibinfo {author} {\bibfnamefont {A.}~\bibnamefont {Greljo}}, \bibinfo
  {author} {\bibfnamefont {J.~F.}\ \bibnamefont {Kamenik}}, \ and\ \bibinfo
  {author} {\bibfnamefont {N.}~\bibnamefont {Ko\v{s}nik}},\ }\bibfield  {title}
  {\enquote {\bibinfo {title} {{Physics of leptoquarks in precision experiments
  and at particle colliders}},}\ }\href {\doibase
  10.1016/j.physrep.2016.06.001} {\bibfield  {journal} {\bibinfo  {journal}
  {Phys. Rept.}\ }\textbf {\bibinfo {volume} {641}},\ \bibinfo {pages} {1--68}
  (\bibinfo {year} {2016})},\ \Eprint {http://arxiv.org/abs/1603.04993}
  {arXiv:1603.04993 [hep-ph]} \BibitemShut {NoStop}%
\bibitem [{\citenamefont {Gon\c{c}alves}\ \emph {et~al.}(2023)\citenamefont
  {Gon\c{c}alves}, \citenamefont {Morais}, \citenamefont {Onofre},\ and\
  \citenamefont {Pasechnik}}]{Goncalves:2023qpz}%
  \BibitemOpen
  \bibfield  {author} {\bibinfo {author} {\bibfnamefont {Jo\~ao}\ \bibnamefont
  {Gon\c{c}alves}}, \bibinfo {author} {\bibfnamefont {Ant\'onio~P.}\
  \bibnamefont {Morais}}, \bibinfo {author} {\bibfnamefont {Ant\'onio}\
  \bibnamefont {Onofre}}, \ and\ \bibinfo {author} {\bibfnamefont {Roman}\
  \bibnamefont {Pasechnik}},\ }\bibfield  {title} {\enquote {\bibinfo {title}
  {{Exploring mixed lepton-quark interactions in non-resonant leptoquark
  production at the LHC}},}\ }\href {\doibase 10.1007/JHEP11(2023)147}
  {\bibfield  {journal} {\bibinfo  {journal} {JHEP}\ }\textbf {\bibinfo
  {volume} {11}},\ \bibinfo {pages} {147} (\bibinfo {year} {2023})},\ \Eprint
  {http://arxiv.org/abs/2306.15460} {arXiv:2306.15460 [hep-ph]} \BibitemShut
  {NoStop}%
\bibitem [{\citenamefont {Da~Rold}(2023)}]{DaRold:2023hmx}%
  \BibitemOpen
  \bibfield  {author} {\bibinfo {author} {\bibfnamefont {Leandro}\ \bibnamefont
  {Da~Rold}},\ }\bibfield  {title} {\enquote {\bibinfo {title} {{$b\to
  c\tau\bar\nu$ anomaly from a minimal composite $S_1$ leptoquark}},}\
  }\href@noop {} {\  (\bibinfo {year} {2023})},\ \Eprint
  {http://arxiv.org/abs/2311.04062} {arXiv:2311.04062 [hep-ph]} \BibitemShut
  {NoStop}%
\bibitem [{ATL()}]{ATLASPublic}%
  \BibitemOpen
  \href@noop {} {\enquote {\bibinfo {title} {{ATLAS Public Results}},}\
  }\bibinfo {howpublished}
  {\url{https://twiki.cern.ch/twiki/bin/view/AtlasPublic}}\BibitemShut
  {NoStop}%
\bibitem [{CMS()}]{CMSPublic}%
  \BibitemOpen
  \href@noop {} {\enquote {\bibinfo {title} {{CMS Exotica Public Physics
  Results}},}\ }\bibinfo {howpublished}
  {\url{https://twiki.cern.ch/twiki/bin/view/CMSPublic/PhysicsResultsEXO}}\BibitemShut
  {NoStop}%
\bibitem [{\citenamefont {Blumlein}\ \emph {et~al.}(1997)\citenamefont
  {Blumlein}, \citenamefont {Boos},\ and\ \citenamefont
  {Kryukov}}]{Blumlein:1996qp}%
  \BibitemOpen
  \bibfield  {author} {\bibinfo {author} {\bibfnamefont {Johannes}\
  \bibnamefont {Blumlein}}, \bibinfo {author} {\bibfnamefont {Edward}\
  \bibnamefont {Boos}}, \ and\ \bibinfo {author} {\bibfnamefont {Alexander}\
  \bibnamefont {Kryukov}},\ }\bibfield  {title} {\enquote {\bibinfo {title}
  {{Leptoquark pair production in hadronic interactions}},}\ }\href {\doibase
  10.1007/s002880050538} {\bibfield  {journal} {\bibinfo  {journal} {Z. Phys.
  C}\ }\textbf {\bibinfo {volume} {76}},\ \bibinfo {pages} {137--153} (\bibinfo
  {year} {1997})},\ \Eprint {http://arxiv.org/abs/hep-ph/9610408}
  {arXiv:hep-ph/9610408} \BibitemShut {NoStop}%
\bibitem [{\citenamefont {Dorsner}\ \emph {et~al.}(2014)\citenamefont
  {Dorsner}, \citenamefont {Fajfer},\ and\ \citenamefont
  {Greljo}}]{Dorsner:2014axa}%
  \BibitemOpen
  \bibfield  {author} {\bibinfo {author} {\bibfnamefont {Ilja}\ \bibnamefont
  {Dorsner}}, \bibinfo {author} {\bibfnamefont {Svjetlana}\ \bibnamefont
  {Fajfer}}, \ and\ \bibinfo {author} {\bibfnamefont {Admir}\ \bibnamefont
  {Greljo}},\ }\bibfield  {title} {\enquote {\bibinfo {title} {{Cornering
  Scalar Leptoquarks at LHC}},}\ }\href {\doibase 10.1007/JHEP10(2014)154}
  {\bibfield  {journal} {\bibinfo  {journal} {JHEP}\ }\textbf {\bibinfo
  {volume} {10}},\ \bibinfo {pages} {154} (\bibinfo {year} {2014})},\ \Eprint
  {http://arxiv.org/abs/1406.4831} {arXiv:1406.4831 [hep-ph]} \BibitemShut
  {NoStop}%
\bibitem [{\citenamefont {Diaz}\ \emph {et~al.}(2017)\citenamefont {Diaz},
  \citenamefont {Schmaltz},\ and\ \citenamefont {Zhong}}]{Diaz:2017lit}%
  \BibitemOpen
  \bibfield  {author} {\bibinfo {author} {\bibfnamefont {Bastian}\ \bibnamefont
  {Diaz}}, \bibinfo {author} {\bibfnamefont {Martin}\ \bibnamefont {Schmaltz}},
  \ and\ \bibinfo {author} {\bibfnamefont {Yi-Ming}\ \bibnamefont {Zhong}},\
  }\bibfield  {title} {\enquote {\bibinfo {title} {{The leptoquark
  Hunter\textquoteright{}s guide: Pair production}},}\ }\href {\doibase
  10.1007/JHEP10(2017)097} {\bibfield  {journal} {\bibinfo  {journal} {JHEP}\
  }\textbf {\bibinfo {volume} {10}},\ \bibinfo {pages} {097} (\bibinfo {year}
  {2017})},\ \Eprint {http://arxiv.org/abs/1706.05033} {arXiv:1706.05033
  [hep-ph]} \BibitemShut {NoStop}%
\bibitem [{\citenamefont {Dey}\ \emph {et~al.}(2018)\citenamefont {Dey},
  \citenamefont {Kar}, \citenamefont {Mitra}, \citenamefont {Spannowsky},\ and\
  \citenamefont {Vincent}}]{Dey:2017ede}%
  \BibitemOpen
  \bibfield  {author} {\bibinfo {author} {\bibfnamefont {Ujjal~Kumar}\
  \bibnamefont {Dey}}, \bibinfo {author} {\bibfnamefont {Deepak}\ \bibnamefont
  {Kar}}, \bibinfo {author} {\bibfnamefont {Manimala}\ \bibnamefont {Mitra}},
  \bibinfo {author} {\bibfnamefont {Michael}\ \bibnamefont {Spannowsky}}, \
  and\ \bibinfo {author} {\bibfnamefont {Aaron~C.}\ \bibnamefont {Vincent}},\
  }\bibfield  {title} {\enquote {\bibinfo {title} {{Searching for Leptoquarks
  at IceCube and the LHC}},}\ }\href {\doibase 10.1103/PhysRevD.98.035014}
  {\bibfield  {journal} {\bibinfo  {journal} {Phys. Rev. D}\ }\textbf {\bibinfo
  {volume} {98}},\ \bibinfo {pages} {035014} (\bibinfo {year} {2018})},\
  \Eprint {http://arxiv.org/abs/1709.02009} {arXiv:1709.02009 [hep-ph]}
  \BibitemShut {NoStop}%
\bibitem [{\citenamefont {Bandyopadhyay}\ and\ \citenamefont
  {Mandal}(2018)}]{Bandyopadhyay:2018syt}%
  \BibitemOpen
  \bibfield  {author} {\bibinfo {author} {\bibfnamefont {Priyotosh}\
  \bibnamefont {Bandyopadhyay}}\ and\ \bibinfo {author} {\bibfnamefont {Rusa}\
  \bibnamefont {Mandal}},\ }\bibfield  {title} {\enquote {\bibinfo {title}
  {{Revisiting scalar leptoquark at the LHC}},}\ }\href {\doibase
  10.1140/epjc/s10052-018-5959-x} {\bibfield  {journal} {\bibinfo  {journal}
  {Eur. Phys. J. C}\ }\textbf {\bibinfo {volume} {78}},\ \bibinfo {pages} {491}
  (\bibinfo {year} {2018})},\ \Eprint {http://arxiv.org/abs/1801.04253}
  {arXiv:1801.04253 [hep-ph]} \BibitemShut {NoStop}%
\bibitem [{\citenamefont {Schmaltz}\ and\ \citenamefont
  {Zhong}(2019)}]{Schmaltz:2018nls}%
  \BibitemOpen
  \bibfield  {author} {\bibinfo {author} {\bibfnamefont {Martin}\ \bibnamefont
  {Schmaltz}}\ and\ \bibinfo {author} {\bibfnamefont {Yi-Ming}\ \bibnamefont
  {Zhong}},\ }\bibfield  {title} {\enquote {\bibinfo {title} {{The leptoquark
  Hunter\textquoteright{}s guide: large coupling}},}\ }\href {\doibase
  10.1007/JHEP01(2019)132} {\bibfield  {journal} {\bibinfo  {journal} {JHEP}\
  }\textbf {\bibinfo {volume} {01}},\ \bibinfo {pages} {132} (\bibinfo {year}
  {2019})},\ \Eprint {http://arxiv.org/abs/1810.10017} {arXiv:1810.10017
  [hep-ph]} \BibitemShut {NoStop}%
\bibitem [{\citenamefont {Bhaskar}\ \emph {et~al.}(2020)\citenamefont
  {Bhaskar}, \citenamefont {Das}, \citenamefont {De},\ and\ \citenamefont
  {Mitra}}]{Bhaskar:2020kdr}%
  \BibitemOpen
  \bibfield  {author} {\bibinfo {author} {\bibfnamefont {Arvind}\ \bibnamefont
  {Bhaskar}}, \bibinfo {author} {\bibfnamefont {Debottam}\ \bibnamefont {Das}},
  \bibinfo {author} {\bibfnamefont {Bibhabasu}\ \bibnamefont {De}}, \ and\
  \bibinfo {author} {\bibfnamefont {Subhadip}\ \bibnamefont {Mitra}},\
  }\bibfield  {title} {\enquote {\bibinfo {title} {{Enhancing scalar
  productions with leptoquarks at the LHC}},}\ }\href {\doibase
  10.1103/PhysRevD.102.035002} {\bibfield  {journal} {\bibinfo  {journal}
  {Phys. Rev. D}\ }\textbf {\bibinfo {volume} {102}},\ \bibinfo {pages}
  {035002} (\bibinfo {year} {2020})},\ \Eprint
  {http://arxiv.org/abs/2002.12571} {arXiv:2002.12571 [hep-ph]} \BibitemShut
  {NoStop}%
\bibitem [{\citenamefont {Greljo}\ and\ \citenamefont
  {Selimovic}(2021)}]{Greljo:2020tgv}%
  \BibitemOpen
  \bibfield  {author} {\bibinfo {author} {\bibfnamefont {Admir}\ \bibnamefont
  {Greljo}}\ and\ \bibinfo {author} {\bibfnamefont {Nudzeim}\ \bibnamefont
  {Selimovic}},\ }\bibfield  {title} {\enquote {\bibinfo {title} {{Lepton-Quark
  Fusion at Hadron Colliders, precisely}},}\ }\href {\doibase
  10.1007/JHEP03(2021)279} {\bibfield  {journal} {\bibinfo  {journal} {JHEP}\
  }\textbf {\bibinfo {volume} {03}},\ \bibinfo {pages} {279} (\bibinfo {year}
  {2021})},\ \Eprint {http://arxiv.org/abs/2012.02092} {arXiv:2012.02092
  [hep-ph]} \BibitemShut {NoStop}%
\bibitem [{\citenamefont {Dor\v{s}ner}\ \emph {et~al.}(2021)\citenamefont
  {Dor\v{s}ner}, \citenamefont {Fajfer},\ and\ \citenamefont
  {Lejli\'c}}]{Dorsner:2021chv}%
  \BibitemOpen
  \bibfield  {author} {\bibinfo {author} {\bibfnamefont {Ilja}\ \bibnamefont
  {Dor\v{s}ner}}, \bibinfo {author} {\bibfnamefont {Svjetlana}\ \bibnamefont
  {Fajfer}}, \ and\ \bibinfo {author} {\bibfnamefont {Ajla}\ \bibnamefont
  {Lejli\'c}},\ }\bibfield  {title} {\enquote {\bibinfo {title} {{Novel
  Leptoquark Pair Production at LHC}},}\ }\href {\doibase
  10.1007/JHEP05(2021)167} {\bibfield  {journal} {\bibinfo  {journal} {JHEP}\
  }\textbf {\bibinfo {volume} {05}},\ \bibinfo {pages} {167} (\bibinfo {year}
  {2021})},\ \Eprint {http://arxiv.org/abs/2103.11702} {arXiv:2103.11702
  [hep-ph]} \BibitemShut {NoStop}%
\bibitem [{\citenamefont {Crivellin}\ \emph {et~al.}(2021)\citenamefont
  {Crivellin}, \citenamefont {M\"{u}ller},\ and\ \citenamefont
  {Schnell}}]{Crivellin:2021egp}%
  \BibitemOpen
  \bibfield  {author} {\bibinfo {author} {\bibfnamefont {Andreas}\ \bibnamefont
  {Crivellin}}, \bibinfo {author} {\bibfnamefont {Dario}\ \bibnamefont
  {M\"{u}ller}}, \ and\ \bibinfo {author} {\bibfnamefont {Luc}\ \bibnamefont
  {Schnell}},\ }\bibfield  {title} {\enquote {\bibinfo {title} {{Combined
  constraints on first generation leptoquarks}},}\ }\href {\doibase
  10.1103/PhysRevD.103.115023} {\bibfield  {journal} {\bibinfo  {journal}
  {Phys. Rev. D}\ }\textbf {\bibinfo {volume} {103}},\ \bibinfo {pages}
  {115023} (\bibinfo {year} {2021})},\ \bibinfo {note} {[Addendum: Phys.Rev.D
  104, 055020 (2021)]},\ \Eprint {http://arxiv.org/abs/2104.06417}
  {arXiv:2104.06417 [hep-ph]} \BibitemShut {NoStop}%
\bibitem [{\citenamefont {Bandyopadhyay}\ \emph {et~al.}(2022)\citenamefont
  {Bandyopadhyay}, \citenamefont {Karan}, \citenamefont {Mandal},\ and\
  \citenamefont {Parashar}}]{Bandyopadhyay:2021pld}%
  \BibitemOpen
  \bibfield  {author} {\bibinfo {author} {\bibfnamefont {Priyotosh}\
  \bibnamefont {Bandyopadhyay}}, \bibinfo {author} {\bibfnamefont {Anirban}\
  \bibnamefont {Karan}}, \bibinfo {author} {\bibfnamefont {Rusa}\ \bibnamefont
  {Mandal}}, \ and\ \bibinfo {author} {\bibfnamefont {Snehashis}\ \bibnamefont
  {Parashar}},\ }\bibfield  {title} {\enquote {\bibinfo {title}
  {{Distinguishing signatures of scalar leptoquarks at hadron and muon
  colliders}},}\ }\href {\doibase 10.1140/epjc/s10052-022-10809-9} {\bibfield
  {journal} {\bibinfo  {journal} {Eur. Phys. J. C}\ }\textbf {\bibinfo {volume}
  {82}},\ \bibinfo {pages} {916} (\bibinfo {year} {2022})},\ \Eprint
  {http://arxiv.org/abs/2108.06506} {arXiv:2108.06506 [hep-ph]} \BibitemShut
  {NoStop}%
\bibitem [{\citenamefont {Bhaskar}\ \emph
  {et~al.}(2022{\natexlab{a}})\citenamefont {Bhaskar}, \citenamefont {Das},
  \citenamefont {De}, \citenamefont {Mitra}, \citenamefont {Nayak},\ and\
  \citenamefont {Neeraj}}]{Bhaskar:2022ygp}%
  \BibitemOpen
  \bibfield  {author} {\bibinfo {author} {\bibfnamefont {Arvind}\ \bibnamefont
  {Bhaskar}}, \bibinfo {author} {\bibfnamefont {Debottam}\ \bibnamefont {Das}},
  \bibinfo {author} {\bibfnamefont {Bibhabasu}\ \bibnamefont {De}}, \bibinfo
  {author} {\bibfnamefont {Subhadip}\ \bibnamefont {Mitra}}, \bibinfo {author}
  {\bibfnamefont {Aruna~Kumar}\ \bibnamefont {Nayak}}, \ and\ \bibinfo {author}
  {\bibfnamefont {Cyrin}\ \bibnamefont {Neeraj}},\ }\bibfield  {title}
  {\enquote {\bibinfo {title} {{Leptoquark-assisted singlet-mediated di-Higgs
  production at the LHC}},}\ }\href {\doibase 10.1016/j.physletb.2022.137341}
  {\bibfield  {journal} {\bibinfo  {journal} {Phys. Lett. B}\ }\textbf
  {\bibinfo {volume} {833}},\ \bibinfo {pages} {137341} (\bibinfo {year}
  {2022}{\natexlab{a}})},\ \Eprint {http://arxiv.org/abs/2205.12210}
  {arXiv:2205.12210 [hep-ph]} \BibitemShut {NoStop}%
\bibitem [{\citenamefont {Parashar}\ \emph {et~al.}(2022)\citenamefont
  {Parashar}, \citenamefont {Karan}, \citenamefont {Avnish}, \citenamefont
  {Bandyopadhyay},\ and\ \citenamefont {Ghosh}}]{Parashar:2022wrd}%
  \BibitemOpen
  \bibfield  {author} {\bibinfo {author} {\bibfnamefont {Snehashis}\
  \bibnamefont {Parashar}}, \bibinfo {author} {\bibfnamefont {Anirban}\
  \bibnamefont {Karan}}, \bibinfo {author} {\bibnamefont {Avnish}}, \bibinfo
  {author} {\bibfnamefont {Priyotosh}\ \bibnamefont {Bandyopadhyay}}, \ and\
  \bibinfo {author} {\bibfnamefont {Kirtiman}\ \bibnamefont {Ghosh}},\
  }\bibfield  {title} {\enquote {\bibinfo {title} {{Phenomenology of scalar
  leptoquarks at the LHC in explaining the radiative neutrino masses, muon g-2,
  and lepton flavor violating observables}},}\ }\href {\doibase
  10.1103/PhysRevD.106.095040} {\bibfield  {journal} {\bibinfo  {journal}
  {Phys. Rev. D}\ }\textbf {\bibinfo {volume} {106}},\ \bibinfo {pages}
  {095040} (\bibinfo {year} {2022})},\ \Eprint
  {http://arxiv.org/abs/2209.05890} {arXiv:2209.05890 [hep-ph]} \BibitemShut
  {NoStop}%
\bibitem [{\citenamefont {Ghosh}\ \emph
  {et~al.}(2023{\natexlab{a}})\citenamefont {Ghosh}, \citenamefont {Konar},
  \citenamefont {Saha},\ and\ \citenamefont {Seth}}]{Ghosh:2023ocz}%
  \BibitemOpen
  \bibfield  {author} {\bibinfo {author} {\bibfnamefont {Anupam}\ \bibnamefont
  {Ghosh}}, \bibinfo {author} {\bibfnamefont {Partha}\ \bibnamefont {Konar}},
  \bibinfo {author} {\bibfnamefont {Debashis}\ \bibnamefont {Saha}}, \ and\
  \bibinfo {author} {\bibfnamefont {Satyajit}\ \bibnamefont {Seth}},\
  }\bibfield  {title} {\enquote {\bibinfo {title} {{Precise probing and
  discrimination of third-generation scalar leptoquarks}},}\ }\href {\doibase
  10.1103/PhysRevD.108.035030} {\bibfield  {journal} {\bibinfo  {journal}
  {Phys. Rev. D}\ }\textbf {\bibinfo {volume} {108}},\ \bibinfo {pages}
  {035030} (\bibinfo {year} {2023}{\natexlab{a}})},\ \Eprint
  {http://arxiv.org/abs/2304.02890} {arXiv:2304.02890 [hep-ph]} \BibitemShut
  {NoStop}%
\bibitem [{\citenamefont {Fl\'orez}\ \emph {et~al.}(2023)\citenamefont
  {Fl\'orez}, \citenamefont {Jones-P\'erez}, \citenamefont {Gurrola},
  \citenamefont {Rodriguez},\ and\ \citenamefont {Pe\~nuela
  Parra}}]{Florez:2023jdb}%
  \BibitemOpen
  \bibfield  {author} {\bibinfo {author} {\bibfnamefont {A.}~\bibnamefont
  {Fl\'orez}}, \bibinfo {author} {\bibfnamefont {J.}~\bibnamefont
  {Jones-P\'erez}}, \bibinfo {author} {\bibfnamefont {A.}~\bibnamefont
  {Gurrola}}, \bibinfo {author} {\bibfnamefont {C.}~\bibnamefont {Rodriguez}},
  \ and\ \bibinfo {author} {\bibfnamefont {J.}~\bibnamefont {Pe\~nuela
  Parra}},\ }\bibfield  {title} {\enquote {\bibinfo {title} {{On the
  sensitivity reach of $\textrm{LQ}$ production with preferential couplings to
  third generation fermions at the LHC}},}\ }\href {\doibase
  10.1140/epjc/s10052-023-12177-4} {\bibfield  {journal} {\bibinfo  {journal}
  {Eur. Phys. J. C}\ }\textbf {\bibinfo {volume} {83}},\ \bibinfo {pages}
  {1023} (\bibinfo {year} {2023})},\ \Eprint {http://arxiv.org/abs/2307.11070}
  {arXiv:2307.11070 [hep-ph]} \BibitemShut {NoStop}%
\bibitem [{\citenamefont {Arganda}\ \emph {et~al.}(2023)\citenamefont
  {Arganda}, \citenamefont {D\'\i{}az}, \citenamefont {Perez}, \citenamefont
  {Sand\'a~Seoane},\ and\ \citenamefont {Szynkman}}]{Arganda:2023qni}%
  \BibitemOpen
  \bibfield  {author} {\bibinfo {author} {\bibfnamefont {Ernesto}\ \bibnamefont
  {Arganda}}, \bibinfo {author} {\bibfnamefont {Daniel~A.}\ \bibnamefont
  {D\'\i{}az}}, \bibinfo {author} {\bibfnamefont {Andres~D.}\ \bibnamefont
  {Perez}}, \bibinfo {author} {\bibfnamefont {Rosa~M.}\ \bibnamefont
  {Sand\'a~Seoane}}, \ and\ \bibinfo {author} {\bibfnamefont {Alejandro}\
  \bibnamefont {Szynkman}},\ }\bibfield  {title} {\enquote {\bibinfo {title}
  {{LHC Study of Third-Generation Scalar Leptoquarks with Machine-Learned
  Likelihoods}},}\ }\href@noop {} {\  (\bibinfo {year} {2023})},\ \Eprint
  {http://arxiv.org/abs/2309.05407} {arXiv:2309.05407 [hep-ph]} \BibitemShut
  {NoStop}%
\bibitem [{\citenamefont {Bhaskar}\ \emph
  {et~al.}(2024{\natexlab{a}})\citenamefont {Bhaskar}, \citenamefont {Das},
  \citenamefont {Mandal}, \citenamefont {Mitra},\ and\ \citenamefont
  {Sharma}}]{Bhaskar:2023ftn}%
  \BibitemOpen
  \bibfield  {author} {\bibinfo {author} {\bibfnamefont {Arvind}\ \bibnamefont
  {Bhaskar}}, \bibinfo {author} {\bibfnamefont {Arijit}\ \bibnamefont {Das}},
  \bibinfo {author} {\bibfnamefont {Tanumoy}\ \bibnamefont {Mandal}}, \bibinfo
  {author} {\bibfnamefont {Subhadip}\ \bibnamefont {Mitra}}, \ and\ \bibinfo
  {author} {\bibfnamefont {Rachit}\ \bibnamefont {Sharma}},\ }\bibfield
  {title} {\enquote {\bibinfo {title} {{Fresh look at the LHC limits on scalar
  leptoquarks}},}\ }\href {\doibase 10.1103/PhysRevD.109.055018} {\bibfield
  {journal} {\bibinfo  {journal} {Phys. Rev. D}\ }\textbf {\bibinfo {volume}
  {109}},\ \bibinfo {pages} {055018} (\bibinfo {year} {2024}{\natexlab{a}})},\
  \Eprint {http://arxiv.org/abs/2312.09855} {arXiv:2312.09855 [hep-ph]}
  \BibitemShut {NoStop}%
\bibitem [{\citenamefont {Bertenstam}\ \emph {et~al.}(2025)\citenamefont
  {Bertenstam}, \citenamefont {Finetti}, \citenamefont {Morais}, \citenamefont
  {Pasechnik},\ and\ \citenamefont {Rathsman}}]{Bertenstam:2025jvd}%
  \BibitemOpen
  \bibfield  {author} {\bibinfo {author} {\bibfnamefont {M{\r{a}}rten}\
  \bibnamefont {Bertenstam}}, \bibinfo {author} {\bibfnamefont {Marco}\
  \bibnamefont {Finetti}}, \bibinfo {author} {\bibfnamefont {Ant{\'o}nio~P.}\
  \bibnamefont {Morais}}, \bibinfo {author} {\bibfnamefont {Roman}\
  \bibnamefont {Pasechnik}}, \ and\ \bibinfo {author} {\bibfnamefont {Johan}\
  \bibnamefont {Rathsman}},\ }\bibfield  {title} {\enquote {\bibinfo {title}
  {{Gravitational waves from color restoration in a leptoquark model of
  radiative neutrino masses}},}\ }\href@noop {} {\  (\bibinfo {year} {2025})},\
  \Eprint {http://arxiv.org/abs/2501.01286} {arXiv:2501.01286 [hep-ph]}
  \BibitemShut {NoStop}%
\bibitem [{\citenamefont {Ghosh}\ \emph {et~al.}(2025)\citenamefont {Ghosh},
  \citenamefont {Konar}, \citenamefont {Samui},\ and\ \citenamefont
  {Singh}}]{Ghosh:2025gue}%
  \BibitemOpen
  \bibfield  {author} {\bibinfo {author} {\bibfnamefont {Anupam}\ \bibnamefont
  {Ghosh}}, \bibinfo {author} {\bibfnamefont {Partha}\ \bibnamefont {Konar}},
  \bibinfo {author} {\bibfnamefont {Tousik}\ \bibnamefont {Samui}}, \ and\
  \bibinfo {author} {\bibfnamefont {Ritesh~K.}\ \bibnamefont {Singh}},\
  }\bibfield  {title} {\enquote {\bibinfo {title} {{Jet substructure probe on
  scalar leptoquark models via top polarization}},}\ }\href {\doibase
  10.1007/JHEP07(2025)145} {\bibfield  {journal} {\bibinfo  {journal} {JHEP}\
  }\textbf {\bibinfo {volume} {07}},\ \bibinfo {pages} {145} (\bibinfo {year}
  {2025})},\ \Eprint {http://arxiv.org/abs/2505.16328} {arXiv:2505.16328
  [hep-ph]} \BibitemShut {NoStop}%
\bibitem [{\citenamefont {Athron}\ \emph {et~al.}(2021)\citenamefont {Athron},
  \citenamefont {Bal\'azs}, \citenamefont {Jacob}, \citenamefont {Kotlarski},
  \citenamefont {St\"ockinger},\ and\ \citenamefont
  {St\"ockinger-Kim}}]{Athron:2021iuf}%
  \BibitemOpen
  \bibfield  {author} {\bibinfo {author} {\bibfnamefont {Peter}\ \bibnamefont
  {Athron}}, \bibinfo {author} {\bibfnamefont {Csaba}\ \bibnamefont
  {Bal\'azs}}, \bibinfo {author} {\bibfnamefont {Douglas H.~J.}\ \bibnamefont
  {Jacob}}, \bibinfo {author} {\bibfnamefont {Wojciech}\ \bibnamefont
  {Kotlarski}}, \bibinfo {author} {\bibfnamefont {Dominik}\ \bibnamefont
  {St\"ockinger}}, \ and\ \bibinfo {author} {\bibfnamefont {Hyejung}\
  \bibnamefont {St\"ockinger-Kim}},\ }\bibfield  {title} {\enquote {\bibinfo
  {title} {{New physics explanations of $a_{\mu}$ in light of the FNAL muon
  $g-2$ measurement}},}\ }\href {\doibase 10.1007/JHEP09(2021)080} {\bibfield
  {journal} {\bibinfo  {journal} {JHEP}\ }\textbf {\bibinfo {volume} {09}},\
  \bibinfo {pages} {080} (\bibinfo {year} {2021})},\ \Eprint
  {http://arxiv.org/abs/2104.03691} {arXiv:2104.03691 [hep-ph]} \BibitemShut
  {NoStop}%
\bibitem [{\citenamefont {Albahri}\ \emph {et~al.}(2021)\citenamefont {Albahri}
  \emph {et~al.}}]{Muong-2:2021vma}%
  \BibitemOpen
  \bibfield  {author} {\bibinfo {author} {\bibfnamefont {T.}~\bibnamefont
  {Albahri}} \emph {et~al.} (\bibinfo {collaboration} {Muon g-2}),\ }\bibfield
  {title} {\enquote {\bibinfo {title} {{Measurement of the anomalous precession
  frequency of the muon in the Fermilab Muon $g-2$ Experiment}},}\ }\href
  {\doibase 10.1103/PhysRevD.103.072002} {\bibfield  {journal} {\bibinfo
  {journal} {Phys. Rev. D}\ }\textbf {\bibinfo {volume} {103}},\ \bibinfo
  {pages} {072002} (\bibinfo {year} {2021})},\ \Eprint
  {http://arxiv.org/abs/2104.03247} {arXiv:2104.03247 [hep-ex]} \BibitemShut
  {NoStop}%
\bibitem [{\citenamefont {Abi}\ \emph {et~al.}(2021)\citenamefont {Abi} \emph
  {et~al.}}]{Muong-2:2021ojo}%
  \BibitemOpen
  \bibfield  {author} {\bibinfo {author} {\bibfnamefont {B.}~\bibnamefont
  {Abi}} \emph {et~al.} (\bibinfo {collaboration} {Muon g-2}),\ }\bibfield
  {title} {\enquote {\bibinfo {title} {{Measurement of the Positive Muon
  Anomalous Magnetic Moment to $0.46$~ppm}},}\ }\href {\doibase
  10.1103/PhysRevLett.126.141801} {\bibfield  {journal} {\bibinfo  {journal}
  {Phys. Rev. Lett.}\ }\textbf {\bibinfo {volume} {126}},\ \bibinfo {pages}
  {141801} (\bibinfo {year} {2021})},\ \Eprint
  {http://arxiv.org/abs/2104.03281} {arXiv:2104.03281 [hep-ex]} \BibitemShut
  {NoStop}%
\bibitem [{\citenamefont {Aguillard}\ \emph {et~al.}(2023)\citenamefont
  {Aguillard} \emph {et~al.}}]{Muong-2:2023cdq}%
  \BibitemOpen
  \bibfield  {author} {\bibinfo {author} {\bibfnamefont {D.~P.}\ \bibnamefont
  {Aguillard}} \emph {et~al.} (\bibinfo {collaboration} {Muon g-2}),\
  }\bibfield  {title} {\enquote {\bibinfo {title} {{Measurement of the Positive
  Muon Anomalous Magnetic Moment to $0.20$~ppm}},}\ }\href {\doibase
  10.1103/PhysRevLett.131.161802} {\bibfield  {journal} {\bibinfo  {journal}
  {Phys. Rev. Lett.}\ }\textbf {\bibinfo {volume} {131}},\ \bibinfo {pages}
  {161802} (\bibinfo {year} {2023})},\ \Eprint
  {http://arxiv.org/abs/2308.06230} {arXiv:2308.06230 [hep-ex]} \BibitemShut
  {NoStop}%
\bibitem [{\citenamefont {Mandal}\ \emph {et~al.}(2015)\citenamefont {Mandal},
  \citenamefont {Mitra},\ and\ \citenamefont {Seth}}]{Mandal:2015vfa}%
  \BibitemOpen
  \bibfield  {author} {\bibinfo {author} {\bibfnamefont {Tanumoy}\ \bibnamefont
  {Mandal}}, \bibinfo {author} {\bibfnamefont {Subhadip}\ \bibnamefont
  {Mitra}}, \ and\ \bibinfo {author} {\bibfnamefont {Satyajit}\ \bibnamefont
  {Seth}},\ }\bibfield  {title} {\enquote {\bibinfo {title} {{Single
  Productions of Colored Particles at the LHC: An Example with Scalar
  Leptoquarks}},}\ }\href {\doibase 10.1007/JHEP07(2015)028} {\bibfield
  {journal} {\bibinfo  {journal} {JHEP}\ }\textbf {\bibinfo {volume} {07}},\
  \bibinfo {pages} {028} (\bibinfo {year} {2015})},\ \Eprint
  {http://arxiv.org/abs/1503.04689} {arXiv:1503.04689 [hep-ph]} \BibitemShut
  {NoStop}%
\bibitem [{\citenamefont {Mandal}\ \emph {et~al.}(2019)\citenamefont {Mandal},
  \citenamefont {Mitra},\ and\ \citenamefont {Raz}}]{Mandal:2018kau}%
  \BibitemOpen
  \bibfield  {author} {\bibinfo {author} {\bibfnamefont {Tanumoy}\ \bibnamefont
  {Mandal}}, \bibinfo {author} {\bibfnamefont {Subhadip}\ \bibnamefont
  {Mitra}}, \ and\ \bibinfo {author} {\bibfnamefont {Swapnil}\ \bibnamefont
  {Raz}},\ }\bibfield  {title} {\enquote {\bibinfo {title} {{$R_{D^{(*)}}$
  motivated $\mathcal{S}_1$ leptoquark scenarios: Impact of interference on the
  exclusion limits from LHC data}},}\ }\href {\doibase
  10.1103/PhysRevD.99.055028} {\bibfield  {journal} {\bibinfo  {journal} {Phys.
  Rev. D}\ }\textbf {\bibinfo {volume} {99}},\ \bibinfo {pages} {055028}
  (\bibinfo {year} {2019})},\ \Eprint {http://arxiv.org/abs/1811.03561}
  {arXiv:1811.03561 [hep-ph]} \BibitemShut {NoStop}%
\bibitem [{\citenamefont {Aydemir}\ \emph {et~al.}(2020)\citenamefont
  {Aydemir}, \citenamefont {Mandal},\ and\ \citenamefont
  {Mitra}}]{Aydemir:2019ynb}%
  \BibitemOpen
  \bibfield  {author} {\bibinfo {author} {\bibfnamefont {Ufuk}\ \bibnamefont
  {Aydemir}}, \bibinfo {author} {\bibfnamefont {Tanumoy}\ \bibnamefont
  {Mandal}}, \ and\ \bibinfo {author} {\bibfnamefont {Subhadip}\ \bibnamefont
  {Mitra}},\ }\bibfield  {title} {\enquote {\bibinfo {title} {{Addressing the
  ${\mathbf R_{D^{(*)}}}$ anomalies with an ${\mathbf S_1}$ leptoquark from
  $\mathbf{SO(10)}$ grand unification}},}\ }\href {\doibase
  10.1103/PhysRevD.101.015011} {\bibfield  {journal} {\bibinfo  {journal}
  {Phys. Rev. D}\ }\textbf {\bibinfo {volume} {101}},\ \bibinfo {pages}
  {015011} (\bibinfo {year} {2020})},\ \Eprint
  {http://arxiv.org/abs/1902.08108} {arXiv:1902.08108 [hep-ph]} \BibitemShut
  {NoStop}%
\bibitem [{\citenamefont {Chandak}\ \emph {et~al.}(2019)\citenamefont
  {Chandak}, \citenamefont {Mandal},\ and\ \citenamefont
  {Mitra}}]{Chandak:2019iwj}%
  \BibitemOpen
  \bibfield  {author} {\bibinfo {author} {\bibfnamefont {Kushagra}\
  \bibnamefont {Chandak}}, \bibinfo {author} {\bibfnamefont {Tanumoy}\
  \bibnamefont {Mandal}}, \ and\ \bibinfo {author} {\bibfnamefont {Subhadip}\
  \bibnamefont {Mitra}},\ }\bibfield  {title} {\enquote {\bibinfo {title}
  {{Hunting for scalar leptoquarks with boosted tops and light leptons}},}\
  }\href {\doibase 10.1103/PhysRevD.100.075019} {\bibfield  {journal} {\bibinfo
   {journal} {Phys. Rev. D}\ }\textbf {\bibinfo {volume} {100}},\ \bibinfo
  {pages} {075019} (\bibinfo {year} {2019})},\ \Eprint
  {http://arxiv.org/abs/1907.11194} {arXiv:1907.11194 [hep-ph]} \BibitemShut
  {NoStop}%
\bibitem [{\citenamefont {Bhaskar}\ \emph
  {et~al.}(2021{\natexlab{a}})\citenamefont {Bhaskar}, \citenamefont {Das},
  \citenamefont {Mandal}, \citenamefont {Mitra},\ and\ \citenamefont
  {Neeraj}}]{Bhaskar:2021pml}%
  \BibitemOpen
  \bibfield  {author} {\bibinfo {author} {\bibfnamefont {Arvind}\ \bibnamefont
  {Bhaskar}}, \bibinfo {author} {\bibfnamefont {Diganta}\ \bibnamefont {Das}},
  \bibinfo {author} {\bibfnamefont {Tanumoy}\ \bibnamefont {Mandal}}, \bibinfo
  {author} {\bibfnamefont {Subhadip}\ \bibnamefont {Mitra}}, \ and\ \bibinfo
  {author} {\bibfnamefont {Cyrin}\ \bibnamefont {Neeraj}},\ }\bibfield  {title}
  {\enquote {\bibinfo {title} {{Precise limits on the charge-2/3 U1 vector
  leptoquark}},}\ }\href {\doibase 10.1103/PhysRevD.104.035016} {\bibfield
  {journal} {\bibinfo  {journal} {Phys. Rev. D}\ }\textbf {\bibinfo {volume}
  {104}},\ \bibinfo {pages} {035016} (\bibinfo {year} {2021}{\natexlab{a}})},\
  \Eprint {http://arxiv.org/abs/2101.12069} {arXiv:2101.12069 [hep-ph]}
  \BibitemShut {NoStop}%
\bibitem [{\citenamefont {Bhaskar}\ \emph
  {et~al.}(2021{\natexlab{b}})\citenamefont {Bhaskar}, \citenamefont {Mandal},
  \citenamefont {Mitra},\ and\ \citenamefont {Sharma}}]{Bhaskar:2021gsy}%
  \BibitemOpen
  \bibfield  {author} {\bibinfo {author} {\bibfnamefont {Arvind}\ \bibnamefont
  {Bhaskar}}, \bibinfo {author} {\bibfnamefont {Tanumoy}\ \bibnamefont
  {Mandal}}, \bibinfo {author} {\bibfnamefont {Subhadip}\ \bibnamefont
  {Mitra}}, \ and\ \bibinfo {author} {\bibfnamefont {Mohit}\ \bibnamefont
  {Sharma}},\ }\bibfield  {title} {\enquote {\bibinfo {title} {{Improving
  third-generation leptoquark searches with combined signals and boosted top
  quarks}},}\ }\href {\doibase 10.1103/PhysRevD.104.075037} {\bibfield
  {journal} {\bibinfo  {journal} {Phys. Rev. D}\ }\textbf {\bibinfo {volume}
  {104}},\ \bibinfo {pages} {075037} (\bibinfo {year} {2021}{\natexlab{b}})},\
  \Eprint {http://arxiv.org/abs/2106.07605} {arXiv:2106.07605 [hep-ph]}
  \BibitemShut {NoStop}%
\bibitem [{\citenamefont {Bhaskar}\ \emph
  {et~al.}(2022{\natexlab{b}})\citenamefont {Bhaskar}, \citenamefont
  {Madathil}, \citenamefont {Mandal},\ and\ \citenamefont
  {Mitra}}]{Bhaskar:2022vgk}%
  \BibitemOpen
  \bibfield  {author} {\bibinfo {author} {\bibfnamefont {Arvind}\ \bibnamefont
  {Bhaskar}}, \bibinfo {author} {\bibfnamefont {Anirudhan~A.}\ \bibnamefont
  {Madathil}}, \bibinfo {author} {\bibfnamefont {Tanumoy}\ \bibnamefont
  {Mandal}}, \ and\ \bibinfo {author} {\bibfnamefont {Subhadip}\ \bibnamefont
  {Mitra}},\ }\bibfield  {title} {\enquote {\bibinfo {title} {{Combined
  explanation of W-mass, muon g-2, RK(*) and RD(*) anomalies in a
  singlet-triplet scalar leptoquark model}},}\ }\href {\doibase
  10.1103/PhysRevD.106.115009} {\bibfield  {journal} {\bibinfo  {journal}
  {Phys. Rev. D}\ }\textbf {\bibinfo {volume} {106}},\ \bibinfo {pages}
  {115009} (\bibinfo {year} {2022}{\natexlab{b}})},\ \Eprint
  {http://arxiv.org/abs/2204.09031} {arXiv:2204.09031 [hep-ph]} \BibitemShut
  {NoStop}%
\bibitem [{\citenamefont {Aydemir}\ \emph {et~al.}(2022)\citenamefont
  {Aydemir}, \citenamefont {Mandal}, \citenamefont {Mitra},\ and\ \citenamefont
  {Munir}}]{Aydemir:2022lrq}%
  \BibitemOpen
  \bibfield  {author} {\bibinfo {author} {\bibfnamefont {Ufuk}\ \bibnamefont
  {Aydemir}}, \bibinfo {author} {\bibfnamefont {Tanumoy}\ \bibnamefont
  {Mandal}}, \bibinfo {author} {\bibfnamefont {Subhadip}\ \bibnamefont
  {Mitra}}, \ and\ \bibinfo {author} {\bibfnamefont {Shoaib}\ \bibnamefont
  {Munir}},\ }\bibfield  {title} {\enquote {\bibinfo {title} {{An economical
  model for $B$-flavour and $a_\mu$ anomalies from SO(10) grand
  unification}},}\ }\href@noop {} {\  (\bibinfo {year} {2022})},\ \Eprint
  {http://arxiv.org/abs/2209.04705} {arXiv:2209.04705 [hep-ph]} \BibitemShut
  {NoStop}%
\bibitem [{\citenamefont {Bhaskar}\ and\ \citenamefont
  {Mitra}(2025)}]{Bhaskar:2024snl}%
  \BibitemOpen
  \bibfield  {author} {\bibinfo {author} {\bibfnamefont {Arvind}\ \bibnamefont
  {Bhaskar}}\ and\ \bibinfo {author} {\bibfnamefont {Manimala}\ \bibnamefont
  {Mitra}},\ }\bibfield  {title} {\enquote {\bibinfo {title} {{Boosted top
  quark inspired leptoquark searches at the muon collider}},}\ }\href {\doibase
  10.1016/j.physletb.2025.139656} {\bibfield  {journal} {\bibinfo  {journal}
  {Phys. Lett. B}\ }\textbf {\bibinfo {volume} {868}},\ \bibinfo {pages}
  {139656} (\bibinfo {year} {2025})},\ \Eprint
  {http://arxiv.org/abs/2409.15992} {arXiv:2409.15992 [hep-ph]} \BibitemShut
  {NoStop}%
\bibitem [{\citenamefont {Ghosh}\ \emph
  {et~al.}(2023{\natexlab{b}})\citenamefont {Ghosh}, \citenamefont {Rai},\ and\
  \citenamefont {Samui}}]{Ghosh:2022vpb}%
  \BibitemOpen
  \bibfield  {author} {\bibinfo {author} {\bibfnamefont {Nivedita}\
  \bibnamefont {Ghosh}}, \bibinfo {author} {\bibfnamefont {Santosh~Kumar}\
  \bibnamefont {Rai}}, \ and\ \bibinfo {author} {\bibfnamefont {Tousik}\
  \bibnamefont {Samui}},\ }\bibfield  {title} {\enquote {\bibinfo {title}
  {{Collider signatures of a scalar leptoquark and vectorlike lepton in light
  of muon anomaly}},}\ }\href {\doibase 10.1103/PhysRevD.107.035028} {\bibfield
   {journal} {\bibinfo  {journal} {Phys. Rev. D}\ }\textbf {\bibinfo {volume}
  {107}},\ \bibinfo {pages} {035028} (\bibinfo {year} {2023}{\natexlab{b}})},\
  \Eprint {http://arxiv.org/abs/2206.11718} {arXiv:2206.11718 [hep-ph]}
  \BibitemShut {NoStop}%
\bibitem [{\citenamefont {Bhaskar}\ \emph {et~al.}(2023)\citenamefont
  {Bhaskar}, \citenamefont {Chaurasia}, \citenamefont {Deka}, \citenamefont
  {Mandal}, \citenamefont {Mitra},\ and\ \citenamefont
  {Mukherjee}}]{Bhaskar:2023xkm}%
  \BibitemOpen
  \bibfield  {author} {\bibinfo {author} {\bibfnamefont {Arvind}\ \bibnamefont
  {Bhaskar}}, \bibinfo {author} {\bibfnamefont {Yash}\ \bibnamefont
  {Chaurasia}}, \bibinfo {author} {\bibfnamefont {Kuldeep}\ \bibnamefont
  {Deka}}, \bibinfo {author} {\bibfnamefont {Tanumoy}\ \bibnamefont {Mandal}},
  \bibinfo {author} {\bibfnamefont {Subhadip}\ \bibnamefont {Mitra}}, \ and\
  \bibinfo {author} {\bibfnamefont {Ananya}\ \bibnamefont {Mukherjee}},\
  }\bibfield  {title} {\enquote {\bibinfo {title} {{Right-handed neutrino pair
  production via second-generation leptoquarks}},}\ }\href {\doibase
  10.1016/j.physletb.2023.138039} {\bibfield  {journal} {\bibinfo  {journal}
  {Phys. Lett. B}\ }\textbf {\bibinfo {volume} {843}},\ \bibinfo {pages}
  {138039} (\bibinfo {year} {2023})},\ \Eprint
  {http://arxiv.org/abs/2301.11889} {arXiv:2301.11889 [hep-ph]} \BibitemShut
  {NoStop}%
\bibitem [{\citenamefont {Duraikandan}\ \emph {et~al.}(2025)\citenamefont
  {Duraikandan}, \citenamefont {Khanna}, \citenamefont {Mandal}, \citenamefont
  {Mitra},\ and\ \citenamefont {Sharma}}]{Duraikandan:2024kcy}%
  \BibitemOpen
  \bibfield  {author} {\bibinfo {author} {\bibfnamefont {Gokul}\ \bibnamefont
  {Duraikandan}}, \bibinfo {author} {\bibfnamefont {Rishabh}\ \bibnamefont
  {Khanna}}, \bibinfo {author} {\bibfnamefont {Tanumoy}\ \bibnamefont
  {Mandal}}, \bibinfo {author} {\bibfnamefont {Subhadip}\ \bibnamefont
  {Mitra}}, \ and\ \bibinfo {author} {\bibfnamefont {Rachit}\ \bibnamefont
  {Sharma}},\ }\bibfield  {title} {\enquote {\bibinfo {title} {{Right-handed
  neutrino production through first-generation leptoquarks}},}\ }\href
  {\doibase 10.1103/PhysRevD.111.075032} {\bibfield  {journal} {\bibinfo
  {journal} {Phys. Rev. D}\ }\textbf {\bibinfo {volume} {111}},\ \bibinfo
  {pages} {075032} (\bibinfo {year} {2025})},\ \Eprint
  {http://arxiv.org/abs/2412.19751} {arXiv:2412.19751 [hep-ph]} \BibitemShut
  {NoStop}%
\bibitem [{\citenamefont {De}(2024)}]{De:2024foq}%
  \BibitemOpen
  \bibfield  {author} {\bibinfo {author} {\bibfnamefont {Bibhabasu}\
  \bibnamefont {De}},\ }\bibfield  {title} {\enquote {\bibinfo {title}
  {{Leptoquark-induced CLFV decays with a light SM-singlet scalar}},}\ }\href
  {\doibase 10.1016/j.physletb.2024.138784} {\bibfield  {journal} {\bibinfo
  {journal} {Phys. Lett. B}\ }\textbf {\bibinfo {volume} {855}},\ \bibinfo
  {pages} {138784} (\bibinfo {year} {2024})},\ \Eprint
  {http://arxiv.org/abs/2405.06970} {arXiv:2405.06970 [hep-ph]} \BibitemShut
  {NoStop}%
\bibitem [{\citenamefont {Kumar}\ and\ \citenamefont
  {Srivastava}(2025)}]{Kumar:2025aek}%
  \BibitemOpen
  \bibfield  {author} {\bibinfo {author} {\bibfnamefont {Ranjeet}\ \bibnamefont
  {Kumar}}\ and\ \bibinfo {author} {\bibfnamefont {Rahul}\ \bibnamefont
  {Srivastava}},\ }\bibfield  {title} {\enquote {\bibinfo {title} {{Dark Matter
  Induced Proton Decays}},}\ }\href@noop {} {\  (\bibinfo {year} {2025})},\
  \Eprint {http://arxiv.org/abs/2506.04370} {arXiv:2506.04370 [hep-ph]}
  \BibitemShut {NoStop}%
\bibitem [{ATL(2024)}]{ATL-PHYS-PUB-2024-012}%
  \BibitemOpen
  \href {https://cds.cern.ch/record/2903898} {\emph {\bibinfo {title}
  {{Leptoquark summary plot for scalar or vector models}}}},\ \bibinfo {type}
  {Tech. Rep.}\ (\bibinfo  {institution} {CERN},\ \bibinfo {address} {Geneva},\
  \bibinfo {year} {2024})\ \bibinfo {note}
  {https://cds.cern.ch/record/2903898/files/ATL-PHYS-PUB-2024-012.pdf}\BibitemShut
  {NoStop}%
\bibitem [{CMS(2025)}]{CMSPlot}%
  \BibitemOpen
  \href@noop {} {\enquote {\bibinfo {title} {{CMS Exotica Summary plots for 13
  TeV data: Leptoquark summary plot}},}\ }\bibinfo {howpublished}
  {\url{https://twiki.cern.ch/twiki/pub/CMSPublic/SummaryPlotsEXO13TeV/barplot_QE_QMU_QTAU_QNU_2025March.pdf}}
  (\bibinfo {year} {2025})\BibitemShut {NoStop}%
\bibitem [{\citenamefont {Mandal}\ and\ \citenamefont
  {Mitra}(2013)}]{Mandal:2012rx}%
  \BibitemOpen
  \bibfield  {author} {\bibinfo {author} {\bibfnamefont {Tanumoy}\ \bibnamefont
  {Mandal}}\ and\ \bibinfo {author} {\bibfnamefont {Subhadip}\ \bibnamefont
  {Mitra}},\ }\bibfield  {title} {\enquote {\bibinfo {title} {{Probing Color
  Octet Electrons at the LHC}},}\ }\href {\doibase 10.1103/PhysRevD.87.095008}
  {\bibfield  {journal} {\bibinfo  {journal} {Phys. Rev. D}\ }\textbf {\bibinfo
  {volume} {87}},\ \bibinfo {pages} {095008} (\bibinfo {year} {2013})},\
  \Eprint {http://arxiv.org/abs/1211.6394} {arXiv:1211.6394 [hep-ph]}
  \BibitemShut {NoStop}%
\bibitem [{\citenamefont {Mandal}\ \emph {et~al.}(2016)\citenamefont {Mandal},
  \citenamefont {Mitra},\ and\ \citenamefont {Seth}}]{Mandal:2016csb}%
  \BibitemOpen
  \bibfield  {author} {\bibinfo {author} {\bibfnamefont {Tanumoy}\ \bibnamefont
  {Mandal}}, \bibinfo {author} {\bibfnamefont {Subhadip}\ \bibnamefont
  {Mitra}}, \ and\ \bibinfo {author} {\bibfnamefont {Satyajit}\ \bibnamefont
  {Seth}},\ }\bibfield  {title} {\enquote {\bibinfo {title} {{Probing
  Compositeness with the CMS $eejj$ and $eej$ Data}},}\ }\href {\doibase
  10.1016/j.physletb.2016.05.020} {\bibfield  {journal} {\bibinfo  {journal}
  {Phys. Lett. B}\ }\textbf {\bibinfo {volume} {758}},\ \bibinfo {pages}
  {219--225} (\bibinfo {year} {2016})},\ \Eprint
  {http://arxiv.org/abs/1602.01273} {arXiv:1602.01273 [hep-ph]} \BibitemShut
  {NoStop}%
\bibitem [{\citenamefont {Bhaskar}\ \emph {et~al.}(2025)\citenamefont
  {Bhaskar}, \citenamefont {Das}, \citenamefont {Kundu}, \citenamefont
  {A.~Madathil}, \citenamefont {Mandal},\ and\ \citenamefont
  {Mitra}}]{Bhaskar:2024swq}%
  \BibitemOpen
  \bibfield  {author} {\bibinfo {author} {\bibfnamefont {Arvind}\ \bibnamefont
  {Bhaskar}}, \bibinfo {author} {\bibfnamefont {Diganta}\ \bibnamefont {Das}},
  \bibinfo {author} {\bibfnamefont {Soumyadip}\ \bibnamefont {Kundu}}, \bibinfo
  {author} {\bibfnamefont {Anirudhan}\ \bibnamefont {A.~Madathil}}, \bibinfo
  {author} {\bibfnamefont {Tanumoy}\ \bibnamefont {Mandal}}, \ and\ \bibinfo
  {author} {\bibfnamefont {Subhadip}\ \bibnamefont {Mitra}},\ }\bibfield
  {title} {\enquote {\bibinfo {title} {{Vector leptoquark contributions to
  lepton dipole moments}},}\ }\href {\doibase 10.1103/PhysRevD.111.015045}
  {\bibfield  {journal} {\bibinfo  {journal} {Phys. Rev. D}\ }\textbf {\bibinfo
  {volume} {111}},\ \bibinfo {pages} {015045} (\bibinfo {year} {2025})},\
  \Eprint {http://arxiv.org/abs/2408.11798} {arXiv:2408.11798 [hep-ph]}
  \BibitemShut {NoStop}%
\bibitem [{\citenamefont {Bhaskar}\ \emph
  {et~al.}(2024{\natexlab{b}})\citenamefont {Bhaskar}, \citenamefont
  {Chaurasia}, \citenamefont {Das}, \citenamefont {Kumar}, \citenamefont
  {Mandal}, \citenamefont {Mitra}, \citenamefont {Neeraj},\ and\ \citenamefont
  {Sharma}}]{Bhaskar:2024wic}%
  \BibitemOpen
  \bibfield  {author} {\bibinfo {author} {\bibfnamefont {Arvind}\ \bibnamefont
  {Bhaskar}}, \bibinfo {author} {\bibfnamefont {Yash}\ \bibnamefont
  {Chaurasia}}, \bibinfo {author} {\bibfnamefont {Arijit}\ \bibnamefont {Das}},
  \bibinfo {author} {\bibfnamefont {Atirek}\ \bibnamefont {Kumar}}, \bibinfo
  {author} {\bibfnamefont {Tanumoy}\ \bibnamefont {Mandal}}, \bibinfo {author}
  {\bibfnamefont {Subhadip}\ \bibnamefont {Mitra}}, \bibinfo {author}
  {\bibfnamefont {Cyrin}\ \bibnamefont {Neeraj}}, \ and\ \bibinfo {author}
  {\bibfnamefont {Rachit}\ \bibnamefont {Sharma}},\ }\bibfield  {title}
  {\enquote {\bibinfo {title} {{TooLQit: Leptoquark Models and Limits}},}\
  }\href@noop {} {\  (\bibinfo {year} {2024}{\natexlab{b}})},\ \Eprint
  {http://arxiv.org/abs/2412.19729} {arXiv:2412.19729 [hep-ph]} \BibitemShut
  {NoStop}%
\bibitem [{\citenamefont {Bl\"umlein}\ and\ \citenamefont
  {Boos}(1994)}]{Blumlein:1994tu}%
  \BibitemOpen
  \bibfield  {author} {\bibinfo {author} {\bibfnamefont {J.}~\bibnamefont
  {Bl\"umlein}}\ and\ \bibinfo {author} {\bibfnamefont {E.}~\bibnamefont
  {Boos}},\ }\bibfield  {title} {\enquote {\bibinfo {title} {{Leptoquark
  production at high energy $e^{+} e^{-}$ colliders}},}\ }\href {\doibase
  10.1016/0920-5632(94)90675-0} {\bibfield  {journal} {\bibinfo  {journal}
  {Nucl. Phys. B Proc. Suppl.}\ }\textbf {\bibinfo {volume} {37}},\ \bibinfo
  {pages} {181--192} (\bibinfo {year} {1994})}\BibitemShut {NoStop}%
\bibitem [{\citenamefont {Bhaskar}\ \emph
  {et~al.}(2024{\natexlab{c}})\citenamefont {Bhaskar}, \citenamefont {Das},
  \citenamefont {Mandal}, \citenamefont {Mitra},\ and\ \citenamefont
  {Sharma}}]{PhysRevD.109.055018}%
  \BibitemOpen
  \bibfield  {author} {\bibinfo {author} {\bibfnamefont {Arvind}\ \bibnamefont
  {Bhaskar}}, \bibinfo {author} {\bibfnamefont {Arijit}\ \bibnamefont {Das}},
  \bibinfo {author} {\bibfnamefont {Tanumoy}\ \bibnamefont {Mandal}}, \bibinfo
  {author} {\bibfnamefont {Subhadip}\ \bibnamefont {Mitra}}, \ and\ \bibinfo
  {author} {\bibfnamefont {Rachit}\ \bibnamefont {Sharma}},\ }\bibfield
  {title} {\enquote {\bibinfo {title} {Fresh look at the lhc limits on scalar
  leptoquarks},}\ }\href {\doibase 10.1103/PhysRevD.109.055018} {\bibfield
  {journal} {\bibinfo  {journal} {Phys. Rev. D}\ }\textbf {\bibinfo {volume}
  {109}},\ \bibinfo {pages} {055018} (\bibinfo {year}
  {2024}{\natexlab{c}})}\BibitemShut {NoStop}%
\bibitem [{\citenamefont {Aad}\ \emph {et~al.}(2020)\citenamefont {Aad} \emph
  {et~al.}}]{ATLAS:2020dsk}%
  \BibitemOpen
  \bibfield  {author} {\bibinfo {author} {\bibfnamefont {Georges}\ \bibnamefont
  {Aad}} \emph {et~al.} (\bibinfo {collaboration} {ATLAS}),\ }\bibfield
  {title} {\enquote {\bibinfo {title} {{Search for pairs of scalar leptoquarks
  decaying into quarks and electrons or muons in $ \sqrt{s} $ = 13 TeV $pp$
  collisions with the ATLAS detector}},}\ }\href {\doibase
  10.1007/JHEP10(2020)112} {\bibfield  {journal} {\bibinfo  {journal} {JHEP}\
  }\textbf {\bibinfo {volume} {10}},\ \bibinfo {pages} {112} (\bibinfo {year}
  {2020})},\ \Eprint {http://arxiv.org/abs/2006.05872} {arXiv:2006.05872
  [hep-ex]} \BibitemShut {NoStop}%
\bibitem [{\citenamefont {Sirunyan}\ \emph {et~al.}(2019)\citenamefont
  {Sirunyan} \emph {et~al.}}]{CMS:2018lab}%
  \BibitemOpen
  \bibfield  {author} {\bibinfo {author} {\bibfnamefont {Albert~M}\
  \bibnamefont {Sirunyan}} \emph {et~al.} (\bibinfo {collaboration} {CMS}),\
  }\bibfield  {title} {\enquote {\bibinfo {title} {{Search for pair production
  of second-generation leptoquarks at $\sqrt{s}=$ 13 TeV}},}\ }\href {\doibase
  10.1103/PhysRevD.99.032014} {\bibfield  {journal} {\bibinfo  {journal} {Phys.
  Rev. D}\ }\textbf {\bibinfo {volume} {99}},\ \bibinfo {pages} {032014}
  (\bibinfo {year} {2019})},\ \Eprint {http://arxiv.org/abs/1808.05082}
  {arXiv:1808.05082 [hep-ex]} \BibitemShut {NoStop}%
\bibitem [{\citenamefont {Alloul}\ \emph {et~al.}(2014)\citenamefont {Alloul},
  \citenamefont {Christensen}, \citenamefont {Degrande}, \citenamefont {Duhr},\
  and\ \citenamefont {Fuks}}]{Alloul:2013bka}%
  \BibitemOpen
  \bibfield  {author} {\bibinfo {author} {\bibfnamefont {Adam}\ \bibnamefont
  {Alloul}}, \bibinfo {author} {\bibfnamefont {Neil~D.}\ \bibnamefont
  {Christensen}}, \bibinfo {author} {\bibfnamefont {C\'eline}\ \bibnamefont
  {Degrande}}, \bibinfo {author} {\bibfnamefont {Claude}\ \bibnamefont {Duhr}},
  \ and\ \bibinfo {author} {\bibfnamefont {Benjamin}\ \bibnamefont {Fuks}},\
  }\bibfield  {title} {\enquote {\bibinfo {title} {{FeynRules 2.0 - A complete
  toolbox for tree-level phenomenology}},}\ }\href {\doibase
  10.1016/j.cpc.2014.04.012} {\bibfield  {journal} {\bibinfo  {journal}
  {Comput. Phys. Commun.}\ }\textbf {\bibinfo {volume} {185}},\ \bibinfo
  {pages} {2250--2300} (\bibinfo {year} {2014})},\ \Eprint
  {http://arxiv.org/abs/1310.1921} {arXiv:1310.1921 [hep-ph]} \BibitemShut
  {NoStop}%
\bibitem [{\citenamefont {Degrande}\ \emph {et~al.}(2012)\citenamefont
  {Degrande}, \citenamefont {Duhr}, \citenamefont {Fuks}, \citenamefont
  {Grellscheid}, \citenamefont {Mattelaer},\ and\ \citenamefont
  {Reiter}}]{Degrande:2011ua}%
  \BibitemOpen
  \bibfield  {author} {\bibinfo {author} {\bibfnamefont {Celine}\ \bibnamefont
  {Degrande}}, \bibinfo {author} {\bibfnamefont {Claude}\ \bibnamefont {Duhr}},
  \bibinfo {author} {\bibfnamefont {Benjamin}\ \bibnamefont {Fuks}}, \bibinfo
  {author} {\bibfnamefont {David}\ \bibnamefont {Grellscheid}}, \bibinfo
  {author} {\bibfnamefont {Olivier}\ \bibnamefont {Mattelaer}}, \ and\ \bibinfo
  {author} {\bibfnamefont {Thomas}\ \bibnamefont {Reiter}},\ }\bibfield
  {title} {\enquote {\bibinfo {title} {{UFO - The Universal FeynRules
  Output}},}\ }\href {\doibase 10.1016/j.cpc.2012.01.022} {\bibfield  {journal}
  {\bibinfo  {journal} {Comput. Phys. Commun.}\ }\textbf {\bibinfo {volume}
  {183}},\ \bibinfo {pages} {1201--1214} (\bibinfo {year} {2012})},\ \Eprint
  {http://arxiv.org/abs/1108.2040} {arXiv:1108.2040 [hep-ph]} \BibitemShut
  {NoStop}%
\bibitem [{\citenamefont {Alwall}\ \emph {et~al.}(2014)\citenamefont {Alwall},
  \citenamefont {Frederix}, \citenamefont {Frixione}, \citenamefont {Hirschi},
  \citenamefont {Maltoni}, \citenamefont {Mattelaer}, \citenamefont {Shao},
  \citenamefont {Stelzer}, \citenamefont {Torrielli},\ and\ \citenamefont
  {Zaro}}]{Alwall:2014hca}%
  \BibitemOpen
  \bibfield  {author} {\bibinfo {author} {\bibfnamefont {J.}~\bibnamefont
  {Alwall}}, \bibinfo {author} {\bibfnamefont {R.}~\bibnamefont {Frederix}},
  \bibinfo {author} {\bibfnamefont {S.}~\bibnamefont {Frixione}}, \bibinfo
  {author} {\bibfnamefont {V.}~\bibnamefont {Hirschi}}, \bibinfo {author}
  {\bibfnamefont {F.}~\bibnamefont {Maltoni}}, \bibinfo {author} {\bibfnamefont
  {O.}~\bibnamefont {Mattelaer}}, \bibinfo {author} {\bibfnamefont {H.~S.}\
  \bibnamefont {Shao}}, \bibinfo {author} {\bibfnamefont {T.}~\bibnamefont
  {Stelzer}}, \bibinfo {author} {\bibfnamefont {P.}~\bibnamefont {Torrielli}},
  \ and\ \bibinfo {author} {\bibfnamefont {M.}~\bibnamefont {Zaro}},\
  }\bibfield  {title} {\enquote {\bibinfo {title} {{The automated computation
  of tree-level and next-to-leading order differential cross sections, and
  their matching to parton shower simulations}},}\ }\href {\doibase
  10.1007/JHEP07(2014)079} {\bibfield  {journal} {\bibinfo  {journal} {JHEP}\
  }\textbf {\bibinfo {volume} {07}},\ \bibinfo {pages} {079} (\bibinfo {year}
  {2014})},\ \Eprint {http://arxiv.org/abs/1405.0301} {arXiv:1405.0301
  [hep-ph]} \BibitemShut {NoStop}%
\bibitem [{\citenamefont {Ball}\ \emph {et~al.}(2021)\citenamefont {Ball} \emph
  {et~al.}}]{NNPDF:2021uiq}%
  \BibitemOpen
  \bibfield  {author} {\bibinfo {author} {\bibfnamefont {Richard~D.}\
  \bibnamefont {Ball}} \emph {et~al.} (\bibinfo {collaboration} {NNPDF}),\
  }\bibfield  {title} {\enquote {\bibinfo {title} {{An open-source machine
  learning framework for global analyses of parton distributions}},}\ }\href
  {\doibase 10.1140/epjc/s10052-021-09747-9} {\bibfield  {journal} {\bibinfo
  {journal} {Eur. Phys. J. C}\ }\textbf {\bibinfo {volume} {81}},\ \bibinfo
  {pages} {958} (\bibinfo {year} {2021})},\ \Eprint
  {http://arxiv.org/abs/2109.02671} {arXiv:2109.02671 [hep-ph]} \BibitemShut
  {NoStop}%
\bibitem [{\citenamefont {Bierlich}\ \emph {et~al.}(2022)\citenamefont
  {Bierlich} \emph {et~al.}}]{Bierlich:2022pfr}%
  \BibitemOpen
  \bibfield  {author} {\bibinfo {author} {\bibfnamefont {Christian}\
  \bibnamefont {Bierlich}} \emph {et~al.},\ }\bibfield  {title} {\enquote
  {\bibinfo {title} {{A comprehensive guide to the physics and usage of PYTHIA
  8.3}},}\ }\href {\doibase 10.21468/SciPostPhysCodeb.8} {\  (\bibinfo {year}
  {2022}),\ 10.21468/SciPostPhysCodeb.8},\ \Eprint
  {http://arxiv.org/abs/2203.11601} {arXiv:2203.11601 [hep-ph]} \BibitemShut
  {NoStop}%
\bibitem [{\citenamefont {de~Favereau}\ \emph {et~al.}(2014)\citenamefont
  {de~Favereau}, \citenamefont {Delaere}, \citenamefont {Demin}, \citenamefont
  {Giammanco}, \citenamefont {Lema\^\i{}tre}, \citenamefont {Mertens},\ and\
  \citenamefont {Selvaggi}}]{deFavereau:2013fsa}%
  \BibitemOpen
  \bibfield  {author} {\bibinfo {author} {\bibfnamefont {J.}~\bibnamefont
  {de~Favereau}}, \bibinfo {author} {\bibfnamefont {C.}~\bibnamefont
  {Delaere}}, \bibinfo {author} {\bibfnamefont {P.}~\bibnamefont {Demin}},
  \bibinfo {author} {\bibfnamefont {A.}~\bibnamefont {Giammanco}}, \bibinfo
  {author} {\bibfnamefont {V.}~\bibnamefont {Lema\^\i{}tre}}, \bibinfo {author}
  {\bibfnamefont {A.}~\bibnamefont {Mertens}}, \ and\ \bibinfo {author}
  {\bibfnamefont {M.}~\bibnamefont {Selvaggi}} (\bibinfo {collaboration}
  {DELPHES 3}),\ }\bibfield  {title} {\enquote {\bibinfo {title} {{DELPHES 3, A
  modular framework for fast simulation of a generic collider experiment}},}\
  }\href {\doibase 10.1007/JHEP02(2014)057} {\bibfield  {journal} {\bibinfo
  {journal} {JHEP}\ }\textbf {\bibinfo {volume} {02}},\ \bibinfo {pages} {057}
  (\bibinfo {year} {2014})},\ \Eprint {http://arxiv.org/abs/1307.6346}
  {arXiv:1307.6346 [hep-ex]} \BibitemShut {NoStop}%
\bibitem [{\citenamefont {Cacciari}\ \emph {et~al.}(2008)\citenamefont
  {Cacciari}, \citenamefont {Salam},\ and\ \citenamefont
  {Soyez}}]{Cacciari:2008gp}%
  \BibitemOpen
  \bibfield  {author} {\bibinfo {author} {\bibfnamefont {Matteo}\ \bibnamefont
  {Cacciari}}, \bibinfo {author} {\bibfnamefont {Gavin~P.}\ \bibnamefont
  {Salam}}, \ and\ \bibinfo {author} {\bibfnamefont {Gregory}\ \bibnamefont
  {Soyez}},\ }\bibfield  {title} {\enquote {\bibinfo {title} {{The anti-$k_t$
  jet clustering algorithm}},}\ }\href {\doibase 10.1088/1126-6708/2008/04/063}
  {\bibfield  {journal} {\bibinfo  {journal} {JHEP}\ }\textbf {\bibinfo
  {volume} {04}},\ \bibinfo {pages} {063} (\bibinfo {year} {2008})},\ \Eprint
  {http://arxiv.org/abs/0802.1189} {arXiv:0802.1189 [hep-ph]} \BibitemShut
  {NoStop}%
\bibitem [{\citenamefont {Cacciari}\ \emph {et~al.}(2012)\citenamefont
  {Cacciari}, \citenamefont {Salam},\ and\ \citenamefont
  {Soyez}}]{Cacciari:2011ma}%
  \BibitemOpen
  \bibfield  {author} {\bibinfo {author} {\bibfnamefont {Matteo}\ \bibnamefont
  {Cacciari}}, \bibinfo {author} {\bibfnamefont {Gavin~P.}\ \bibnamefont
  {Salam}}, \ and\ \bibinfo {author} {\bibfnamefont {Gregory}\ \bibnamefont
  {Soyez}},\ }\bibfield  {title} {\enquote {\bibinfo {title} {{FastJet User
  Manual}},}\ }\href {\doibase 10.1140/epjc/s10052-012-1896-2} {\bibfield
  {journal} {\bibinfo  {journal} {Eur. Phys. J. C}\ }\textbf {\bibinfo {volume}
  {72}},\ \bibinfo {pages} {1896} (\bibinfo {year} {2012})},\ \Eprint
  {http://arxiv.org/abs/1111.6097} {arXiv:1111.6097 [hep-ph]} \BibitemShut
  {NoStop}%
\bibitem [{\citenamefont {Sirunyan}\ \emph {et~al.}(2021)\citenamefont
  {Sirunyan} \emph {et~al.}}]{CMS:2021ctt}%
  \BibitemOpen
  \bibfield  {author} {\bibinfo {author} {\bibfnamefont {Albert~M}\
  \bibnamefont {Sirunyan}} \emph {et~al.} (\bibinfo {collaboration} {CMS}),\
  }\bibfield  {title} {\enquote {\bibinfo {title} {{Search for resonant and
  nonresonant new phenomena in high-mass dilepton final states at $ \sqrt{s} $
  = 13 TeV}},}\ }\href {\doibase 10.1007/JHEP07(2021)208} {\bibfield  {journal}
  {\bibinfo  {journal} {JHEP}\ }\textbf {\bibinfo {volume} {07}},\ \bibinfo
  {pages} {208} (\bibinfo {year} {2021})},\ \Eprint
  {http://arxiv.org/abs/2103.02708} {arXiv:2103.02708 [hep-ex]} \BibitemShut
  {NoStop}%
\bibitem [{\citenamefont {Hayrapetyan}\ \emph {et~al.}(2025)\citenamefont
  {Hayrapetyan} \emph {et~al.}}]{CMS:2025iix}%
  \BibitemOpen
  \bibfield  {author} {\bibinfo {author} {\bibfnamefont {Aram}\ \bibnamefont
  {Hayrapetyan}} \emph {et~al.} (\bibinfo {collaboration} {CMS}),\ }\bibfield
  {title} {\enquote {\bibinfo {title} {{Search for $t$-channel scalar and
  vector leptoquark exchange in the high-mass dimuon and dielectron spectra in
  proton-proton collisions at $\sqrt{s}$ = 13 TeV}},}\ }\href@noop {} {\
  (\bibinfo {year} {2025})},\ \Eprint {http://arxiv.org/abs/2503.20023}
  {arXiv:2503.20023 [hep-ex]} \BibitemShut {NoStop}%
\bibitem [{\citenamefont {Hayrapetyan}\ \emph {et~al.}(2024)\citenamefont
  {Hayrapetyan} \emph {et~al.}}]{CMS:2024bnj}%
  \BibitemOpen
  \bibfield  {author} {\bibinfo {author} {\bibfnamefont {Aram}\ \bibnamefont
  {Hayrapetyan}} \emph {et~al.} (\bibinfo {collaboration} {CMS}),\ }\bibfield
  {title} {\enquote {\bibinfo {title} {{Search for pair production of scalar
  and vector leptoquarks decaying to muons and bottom quarks in proton-proton
  collisions at s=13\,\,TeV}},}\ }\href {\doibase 10.1103/PhysRevD.109.112003}
  {\bibfield  {journal} {\bibinfo  {journal} {Phys. Rev. D}\ }\textbf {\bibinfo
  {volume} {109}},\ \bibinfo {pages} {112003} (\bibinfo {year} {2024})},\
  \Eprint {http://arxiv.org/abs/2402.08668} {arXiv:2402.08668 [hep-ex]}
  \BibitemShut {NoStop}%
\bibitem [{\citenamefont {Tumasyan}\ \emph {et~al.}(2022)\citenamefont
  {Tumasyan} \emph {et~al.}}]{CMS:2022nty}%
  \BibitemOpen
  \bibfield  {author} {\bibinfo {author} {\bibfnamefont {Armen}\ \bibnamefont
  {Tumasyan}} \emph {et~al.} (\bibinfo {collaboration} {CMS}),\ }\bibfield
  {title} {\enquote {\bibinfo {title} {{Inclusive nonresonant multilepton
  probes of new phenomena at $\sqrt s$=13\,\,TeV}},}\ }\href {\doibase
  10.1103/PhysRevD.105.112007} {\bibfield  {journal} {\bibinfo  {journal}
  {Phys. Rev. D}\ }\textbf {\bibinfo {volume} {105}},\ \bibinfo {pages}
  {112007} (\bibinfo {year} {2022})},\ \Eprint
  {http://arxiv.org/abs/2202.08676} {arXiv:2202.08676 [hep-ex]} \BibitemShut
  {NoStop}%
\bibitem [{\citenamefont {Allwicher}\ \emph {et~al.}(2021)\citenamefont
  {Allwicher}, \citenamefont {Arnan}, \citenamefont {Barducci},\ and\
  \citenamefont {Nardecchia}}]{Allwicher:2021rtd}%
  \BibitemOpen
  \bibfield  {author} {\bibinfo {author} {\bibfnamefont {Lukas}\ \bibnamefont
  {Allwicher}}, \bibinfo {author} {\bibfnamefont {Pere}\ \bibnamefont {Arnan}},
  \bibinfo {author} {\bibfnamefont {Daniele}\ \bibnamefont {Barducci}}, \ and\
  \bibinfo {author} {\bibfnamefont {Marco}\ \bibnamefont {Nardecchia}},\
  }\bibfield  {title} {\enquote {\bibinfo {title} {{Perturbative unitarity
  constraints on generic Yukawa interactions}},}\ }\href {\doibase
  10.1007/JHEP10(2021)129} {\bibfield  {journal} {\bibinfo  {journal} {JHEP}\
  }\textbf {\bibinfo {volume} {10}},\ \bibinfo {pages} {129} (\bibinfo {year}
  {2021})},\ \Eprint {http://arxiv.org/abs/2108.00013} {arXiv:2108.00013
  [hep-ph]} \BibitemShut {NoStop}%
\bibitem [{\citenamefont {Allwicher}\ \emph {et~al.}(2025)\citenamefont
  {Allwicher}, \citenamefont {Faroughy}, \citenamefont {Martines},
  \citenamefont {Sumensari},\ and\ \citenamefont {Wilsch}}]{Allwicher:2024mzw}%
  \BibitemOpen
  \bibfield  {author} {\bibinfo {author} {\bibfnamefont {Lukas}\ \bibnamefont
  {Allwicher}}, \bibinfo {author} {\bibfnamefont {Darius~A.}\ \bibnamefont
  {Faroughy}}, \bibinfo {author} {\bibfnamefont {Matheus}\ \bibnamefont
  {Martines}}, \bibinfo {author} {\bibfnamefont {Olcyr}\ \bibnamefont
  {Sumensari}}, \ and\ \bibinfo {author} {\bibfnamefont {Felix}\ \bibnamefont
  {Wilsch}},\ }\bibfield  {title} {\enquote {\bibinfo {title} {{On the EFT
  validity for Drell{\textendash}Yan tails at the LHC}},}\ }\href {\doibase
  10.1140/epjc/s10052-025-14171-4} {\bibfield  {journal} {\bibinfo  {journal}
  {Eur. Phys. J. C}\ }\textbf {\bibinfo {volume} {85}},\ \bibinfo {pages} {463}
  (\bibinfo {year} {2025})},\ \Eprint {http://arxiv.org/abs/2412.14162}
  {arXiv:2412.14162 [hep-ph]} \BibitemShut {NoStop}%
\bibitem [{\citenamefont {Aebischer}\ \emph {et~al.}(2025)\citenamefont
  {Aebischer}, \citenamefont {Buras},\ and\ \citenamefont
  {Kumar}}]{Aebischer:2025qhh}%
  \BibitemOpen
  \bibfield  {author} {\bibinfo {author} {\bibfnamefont {Jason}\ \bibnamefont
  {Aebischer}}, \bibinfo {author} {\bibfnamefont {Andrzej~J.}\ \bibnamefont
  {Buras}}, \ and\ \bibinfo {author} {\bibfnamefont {Jacky}\ \bibnamefont
  {Kumar}},\ }\bibfield  {title} {\enquote {\bibinfo {title} {{SMEFT ATLAS: The
  Landscape Beyond the Standard Model}},}\ }\href@noop {} {\  (\bibinfo {year}
  {2025})},\ \Eprint {http://arxiv.org/abs/2507.05926} {arXiv:2507.05926
  [hep-ph]} \BibitemShut {NoStop}%
\end{thebibliography}%

\end{document}